\documentclass[acmsmall]{acmart}

\usepackage{tikz}
\usepackage{url}
\usepackage{pdfpages}
\usepackage{afterpage}
\usepackage{color, colortbl}
\usepackage{graphicx}
\usepackage{amsmath}
\usepackage{amsfonts}
\usepackage{import}
\usepackage{pifont}
\usepackage{multirow}
\usepackage{makecell}
\usepackage{subfigure}
\usepackage{footmisc}
\usepackage{mathtools}
\usepackage{multicol,multirow}
\usepackage{fancyhdr} 
\usepackage{algorithm}
\usepackage[noend]{algpseudocode}
\usepackage{amsthm}
\usepackage{bm}
\usepackage{threeparttable}

\theoremstyle{definition}

\def\PROPROC{{%
    \setbox0\hbox{$\prod$}%
    \rlap{\hbox to \wd0{\hss $\times$ \hss}}\box0
}}

\def\HAMPROC{{%
    \setbox0\hbox{$\prod$}%
    \rlap{\hbox to \wd0{\hss $\circ$ \hss}}\box0
}}

\AtBeginDocument{%
  \providecommand\BibTeX{{%
    \normalfont B\kern-0.5em{\scshape i\kern-0.25em b}\kern-0.8em\TeX}}}

\begin{document}

\title{Communication-Efficient Data Parallel Distributed Deep Learning: A Comprehensive Survey}



\author{Zhenheng Tang}
\affiliation{
  \institution{Hong Kong Baptist University}
\city{}
\state{}
\country{}
}
\email{zhtang@comp.hkbu.edu.hk}

\author{Shaohuai Shi}
\affiliation{
  \institution{Harbin Institute of Technology, Shenzhen}
\city{}
\state{}
\country{}
}
\email{shaohuais@hit.edu.cn}

\author{Wei Wang}
\affiliation{
  \institution{The Hong Kong University of Science and Technology}
\city{}
\state{}
\country{}
}
\email{weiwa@cse.ust.hk}

\author{Bo Li}
\affiliation{
 \institution{The Hong Kong University of Science and Technology}
\city{}
\state{}
\country{}
}
 \email{bli@cse.ust.hk}

\author{Xiaowen Chu}
\affiliation{%
  \institution{The Hong Kong University of Science and Technology (Guangzhou)}
\city{}
\state{}
\country{}
}
\email{xwchu@ust.hk}

\renewcommand{\shortauthors}{Z. Tang, S. Shi, W. Wang, B. Li, and X. Chu.}

\begin{abstract}
Distributed deep learning (DL) has become prevalent in recent years to reduce training time by leveraging multiple computing devices (e.g., GPUs/TPUs) due to larger models and datasets. However, system scalability is limited by communication becoming the performance bottleneck. Addressing this communication issue has become a prominent research topic. In this paper, we provide a comprehensive survey of the communication-efficient distributed training algorithms, focusing on both system-level and algorithmic-level optimizations. 
We first propose a taxonomy of data-parallel distributed training algorithms that incorporates four primary dimensions: communication synchronization, system architectures, compression techniques, and parallelism of communication and computing tasks. We then investigate state-of-the-art studies that address problems in these four dimensions. We also compare the convergence rates of different algorithms to understand their convergence speed. Additionally, we conduct extensive experiments to empirically compare the convergence performance of various mainstream distributed training algorithms. Based on our system-level communication cost analysis, theoretical and experimental convergence speed comparison, we provide readers with an understanding of which algorithms are more efficient under specific distributed environments. Our research also extrapolates potential directions for further optimizations.
\end{abstract}




\maketitle


\section{Introduction}\label{sec:intro}
\begin{figure*}[h!]
    \subfigbottomskip=-1pt
    \subfigcapskip=1pt
  \centering
     \subfigure[Memory of GPUs and neural networks.]{\includegraphics[width=0.4\textwidth]{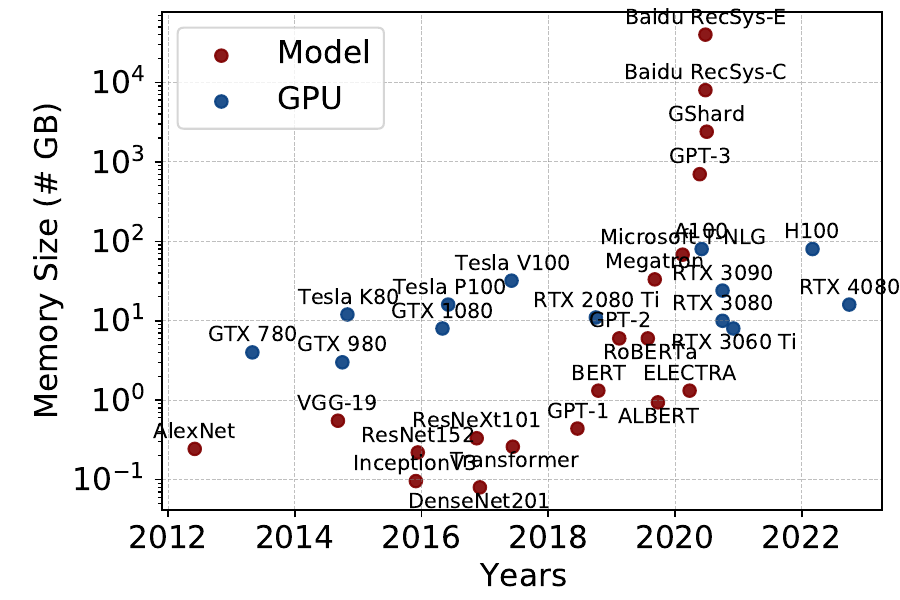}}
     \subfigure[FLOPS of GPU and FLOPs of training models.]{\includegraphics[width=0.4\textwidth]{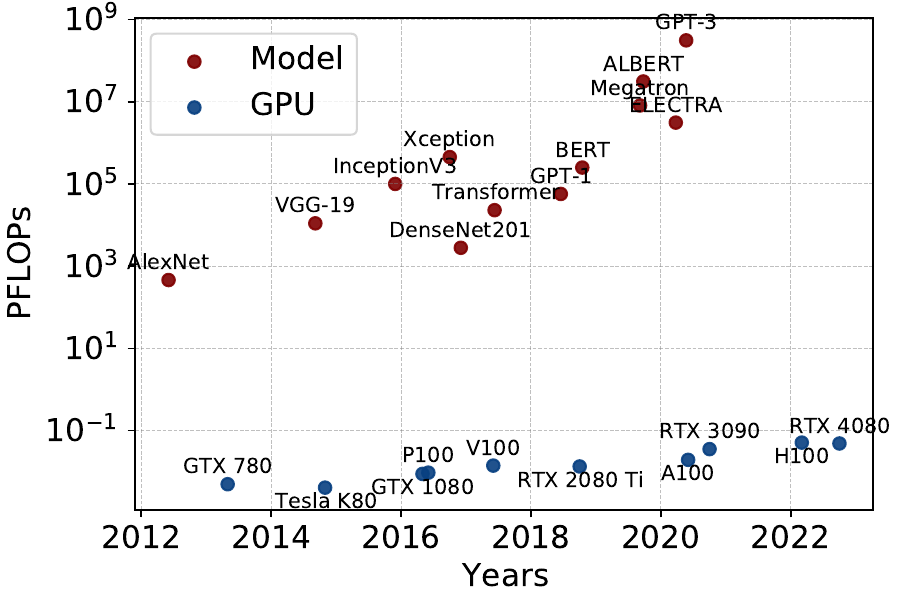}}
    \caption{The trends of GPU and neural networks.}
    \label{fig:TrendsGPUMODEL}
\vspace{-0.0cm}
\end{figure*}

Deep learning (DL) has made significant progress in recent years. Researchers and engineers have applied DL technologies to tackle intricate problems across various fields, including but not limited to computer vision~\citep{resnet}, natural language processing~\citep{attention,bert,yang2020survey}, speech recognition~\citep{deepspeech} and many others. 
DL typically involves increased sizes of training datasets and model parameters in deep neural networks (DNNs) to enhance the predictive performance in different applications, such as accuracy in classification tasks~\citep{ml1991,Russakovsky2015ILS28465472846559,8237359}. However, as data size and model complexity increase, the training process becomes exceedingly computationally intensive and time-consuming. For example, training a state-of-the-art ResNet-50~\citep{resnet} model (in 90 epochs) on the ImageNet dataset~\citep{Imagenet} using a latest Nvidia Tesla V100 GPU requires approximately two days~\citep{wang2019performance}. As Figure~\ref{fig:TrendsGPUMODEL} shows, however, the development speed of GPU FLOPs and memory cannot catch up the development of newly large neural networks, like GPT-3~\citep{gpt3}, GShard~\citep{lepikhin2021gshard} and Baidu RecSys~\citep{zhao2020distributed}. In addition, hyper-parameter tuning is necessary to achieve satisfactory results for certain tasks, which further demands significant time and financial investment.

To mitigate the time-consuming training process, two approaches have gained traction: (1) maximizing the utilization of a single accelerator's computing power by implementing highly optimized software~\citep{chetlur2014cudnn,shi2016benchmarking,xu2017performance,sze2017efficient,yan2020optimizing}, and (2) employing distributed training~\citep{padam,disml,sparknet,PSGD} to accelerate the training process by making use of multiple processors, including CPUs~\citep{you2018imagenet}, GPUs~\citep{goyal2017accurate,shi2018performance,jia2018highly} and TPUs~\citep{you2019large}.
Intuitively, multiple processors working collaboratively in training one model can reduce the overall training time. However, the communication cost between processors typically restricts system scalability~\citep{shi2018performance}. For example, when deploying the high-speed computing devices (e.g., Nvidia A100/H100 GPUs) with low-speed interconnects (e.g., PCIe or 10GbE) to collaboratively train a deep model, where the computation-to-communication ratio is low~\cite{shi2020quantitative}, using multiple processors could result in very low hardware utilization~\citep{DGC}. And the collaborative training between different geographically distributed GPUs suffers from higher communication costs, because the wide-area networks have lower communication bandwidth~\citep{tang2022gossipfl}. Therefore, different parallel algorithms should be carefully designed to leverage the computing power of distributed clusters.


\textbf{Data parallelism}~\cite{dean2012large,goyal2017accurate} is a widely employed distributed training technique. This approach involves replicating the model parameters to all computing workers. During a single iteration, each worker computes the local gradient or model updates by sampling different mini-batches of data. Workers then exchanges the results with the other workers. Following this, aggregation and broadcast operations are executed to obtain the new global model.
 


\textbf{Model parallelism}~\citep{dean2012large,3154842,Mirhoseini2017} is another distributed training approach that involves partitioning model parameters among multiple computing workers. Every worker holds different parameters or layers of the model (i.e., a subset of $x$). In this paper, we focus mainly on techniques associated with data parallelism. Note that some of them could be used in parallel to model parallelism.

The term of \textbf{pipeline parallelism}~\citep{shi2018adag,zhang2017poseidon,MLSYS2022_cedebb6e,harlap2018pipedream,huang2019gpipe,shoeybi2019megatron,li2021chimera} can refer to different techniques under different contexts. In the context of data parallelism, pipelining means to execute the computational tasks and communication tasks simultaneously so as to reduce the iteration time. In the context of model parallelism, it refers to the technique that workers are assigned different stages of model training to execute, and intermediate results are transmitted between workers to hide some training time. In this paper, we mainly focus on the pipelining techniques deployed in data parallelism, which is used to reduce the communication overhead.

This survey provides a comprehensive overview and taxonomy of research works that aim to enhance communication efficiency in distributed training. We deconstruct the distributed training framework into several orthogonal components, including \textit{communication synchronization}, \textit{system architectures}, \textit{compression techniques} and \textit{parallelism of communication and computation}. Additionally, we offer an \textit{overview of convergence analysis} of these algorithms, examining the trade-off between communication efficiency and convergence. Furthermore, we develop a \textit{benchmark framework} of mainstream distributed training algorithms, including different synchronization schemes, communication topology, compression algorithms, and various numbers of workers, providing the practical reference to readers. Through this survey, we hope that readers can gain insight into the advancements made in this field and be inspired to develop new efficient distributed training algorithms and frameworks.

\subsection{Related Work}\label{sec:related}
There exist several surveys providing an introduction and review to distributed machine or deep learning algorithms. Peteiro-Barral et al.~\citep{Peteiro2013} introduced distributed machine learning algorithms for big data. Xing et al.~\citep{XING2016179} mainly focused on different synchronization schemes, scheduling, and balancing workloads and communication typologies. Ben-Nun et al.~\citep{10.1145/3320060} focused on DNN operators and approaches for parallelism. Guo et al.~\citep{DBLP:abs-1808-04752} gave a thorough review of different quantized neural networks. Meanwhile, Zhang et al.~\citep{8644613} provided a brief overview of large-scale distributed DL systems, including parallelism, parameter server architectures, synchronization schemes, related applications, and platforms.

Our article, in contrast to these previous surveys, concentrates specifically on communication-efficient distributed DL training. We present a detailed discussion of communication compression techniques that have not been fully demystified in the previous surveys. Additionally, we provide a quick review of auxiliary technologies and offer a comparison of convergence bounds. Moreover, we conduct extensive experiments to compare the performance of different mainstream distributed training algorithms, which provides readers with an empirical understanding of these algorithms.

\subsection{Organization}
This survey aims to provide a comprehensive analysis of communication-efficient distributed training algorithms with data parallelism, focusing on four key aspects: communication synchronization, system architectures, compression techniques, and scheduling methods. The rest of the paper is organized as follows. In Section \ref{sec:problemformulation}, we illustrate the key issues of distributed training and propose a taxonomy of related research to summarize existing methods. We discuss the synchronous and asynchronous frameworks in Section \ref{sec:async}, followed by an overview of the system architectures that support gradient/model aggregation in Section \ref{sec:cenvsdecen}. In Section \ref{sec:quantization} and Section \ref{sec:sparsification}, we introduce the techniques in reducing the communication traffic with gradient/model compression. The scheduling methods are introduced in Section \ref{sec:scheduling}. We summarize the theoretical convergence bounds in Section \ref{sec:convergence}. Additionally, we present some auxiliary tricks to train deep models with communication-efficient algorithms in Section \ref{sec:auxiliary}. Finally, we conclude the paper in Section \ref{sec:conclusion}.

\subsection{Benchmark Framework and Experiment Configuration}\label{sec:overall-exp-conf}
Our benchmark framework is built upon FedML~\citep{chaoyanghe2020fedml} and MPI~\citep{mpi4py}. Our framework provides users with flexible and scalable application programming interfaces (APIs) for testing and developing novel distributed training algorithms. Multiple algorithms have already been developed with distinct synchronous schemes, communication topologies, and compression techniques. These algorithms can be combined to create new ones that are mostly independent of each other.

We conduct experiments of different distributed training algorithms on two typical DL tasks. One is the image classification on CIFAR-10~\citep{krizhevsky2010cifar} using ResNet-20~\citep{resnet}. The other is Shakespeare, a natural language processing task based on the dataset obtained from \textit{The Complete Works of William Shakespeare}. The model used for Shakespeare is a stacked character-level LSTM language model, proposed by~\citep{FederatedLearning}. We conduct experiments with different workers ($4 \sim 32$) to assess the scalability of different algorithms. Each worker is equipped with a NVIDIA RTX 2080 Ti and Pytorch version is V1.7. Note that the hardware platform doesn't affect the convergence performance and model test accuracy. The results of the experiments are presented at the end of each chapter, providing the readers with an accurate and practical understanding of the algorithms.

\section{Taxonomy of Distributed DL}\label{sec:problemformulation}

The mathematical formulation of training DL models can be defined as an optimization problem
\begin{equation}\label{eq:single_dl}
    \min \limits_{\mathbf{x}\in \mathbb{R}^N }f(\mathbf{x}):=\mathbb{E}_{\xi_i \sim \mathcal{D}} F(\mathbf{x};\xi_i),
\end{equation}
where the random variable $\xi_i$ follows a probability distribution $\mathcal{D}$, which denotes the data samples from a given dataset, $\mathbf{x}$ represents all the parameters of the model, $N$ is the number of parameters, and $F:\mathbb{R}^N \to \mathbb{R}$ denotes the objective function with respect to $\mathbf{x}$ and $\xi_i$, $f$ the expectation of $F$.  

Gradient-based optimization algorithms are commonly used in DL. Due to the high computational complexity of second-order gradient descent methods~\cite{martens2015optimizing,shi2021accelerating,zhang2023eva} with DNNs, the first-order gradient descent methods, particularly the stochastic gradient descent (SGD) with mini-batch\footnote{To simplify, throughout this paper, we use SGD to denote the gradient descent with mini-batch which includes the cases of one sample and more than one samples in each training iteration.} and its variants (e.g., Adam), are commonly used. The update rule of \textbf{single-device} SGD is as follows. 
\begin{equation}
    G_{t}(\mathbf{x}_{t}) = {\nabla}F_{t}(\mathbf{x}_{t};\xi_{t})
\end{equation}
\begin{equation}\label{eq:sgd}
    \mathbf{x}_{t+1} = \mathbf{x}_{t} - {\gamma}G_{t}(\mathbf{x}_{t}),
\end{equation}
where $\mathbf{x}_t \in \mathbb{R}^N$ is the $N$-dimensional model parameter at iteration $t$, $\xi_{t}$ is a randomly sampled mini-batch of data, and $\gamma$ is the learning rate (or step size). SGD is an iterative algorithm and has been proved that it can solve \eqref{eq:single_dl} under the assumptions that $f_s(\mathbf{x})$ is non-convex and is with L-Lipschitzian gradients~\citep{Bottou2016OptimizationMF}. The iterative process generally contains several steps: 1) It samples a mini-batch of \textbf{data} $\xi_{t}$. 2) It performs the \textbf{feed-forward} computations to evaluate the objective function $F_{t}(\mathbf{x}_{t};\xi_{t})$). 3) It performs \textbf{backward propagation} to calculate the gradients, ${\nabla}F_{t}(\mathbf{x}_{t};\xi_{t})$ with respect to model parameters. 4) Finally, it \textbf{updates} model parameters by Eq. \eqref{eq:sgd}. 


Distributed training modifies the above four basic single-device SGD procedures into distributed versions. We firstly describe the procedures of the widely used distributed     training algorithm, \textbf{bulk synchronous parallel SGD} (BSP-SGD)~\citep{BSP7917379181}, and then elucidate how other algorithms differ from it. In BSP-SGD, each worker (indexed as $i$) has an identical global model (e.g.., downloads from the Parameter Server (PS)), and then processes a different subset of data (i.e., $\xi_{i,t}$) to independently compute gradients (${\nabla}F(\mathbf{x}_t;\xi_{i,t})$). These computations are performed simultaneously among all workers. After computing gradients, workers aggregate them through the PS or an all-reduce operation, with a synchronization to update the model parameters, and then proceed to the next iteration. 
The update rule of BSP-SGD can be formulated as: 
\begin{align}
    G_{i,t}(\mathbf{x}_{t})  = {\nabla}F_{i,t}(\mathbf{x}_{t};\xi_{i,t}), \label{eq:ssgd_gra} \\  
    \mathbf{x}_{t+1}  = \mathbf{x}_{t} - {\gamma}\frac{1}{n}\sum_{i=1}^{n}G_{i,t}(\mathbf{x}_{t}), \label{eq:ssgd_upd}
\end{align}
where $G_{i,t}(\mathbf{x}_{t})$ represents the gradient of $F_{i,t}(\mathbf{x}_{t})$ of worker $i$ at iteration $t$, $n$ the number of workers.

\begin{figure*}[t]
	\centering
	\includegraphics[width=0.9\linewidth]{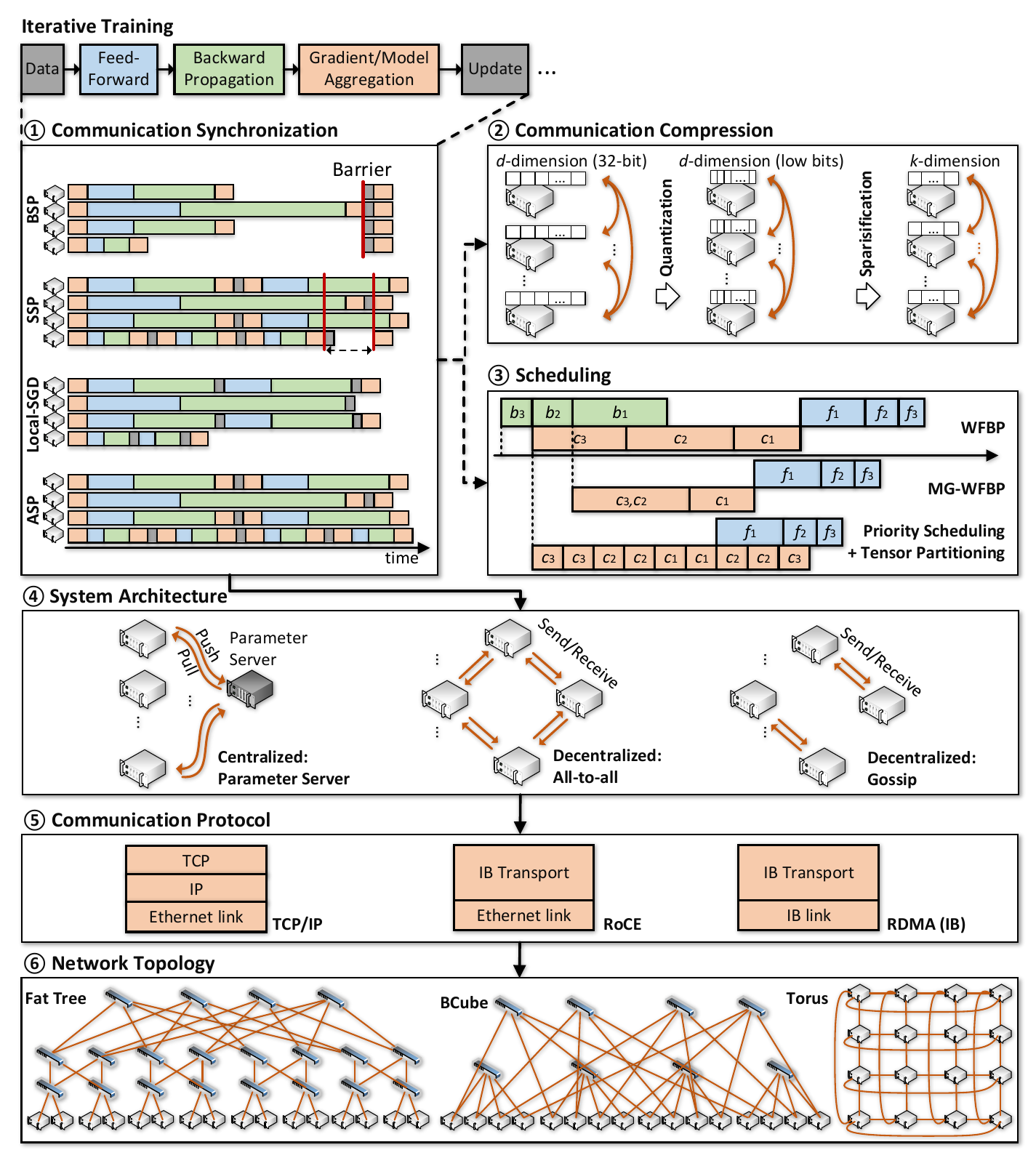}
	\vspace{-10pt}
	\caption{Overview of data-parallel distributed deep learning.}
	\label{fig:overview}
\vspace{-10pt}
\end{figure*}

We summarize and classify algorithms that improve BSP-SGD in four orthogonal dimensions shown in Table~\ref{tab:Taxonomy_DisSGD}. Furthermore, we provide an overview of communication-efficient algorithms for distributed DL in Fig. \ref{fig:overview}, which highlights the main techniques employed in each dimension.
\begin{enumerate}
    \item \textbf{Flexible synchronization} aims to relax the strict synchronization constraints of BSP to reduce the impact of synchronization and the number of communications in the same period (Fig. \ref{fig:overview}.\ding{192} and Section \ref{sec:async}).
    \item \textbf{Different system architectures} propose changing the communication topology to avoid the communication congestion of the PS and workers.  (Fig. \ref{fig:overview}.\ding{195} and Section \ref{sec:cenvsdecen})
    \item \textbf{Compression techniques} explore to compress the communication data, thereby reduce communication traffic and communication time.  (Fig. \ref{fig:overview}.\ding{193}, Section \ref{sec:quantization} and \ref{sec:sparsification})
    \item \textbf{The parallelism of communication and computing} seeks to hide the communication time to achieve shorter iteration time.  (Fig. \ref{fig:overview}.\ding{194} and Section \ref{sec:scheduling})
\end{enumerate}
The communication protocols (Fig. \ref{fig:overview}.\ding{196}) and the network topology (Fig. \ref{fig:overview}.\ding{197}) are also important factors that influence the communication efficiency at the hardware-level. We mainly discuss the algorithm-level methods and protocols in this survey.

\begin{table}[t]
  \centering
  \fontsize{7}{7}\selectfont
  \begin{threeparttable}
  \caption{Taxonomy of Distributed SGD}
  \label{tab:Taxonomy_DisSGD}
    \begin{tabular}{ccc}
    \toprule
    Dimension & Method & Characteristic\cr
    \midrule
    \multirowcell{4}{\bf Communication \\ \bf Synchronization} & Synchronous & Frequent communications and synchronization \cr
    & Stale-Synchronous & Trade-off between Synchronous and Asynchronous \cr
    & Asynchronous & No need of synchronization \cr
    & Local SGD  & Less frequent communications \cr
    \hline
    \multirowcell{3}{\bf System\\\bf Architectures}
    & Parameter-Server  & Centralized topology \cr
    & All-Reduce  & Decentralized topology and collective communication \cr
    & Gossip  & Decentralized topology and peer-to-peer communication \cr
    \hline
    \multirowcell{2}{\bf Compression \\ \bf Techniques} & Quantization  &  Communicate low-precision parameters \cr
    & Sparsification  & Communicate selected parameters\cr
    \hline    
    \multirowcell{2}{\bf Parallelism of\\\bf Communication and Computing} & Pipelining  & Hide the communication or computation time \cr
     & Scheduling  &  Dynamically schedule the computing and communication tasks\cr    
    \bottomrule
    \end{tabular}
    \end{threeparttable}
\vspace{-10pt}
\end{table}
\vspace{-0.1cm}

\section{Synchronous/asynchronous framework}\label{sec:async}
\begin{figure*}[t]
	\centering
	\includegraphics[width=0.9\linewidth]{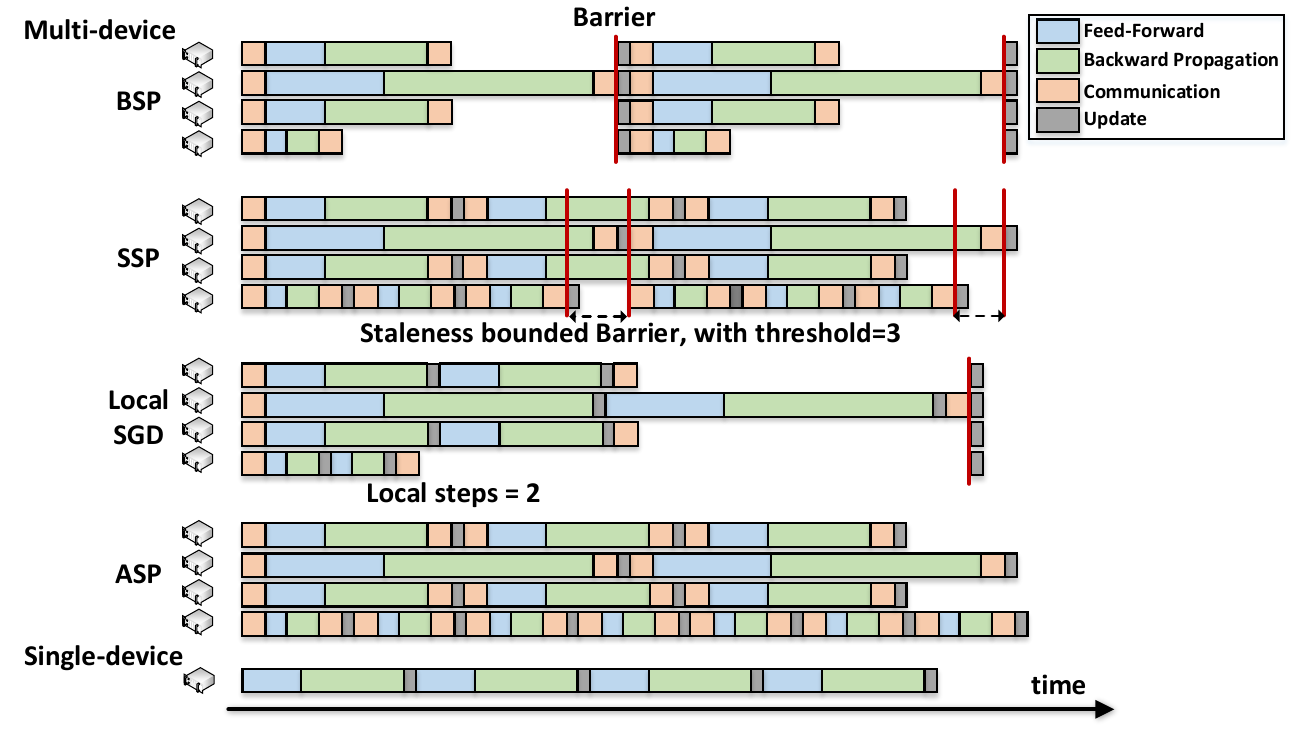}
	\vspace{-15pt}
	\caption{Comparison of training with single device and multiple devices under different communication synchronization with the PS architecture.}
	\label{fig:timeline}
\vspace{-10pt}
\end{figure*}

In the data parallel distributed DL, \textbf{synchronization} in BSP-SGD refers to the process in which all workers should be synchronized to complete the transmission of all parameters or gradients before proceeding to the next training round. Flexible synchronization, which relaxes the strict synchronization of BSP-SGD, affects not only communication traffic but also the performance and convergence of model training. Therefore, there is a trade-off between communication traffic and convergence. Moreover, different synchronization schemes can be combined with different architectures. In this section, we describe four representative synchronization schemes under the PS architecture, which has the most extensive range of applications. The timeline of the four different schemes is shown in Fig. \ref{fig:timeline}.


\subsection{Synchronous Framework}

The classical synchronous framework, namely BSP, as mentioned in section~\ref{sec:problemformulation}, involves the following steps: 1) Data loading; 2) Feed-forward computations; 3) Backward propagation computations; 4) Gradient aggregation with a barrier; and 5) Model updating with aggregated gradients. The step (4) requires synchronization of all workers, leading to the \textit{straggler problem}, where few slow workers significantly affect the system throughput~\citep{AsynDisADMM}. Additionally, the aggregation of all gradients leads to high communication costs, severely limiting the scalability of the system.


In the parameter server (PS) architecture (\S\ref{subsec:parameterservers}), the synchronous barrier is enforced until all workers finish transmitting their parameters to the parameter servers~\citep{Krizhevsky2014OneWT,Bradley2011,552669,375451}. Similarly, in the All-Reduce architecture (\S\ref{subsec:allreduce}), synchronization requires that all workers wait for the completion of the all-reduce operation (i.e., gradient aggregation), ensuring that all workers have the same updated global model~\citep{Zou2014MarianaTD,642949,DGC}. In contrast, for the decentralized architecture (\S\ref{subsec:gossip}), synchronization involves waiting for the completion of communication. Unlike in All-Reduce, workers in the decentralized architecture do not necessarily maintain the same model~\citep{CanDecent,Ram2008DistributedSS}.

\subsection{Stale-synchronous Framework}
The stale-synchronous parallel (SSP) framework~\citep{MoreEffDMLviaStaleSync} aims to mitigate the straggler problem with relaxed synchronization. Specifically, SSP allows faster workers to perform more updates than slower ones, reducing the waiting time of faster workers, as illustrated in Fig. \ref{fig:timeline}. However, to maintain model consistency and ensure convergence, SSP imposes a staleness bounded barrier that limits the iteration gap between the fastest and slowest workers. For a maximum staleness bound $s$, the update formula of worker $i$ at iteration $t+1$ is modified to
\begin{align}
    \mathbf{x}_{i,t+1} =  \mathbf{x}_{0} - {\gamma}\sum_{k=1}^{t-s-1}\sum_{j=1}^{n}G_{j,k}(\mathbf{x}_{j,k})- {\gamma}\sum_{k=t-s}^{t}G_{i,k}(\mathbf{x}_{i,k}) - {\gamma}\sum_{(j,k)\in\mathcal{S}_{i,t+1}}G_{j,k}(\mathbf{x}_{j,k}), \label{eq:SSGSGD}
\end{align}
where $\mathcal{S}_{i,t+1}$ is some subset of the updates from other workers during period $\left[ t-s, t\right]$, and $n$ represents the number of workers. The historical updates added to $\mathbf{x}_{i,t+1}$ consist of three terms: pre-window, read-my-writes, in-window updates, represented by second, third and forth terms in the right side in equation (\ref{eq:SSGSGD}). The pre-window update represents the synchronized gradients which are obtained from all workers, the read-my-writes update is local gradients, the in-window update is the gradients from other workers. Less $s$ means more timely gradient synchronization but less communication efficiency. Determining a proper $s$ is challenging to achieve good end-to-end training performance~\citep{MoreEffDMLviaStaleSync}. To this end, instead of setting a staleness bound, Chen et al.~\citep{RevistSynSGD} proposed the backup worker scheme. Specifically, a subset of workers (called backup workers) are used to compute the mini-batch gradient. The PS updates parameters without waiting for all gradients. The gradients from slowest workers are dropped directly.


In addition, there is a congestion problem in SSP with parameter servers. Chen et al.~\citep{8737587} proposed a Round-Robin Synchronization Parallel (R$^2$SP) method to address it. R$^2$SP staggers worker updates throughout the training process and coordinates workers to update gradients in a fixed round-robin order to evenly distribute the communication load and reduce congestion.


\subsection{Asynchronous Framework}

The asynchronous parallel SGD (ASP-SGD) framework allows the server to update the global model with the updates from a part of workers instead of all workers~\citep{hogwild,muli2013,Tamingwild,AsynDisMLspar,NEURIPS2022_029df12a}. ASP-SGD enables more independent updates of the nodes and reduces one-round data transmission during communication between the workers and the PS as shown in Fig. \ref{fig:timeline}. In ASP-SGD, each worker sends its gradients to the PS after gradient calculation. Then the PS updates the global model without waiting for the other workers. Thus, asynchronous frameworks get rid of straggler problems. Note that ASP-SGD is not friendly to the All-Reduce architecture. The update formula of ASP-SGD can be summarized as
\begin{equation}\label{eq:asyncsgd}
    \mathbf{x}_{t+1} =  \mathbf{x}_{t} - {\gamma}\sum_{i=1}^{n}G_{i,t-\tau_{i,k}}(\mathbf{x}_{i,t-\tau_{k,i}}),
\end{equation}
where the $\tau_{k,i}$ represents the period between when worker $i$ calculates the gradient and when the server conducts SGD.

Distributed Alternating Direction Method of Multipliers (D-ADMM)~\citep{Dadmm,convergeceAsynDisADMM} is an early work that proposed asynchronous updating on different parameters of an optimization problem. However, it requires the maintenance of a global clock, and each group of workers in D-ADMM needs to be aware of each other’s progress. Moreover, D-ADMM heavily depends on the network topology and requires a central node to keep the global model while the central node still needs to wait for all worker nodes to finish their tasks, which is akin to the synchronous framework. Moreover,  ~\citep{Dadmm,convergeceAsynDisADMM} do not consider the optimization of neural networks. 

To address these limitations, Zhang et al. ~\citep{AsynDisADMM} proposed an asynchronous distributed ADMM using the star topology and the PS architecture. Although asynchronous training has been demonstrated to be faster than synchronous training in the absence of slow workers, it tend to have inferior convergence performance than synchronous optimization. Thus, ~\citep{AsynDisADMM} employed partial barrier and bounded delay as two conditions to control the asynchrony, trying to obtain similar convergence of the synchronous optimization.

DistBelief~\citep{dean2012large} is an early asynchronous framework that is capable of harnessing computing clusters with thousands of machines for large-scale model training. It proposes \textit{Downpour SGD}, an ASP-SGD optimization method, which involves multiple workers processing data in parallel to compute their own updates and communicating with the PS. Later,  Li et al.~\citep{communicationDisMLwithPS} proposed an efficient algorithm named Delayed Block Proximal Gradient Method. In this algorithm, only a block of parameters is asynchronously updated per iteration. Consequently, only a portion of the parameters is required to be transmitted between the master and workers, and the waiting is not necessary. To further reduce the communication costs, Grishchenko et al.~\citep{AsynDisMLspar} developed an asynchronous distributed algorithm that incorporates sparsification of upward communications (workers-to-master). Sparsification (\S\ref{sec:sparsification}) is implemented by uniformly sampling a selection of local update entries to enhance communication efficiency.


Asynchronous frameworks provide better system performance by addressing straggler problems. Nevertheless, asynchronous algorithms suffer from inferior convergence performance. Consequently, synchronous SGD remains the state-of-the-art method in the data center setting if workers have uniform hardware and workloads~\citep{RevistSynSGD}.



\subsection{Local SGD}
Local-SGD~\citep{Zhang2014DSM,bijral2016data,Zhang2016ParallelSW,NIPS2019_9288,McDonald2010DistributedTS,NIPS2009_3881,6853589,Zhang2015DLE296,Yu2018ParallelRS,spiridonoff2021communicationefficient} is another set of algorithms that rely on strict synchronization but permit flexible communication frequencies. In Local-SGD, each worker independently executes several or more iterations before averaging all local models to obtain the most recent global model. Model Average~\citep{FederatedLearning} is a similar approach that performs several local iterations and synchronizing the model. The procedure of Local-SGD can be formalized as
\begin{equation}\label{eq:localsgd}
    \mathbf{x}_{i, t+1} = \left \{
    \begin{array}{ll}
    \mathbf{x}_{i,t} - {\gamma}G_{i,t}(\mathbf{x}_{i,t}) ,& \text{if} \ t + 1 \not\in \mathcal{I}_T\\
    \mathbf{x}_{i,t} - {\gamma}\frac{1}{n}\sum_{i=1}^{n}G_{i,t}(\mathbf{x}_{i,t}), & \text{if} \ t + 1 \in \mathcal{I}_T
    \end{array}
    \right .
\end{equation}
where $\mathcal{I}_T$ represents the synchronization timestamps.


While Local-SGD allows for flexible communication frequency to reduce the overall communication overhead, reducing it excessively can lead to a decline in convergence performance. Consequently, Local-SGD often requires an increased number of training iterations to attain comparable model accuracy to that of BSP-SGD, thereby potentially slowing down the training process. Therefore, it is important to find the optimal communication period that strikes a balance between communication overhead and convergence performance. To improve the performance of Local-SGD, Yu et al.~\citep{pmlrv97yu19d} combined distributed momentum SGD and Local-SGD achieving a linear speedup in training. Jiang et al.~\citep{AlinearSpeedupAnalysis} reduced communication complexity by exploiting the quantization method (\S\ref{sec:quantization}) in combination with Local SGD. 

Increasing batch sizes can also reduce the number of iterations needed to reduce communication data. CR-PSGD-Catalyst~\citep{pmlrv97yu19c} proposed to dynamically increase batch sizes after each training iteration, while guaranteeing the same convergence rate of SSP. However, large-batch SGD can lead to decreased generalization performance ~\citep{chen2016scalable,NIPS2019_8452,Hoffer2017TLG32947713294936,DBLPconficlrKeskarMNST17,DBLPjournalscorrabs181103600}. To address this issue, Lin et al.~\citep{Lin2018DontUL} proposed \textit{post-local SGD} to allow one to scale the training onto much more parallel computing devices. This algorithm divides the whole training process into two phases, using mini-batch SGD in the first phase and Local-SGD in the second phase.

Lazy Aggregation (LAG)~\citep{LAG,LENA} delays the gradient uploading if the new gradient of a worker is similar with the old gradient of the last iteration. And the server will use the old gradient of the worker $i$ as the new gradient to conduct aggregation. This can be seen as a special Local-SGD, as it skips the communication round. The 3PC~\citep{3PC} designs a novel compressor which unifies the error-feedback (\S\ref{sec:auxiliary}) and LAG together, and obtains higher compression ratio and faster convergence.

FedAvg~\citep{FederatedLearning,tang2023fedml,tang2022gossipfl,tang2022virtual} is another representative Local-SGD algorithm, which is a fundamental method in federated learning~\cite{FederatedLearning,kairouz2019advances}. In FedAvg, the server randomly selects a part of clients to conduct local training at each communication round. Different from setting a fixed number of iterations in Local-SGD, clients in FedAvg conduct local training on all local samples for one epoch or several epochs. Thus, the number of iterations in FedAvg actually is proportional to the size of local datasets. After local training, the server collects local models for aggregation. 

\subsection{Experimental Comparison}
\begin{table*}[]
\begin{threeparttable}
\centering
\fontsize{8}{8}\selectfont
\caption{Test accuracy [\%] comparison of different communication synchronization schemes.}
\vspace{-0.3cm}
\label{tab:exp-sync}
\begin{tabular}{cc|cccc|cccc}
\toprule
\multirow{2}{*}{Algorithm}           & \multirow{2}{*}{\# of Worker} & \multicolumn{4}{c|}{Resnet20} & 
\multicolumn{4}{c}{RNN} \\ \cline{3-10}
                                &                             &  \makecell[c]{$\gamma=$  \\  0.001}  & \makecell[c]{$\gamma=$  \\ 0.01}  & \makecell[c]{$\gamma=$  \\  0.1}  & \makecell[c]{$\gamma=$  \\ 0.3}  & \makecell[c]{$\gamma=$  \\ 0.03}  & \makecell[c]{$\gamma=$  \\ 0.1} & \makecell[c]{$\gamma=$  \\ 0.3} & \makecell[c]{$\gamma=$  \\  1.0} \\
\toprule
\multirow{2}{*}{BSP-SGD}           & 4                                               & 84.16  & 89.70 & \textbf{91.25} & 90.36 & 52.03 & 55.24 & \textbf{56.28} & 47.70 \\ 
                                & 32                                              & 65.42 & 85.16 & 89.21 & \textbf{89.77} & 53.74 & 53.73 & 52.16 & \textbf{54.34} \\ \cline{2-10} 
\multirow{2}{*}{ASP-SGD}          & 4                                               & 88.27 & \textbf{92.27} & 90.84 & 85.97 & \textbf{53.73} & 50.18 & 52.39 & 43.81 \\
                                & 32                                              & 0.00  & 0.00  & 0.00  & 10.00 & \textbf{53.79} & 55.34 & 3.75  & 3.04  \\ \cline{2-10} 
\multirow{2}{*}{Local-SGD $\tau=2$}  & 4                                               & 84.04 & 88.97 & \textbf{92.01} & 88.86 & \textbf{55.03} & 54.94 & 52.87 & 46.72 \\
                                & 32                                              & 65.41 & 84.96 & 89.35 & \textbf{90.11} & \textbf{54.11} & 51.72 & 52.30 & 46.17 \\ \cline{2-10} 
\multirow{2}{*}{Local-SGD $\tau=4$}  & 4                                               & 83.85 & 89.61 & \textbf{90.90} & 90.00 & 51.92 & 55.05 & \textbf{56.14} & 47.36 \\
                                & 32                                              & 65.55 & 85.09 & 89.44 & \textbf{89.85} & \textbf{54.36} & 51.72 & 53.40 & 47.15 \\ \cline{2-10} 
\multirow{2}{*}{Local-SGD $\tau=8$}  & 4                                               & 84.58 & 89.32 & \textbf{91.20} & 90.86 & \textbf{55.25} & 55.19 & 53.35 & 48.55 \\
                                & 32                                              & 64.48 & 84.82 & 89.41 & \textbf{90.09} & 54.07 & 51.72 & \textbf{54.91} & 47.30 \\ \cline{2-10} 
\multirow{2}{*}{Local-SGD $\tau=16$} & 4                                               & 84.02 & 89.25 & \textbf{90.99} & 90.56 & 55.32 & \textbf{55.41} & 53.43 & 48.84 \\
                                & 32                                              & 64.74 & 84.69 & 89.57 & \textbf{90.15} & 53.85 & 51.72 & \textbf{56.69} & 46.93 \\ \cline{2-10} 
\multirow{2}{*}{FedAvg}         & 4                                               & 62.41 & 84.37 & 89.81 & \textbf{90.10} & 52.20 & \textbf{55.23} & 54.91 & 54.92 \\
                                & 32                                              & 40.23 & 64.42 & 83.45 & \textbf{86.65} & 28.77 & 43.70 & 51.91 & \textbf{52.02} \\
\toprule
\end{tabular}
\begin{tablenotes}
	\item Note: $\tau$ means the number of local iterations of Local SGD. 
\end{tablenotes}
\end{threeparttable}
\vspace{-10pt}
\end{table*}

\begin{table*}[]
\centering
\fontsize{8}{8}\selectfont
\caption{Test accuracy [\%] comparison of different communication synchronization schemes.}
\vspace{-0.3cm}
\label{tab:exp-number-worker}
\begin{tabular}{cc|cccc|cccc}
\toprule
\multirow{2}{*}{Algorithm}           & \multirow{2}{*}{\# of Worker} & \multicolumn{4}{c|}{Resnet20} & 
\multicolumn{4}{c}{RNN} \\ \cline{3-10}
                                &                             &  \makecell[c]{$\gamma=$  \\  0.001}  & \makecell[c]{$\gamma=$  \\ 0.01}  & \makecell[c]{$\gamma=$  \\  0.1}  & \makecell[c]{$\gamma=$  \\ 0.3}  & \makecell[c]{$\gamma=$  \\ 0.03}  & \makecell[c]{$\gamma=$  \\ 0.1} & \makecell[c]{$\gamma=$  \\ 0.3} & \makecell[c]{$\gamma=$  \\  1.0} \\
\toprule
\multirow{4}{*}{BSP-SGD}                 & 4                           & 84.16 & 89.70 & \textbf{91.25} & 90.36 & 52.03 & 55.24 & \textbf{56.28} & 47.70 \\
                                      & 8                           & 78.66 & 88.04 & \textbf{90.53} & 90.48 & 52.92 & \textbf{54.89} & 54.63 & 47.80 \\
                                      & 16                          & 72.51 & 87.03 & \textbf{90.04} & 90.02 & \textbf{54.79} & 53.72 & 52.99 & 47.75 \\
                                      & 32                          & 65.42 & 85.16 & 89.21 & \textbf{89.77} & 53.74 & 53.73 & 52.16 & \textbf{54.34} \\ \cline{2-10} 
\multirow{4}{*}{BSP-EFTopK}      & 4                           & 83.78 & 88.91 & \textbf{90.90} & 89.28 & 51.73 & 55.36 & \textbf{55.74} & 47.66 \\
                                      & 8                           & 79.76 & 88.25 & 89.84 & \textbf{90.09} & 52.64 & \textbf{54.89} & 54.48 & 46.65 \\
                                      & 16                          & 72.23 & 87.01 & \textbf{89.80} & 88.97 & \textbf{54.70} & 53.60 & 52.86 & 47.24 \\
                                      & 32                          & 64.52 & 84.33 & 88.08 & \textbf{88.69} & \textbf{53.98} & 51.62 & 52.14 & 46.88 \\ \cline{2-10} 
\multirow{4}{*}{ASP-SGD}                & 4                           & 88.27 & \textbf{92.27} & 90.84 & 85.97 & \textbf{53.73} & 50.18 & 52.39 & 43.81 \\
                                      & 8                           & 88.15 & \textbf{90.49} & 89.67 & 10.00 & \textbf{53.72} & 43.69 & 52.94 & 44.34 \\
                                      & 16                          & 85.69 & \textbf{85.74} & 81.47 & 10.00 & \textbf{54.01} & 37.22 & 51.60 & 6.24  \\
                                      & 32                          & 0.00  & 0.00  & 0.00  & 10.00 & 53.79 & \textbf{55.34} & 3.75  & 3.04  \\ \cline{2-10} 
\multirow{4}{*}{Local-SGD $\tau=4$}          & 4                           & 83.85 & 89.61 & \textbf{90.90} & 90.00 & 51.92 & 55.05 & \textbf{56.14} & 47.36 \\
                                      & 8                           & 79.06 & 88.42 & 89.95 & \textbf{90.72} & 53.24 & 54.71 & \textbf{55.11} & 47.76 \\
                                      & 16                          & 72.92 & 84.98 & 86.89 & \textbf{90.58} & \textbf{55.12} & 53.45 & 53.33 & 47.22 \\
                                      & 32                          & 65.55 & 85.09 & 89.44 & \textbf{89.85} & \textbf{54.36} & 51.72 & 53.40 & 47.15 \\ \cline{2-10} 
\multirow{4}{*}{Local-SGD $\tau=4$ TopK} & 4                           & 60.73 & 81.18 & \textbf{85.20} & 83.83 & 50.54 & 50.78 & \textbf{52.64} & 50.25 \\
                                      & 8                           & 54.43 & 76.96 & \textbf{85.00} & 84.78 & 48.01 & 49.13 & \textbf{52.28} & 48.89 \\
                                      & 16                          & 46.33 & 70.29 & \textbf{83.54} & 83.38 & 44.48 & 47.47 & \textbf{51.58} & 47.47 \\
                                      & 32                          & 39.75 & 62.86 & 79.11 & \textbf{81.55} & 40.75 & 44.62 & \textbf{50.40} & 46.64 \\
\toprule
\end{tabular}
\vspace{-10pt}
\end{table*}

\begin{table*}[]
\centering
\fontsize{8}{8}\selectfont
\caption{Mean and STD. of the test accuracy [\%] of some experiments with 3 different random seeds.}
\vspace{-0.3cm}
\label{tab:exp-repeat}
\begin{tabular}{c|cccc}
\toprule
\multirow{2}{*}{Algo}  &  \multicolumn{4}{c}{Resnet20} \\ \cline{2-5}
    &  $\gamma=$ 0.001  & $\gamma=$ 0.01  & $\gamma=$ 0.1  & $\gamma=$ 0.3 \\
\toprule
BSP-SGD         & $ 65.43\pm0.54 $ & $ 85.09\pm0.11 $ & $ 89.15\pm0.14 $ &  \textbf{89.28}$\pm0.74 $ \\
BSP-EFTopK  & $ 64.62\pm0.80 $ & $ 84.18\pm0.24 $ & $ 87.81\pm0.23 $ & \textbf{88.14}$\pm0.44 $ \\
DP-SGD (Gossip)       & $64.86 \pm 0.76  $  & $84.83 \pm 0.23$  & $88.83\pm 0.11$ & \textbf{89.11}$\pm$ 0.45 \\
Local-SGD $\tau=4$  & $ 65.28\pm0.67 $ & $ 84.78\pm0.31 $ & $ 89.10\pm0.24 $ &  \textbf{89.49}$\pm$ 0.49 \\
FedAvg      & $ 39.94\pm 0.47 $ & $ 64.03\pm0.81 $ & $ 83.6\pm0.11 $ & \textbf{ 86.61} $\pm$0.25  \\
\toprule
\end{tabular}
\vspace{-10pt}
\end{table*}

\textbf{Different synchronization schemes} have different properties of communication efficiency and convergence rate, some works compare few schemes~\citep{Lin2020Dont,NEURIPS2022_029df12a,jiang2022pisces}. However, to the best our knowledge, there is no works benchmark all of them together in unified experiment settings. We jointly evaluated the performance of BSP-SGD, ASP-SGD, Local-SGD and FedAvg with unified dataset settings and hyper-parameters. Experiment settings are provided in Section~\ref{sec:overall-exp-conf}. For Local-SGD, we compare different number of local iterations, including 2, 4, 8, and 16. Our results, as summarized in Table~\ref{tab:exp-sync}, indicate that BSP-SGD, ASP-SGD, Local-SGD and FedAvg finally obtain the similar test accuracy. And the performance of Local-SGD is not very sensitive to the local iterations $\tau$, suggesting that the communication overheads can be reduced by skipping some communication rounds without sacrificing the model performance.

\textbf{Different number of workers} significantly influence the convergence of algorithms. Table~\ref{tab:exp-number-worker} shows the influence of different workers (4, 8, 16, and 32) to different algorithms. For both two datasets and all algorithms, to achieve the same test accuracy, higher learning rates are required when the number of workers is increased, as it is similar to higher batch size~\citep{you2019large}. However, for most algorithms, increasing the number of workers can lead to a degradation in the final accuracy due to the well-known generalization problem of large batch SGD~\citep{Hoffer2017TLG32947713294936,DBLPconficlrKeskarMNST17}, which is a significant bottleneck for distributed training scalability.


ASP-SGD suffers from a larger drop in the test accuracy, and even diverges when the number of workers increases to 32. This phenomenon is attributed to the higher degree of staleness with larger worker numbers. Additionally, the experimental results suggest that ASP-SGD may require a lower learning rate compared to other algorithms with the same number of workers. This could be because excessively moving towards a stale gradient direction could be futile and biased.

In conclusion, while efficient communication algorithms reduce the exchange of information compared to BSP-SGD, they may compromise the convergence performance of the model. Therefore, the benefits of communication-efficient algorithms come at a cost. To provide a better understanding of the varying impacts of these methods, we present a comparison in Table~\ref{tab:Influences_DisSGD}.


\begin{table*}[htbp]
\fontsize{7}{7}\selectfont
\begin{center}
\caption{Influences of different combinations of architectures and synchronization schemes}
\label{tab:Influences_DisSGD}
\begin{threeparttable}        
\begin{tabular}{cccccc}
\toprule
Architecture & Synchronization & Model Consistency & Communication Frequency & Communication Congestion & Convergence \\ 
\midrule
\multirow{4}{*}{PS} & BSP-SGD & high & high & high & stable \\
 & SSP-SGD & normal & high & normal & normal \\  
 & ASP-SGD & low & high & low & unstable \\ 
 & Local-SGD & normal & low & high & unstable \\   \hline
\multirow{4}{*}{All-Reduce} & BSP-SGD & high & high & low & easy \\  
 & SSP-SGD & - & - & - & - \\  
 & ASP-SGD & - & - & - & - \\ 
 & Local-SGD & normal & low & low & stable \\   \hline
\multirow{4}{*}{Gossip}  & BSP-SGD & low & high & low & stable \\ 
 & SSP-SGD & - & - & - & - \\ 
 & ASP-SGD & low & high & low & unstable \\ 
 & Local-SGD & low & low & low & stable \\ 
 \bottomrule
\end{tabular}
    \begin{tablenotes}    
        \footnotesize              
        \item Notes: Here the different levels only represent a relative description comparing with the other methods. The model consistency measures how local models are different from others, the communication frequency measures how frequently workers communicate with others, and communication congestion measures how heavy the traffic of the central node is.
      \end{tablenotes}          
\end{threeparttable}     
\end{center}
\end{table*}
\vspace{-10pt}

\section{Centralized/Decentralized framework}\label{sec:cenvsdecen}
\begin{figure*}[!h]
	\centering
    \subfigure[PS architecture.]
	{
	\includegraphics[width=0.25\linewidth]{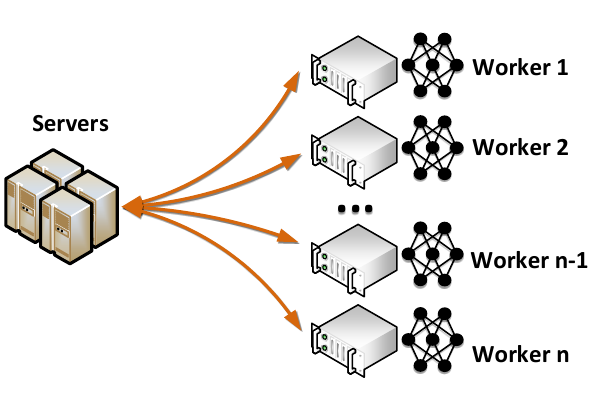}
	}
	\subfigure[All-Reduce architecture.]
	{
	\includegraphics[width=0.3\linewidth]{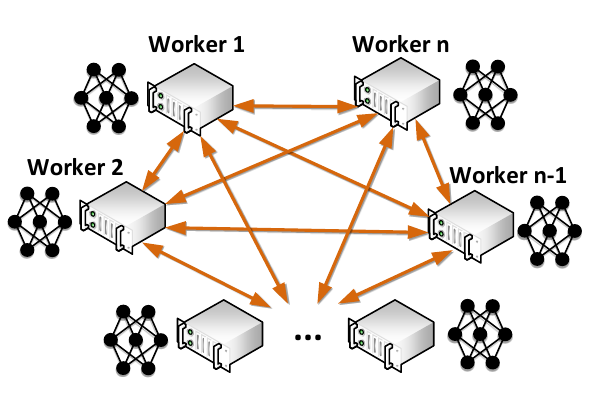}
	}
	\subfigure[Gossip architecture.]
	{
	\includegraphics[width=0.3\linewidth]{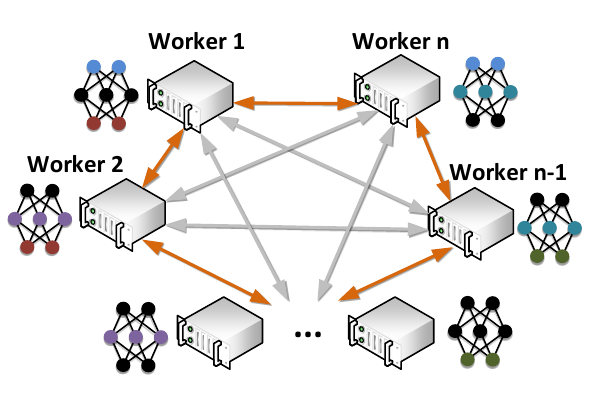}
	}
    \vspace{-0.5cm}
	\caption{Three different system architectures for model/gradient aggregation. 
    In (a) and (b), workers synchronize trained models and conduct local training with the synchronized model. In (c), workers communicate model parameters with some neighbors, and start local training with heterogeneous models parameters.}
	\label{fig:architectures}
\vspace{-0.5cm}
\end{figure*}

Various system architectures have been proposed to support efficient data communication among workers in distributed DL. The choice of the most appropriate architecture depends on the available hardware resources. Regarding how to average model parameters/gradients among distributed workers, system architectures can be categorized into three types: Parameter Server (PS), All-Reduce, and Gossip as shown in Fig.~\ref{fig:architectures}.




\subsection{Parameter Server}\label{subsec:parameterservers}
The Parameter Server (PS) architecture is shown in Fig. \ref{fig:architectures}(a). Servers are responsible for storing the newest global model, and gathering the updating information from the workers. Workers independently load the dataset, pull the newest global model from the server, and then compute the updates, which will be transmitted to the servers~\citep{PSforDisML,ScalDMLwithPS}. The PS architecture is referred to as the centralized framework, commonly used in distributed training ~\citep{PSforDisML,ScalDMLwithPS,dean2012large,MoreEffDMLviaStaleSync,communicationDisMLwithPS,Ooi2015SD}. The major issue with this architecture is the communication congestion in the server~\citep{Zhang2015DLE296,HowToScaleDDL} as it requires to extensively communicate with all workers. 

Early PS based distributed machine learning methods ~\citep{ScalDMLwithPS,DisGraphlab,AsystemLarScalSupML,AnArchforParallel,MoreEffDMLviaStaleSync,PSforDisML} primarily focused on how to implement distributed machine learning at the system level to support a vast number of model parameters. In the pioneering framework DistBelief~\citep{dean2012large}, Dean et al. successfully trained the model with billions of parameters on an extremely large-scale cluster. Initially, for data parallel distributed machine learning algorithms, numerous works focused on CPU clusters ~\citep{Ahmed2012,pmlrv32ahn14,Cui2014EBS,Cui2014IP67,Power2010PBF,7816979}. Later, Cui et al.~\citep{Cui2016} proposed GeePS, a PS architecture to scale deep learning models training across numerous GPUs, addressing the challenge of limited GPU memory when training large deep models.

Subsequently, researchers focused on developing communication-efficient methods to address the communication bottleneck in PS-based distributed machine learning. In an early work by Li et al.~\citep{communicationDisMLwithPS}, controllable synchronization and user-definable filters were used to reduce data communication volumes. Later, numerous methods were proposed to achieve efficient communication with data compression like sparsification~\citep{DorefaNet,1bit,scalableDisDNN,QSGD,ECQSGD,TernGrad,signSGD,EFsignSGD} (\S\ref{sec:sparsification}) and quantization~\citep{SparOLviaTrunc,FLStrategy,GradSparforDisOptim,CommQuantforDataPara,DGC,zhang2017poseidon} (\S\ref{sec:quantization}). 

In addition, Wang et al. ~\citep{icdcsWangWL19} proposed a novel measurement to determine the relevance of a worker. They measured the relevance by comparing the local update with the global update from the previous iteration. If the relevance exceeds a threshold, this update is considered as relevant and transmitted to the server. Another approach to reducing communication overhead is to use programmable switches~\citep{In-network}. By having workers connected to the same switch transmit their model updates over the network and complete the aggregation in the network, communication overheads between machines can be reduced. However, this approach presents challenges such as limited computation and storage of switches, as well as the packet loss. Specialized algorithms need to be designed to address these issues.

\subsection{All-Reduce}\label{subsec:allreduce}
To avoid the communication bottleneck in the PS, practitioners and researchers turned to the All-Reduce architecture for gradient aggregation without central servers~\citep{awan2017s,chu2017efficient,wang2019impact,goyal2017accurate,jia2018highly}. As depicted in Fig. \ref{fig:architectures}(b), all workers communicate with each other without a central node to obtain the gradients from all the other workers. The aggregated gradients are used to update their local models, thereby achieving consistency with other workers (the initialized models in different workers are identical). It is a model-centralized topology since there is a consistent global model attained through synchronization, which allows the updating equation \eqref{eq:ssgd_upd} (BSP-SGD) to remain unchanged. However, it is not suitable for asynchronous communication due to the collective communication nature of All-Reduce. While it is difficult to apply All-Reduce in the asynchronous part of SSP, it is easily applicable in the synchronous part of SSP.

The high-performance computing community has a long history of system optimizations for the All-Reduce collective, as various approaches proposed to improve its performance ~\citep{OptimizationofCollective,Thakur2005,Hoefler:2010,sanders2009two}. Some popular All-Reduce algorithms with different latency and bandwidth costs for an $N$-dimensional vector and an $n$-node cluster are summarized in Table \ref{tab:allreduce}. The communication cost of sending or receiving an $N$-dimensional vector is modeled as $\alpha + \beta N$ ~\citep{sarvotham2001connection,Shi2018MGWFBPED}.


\begin{table}[!t]
	\centering
	\caption{Communication costs of some representative All-Reduce algorithms}
	\label{tab:allreduce}
	\begin{tabular}{lcc}
  \toprule
		Algorithm &  Latency & Bandwidth \\
  \midrule
		Binary tree & $2\alpha \log n$ & $2\beta(\log n)N$ \\
		Recursive doubling& $\alpha \log n$ & $\beta(\log n)N$  \\
		Ring & $2(n-1)\alpha$ & $\frac{2(n-1)}{n}\beta N$  \\
  \bottomrule
	\end{tabular}
\end{table}
Table \ref{tab:allreduce} summarizes various All-Reduce algorithms, which have been optimized by the high-performance computing community over time~\citep{OptimizationofCollective,Thakur2005,Hoefler:2010,sanders2009two}. Among these algorithms, the ring-based All-Reduce is widely used in distributed DL due to its bandwidth optimality (e.g., Gloo\footnote{\url{https://github.com/facebookincubator/gloo}} and earlier versions of NCCL\footnote{\url{https://developer.nvidia.com/nccl}}.). However, the latency of the ring-based All-Reduce is linearly proportional to the number of workers, leading to high communication costs when scaling to large-scale clusters~\cite{you2017scaling,Shi2018MGWFBPED,shi2021mgj,shi2021accelerating}. To address this issue, recent updates of NCCL (started from version 2.4)\footnote{\url{https://devblogs.nvidia.com/massively-scale-deep-learning-training-nccl-2-4/}} have integrated the double binary trees~\citep{sanders2009two} to perform an all-reduction to achieve full bandwidth and a logarithmic latency. The double binary trees require the whole message to be broken down into multiple blocks and the workers to be formed as a binary tree so that different blocks can be executed in parallel. Therefore, for some small messages or small-scale clusters, recursive doubling or ring-based algorithms would be better.

To further reduce the latency term in All-Reduce while preserving the bandwidth optimality, the hierarchical All-Reduce algorithms ~\citep{goyal2017accurate,jia2018highly,ueno2019exhaustive} were also proposed, which can reduce the latency cost several times (related to the number of hierarchies). 2D-Torus All-Reduce~\citep{mikami2018massively,jouppi2017datacenter} can also massively reduce the communication latency using a 2D-Torus topology network. Under different topology networks (e.g., BCube ~\citep{guo2009bcube}), it is also important to carefully design the All-Reduce algorithms to achieve lower latency and higher bandwidth. Wang et al. ~\citep{wang2018bml} proposed BML for the BCube topology to achieve efficient communication. Some topology-aware algorithms (e.g., BLink ~\citep{wang2020blink} and PLink~\citep{luo2020plink}) were designed to be adaptive to distributed environments to highly utilize the network bandwidth with low latency.

\subsection{Gossip}\label{subsec:gossip}



The Gossip architecture has emerged as a mechanism for inter-worker communication and is employed to solve the distributed average problem~\citep{gossip1,gossip2,gossip3,gossip4}. Researchers have exploited this architecture to improve BSP-SGD ~\citep{DisSubGradMultiOptim,AsynDisOptimRandADMM,OptimDisOptim,OptimNonsmoothDisOptim,DualApprochforOptim,MultiagentMirrorDescentDenct,CommEffDenctStoc,CanDecent,DecentTrainingoverDecentData,StocGradPush,CommCompforDecent}.

As illustrated in Fig. \ref{fig:architectures}(c), the gossip architecture does not require global models or parameter servers (represented by different colors of local models). Instead, each worker communicates updates with their neighbors (also named peers). Workers exchange messages via edges chosen in each iteration (the blues ones in Fig. \ref{fig:architectures}(c)), which can be described as a communication graph. Workers are not required to communicate with all others, reducing communication costs. During training, the algorithm does not guarantee parameter consistency across all workers after each communication, but guarantees it (i.e., consensus) at the end of the algorithm. This means that local models are different during every iteration. It should be noted that in asynchronous and SSP with the PS architecture, although local models are also different, a global model is still maintained.


Similar to the All-Reduce architecture, gossip architecture offers the benefit of having no master node, thus eliminating communication issues of the central server \citep{CommCompforDecent}. Compared to All-Reduce architecture, the gossip architecture is more fault-tolerant to the worker failures and it can also be theoretically guaranteed~\citep{CanDecent} with less communication costs than PS or All-Reduce. 

In the gossip architecture, the key issue is ensuring the attainment of identical model parameters across all workers. This problem is referred to as ``consensus'' and has been studied extensively in the literature  ~\citep{7942055DistributedLinearized,1498447Gossipalgorithms,Carli2010,4497789Randomizedconsensus,4118472ConsensusandCooperation,4434671Adistributedconsensus}. Formally, consensus is a state of all workers attaining a same opinion in common. In the gossip distributed SGD, the consensus is the status that all workers have the identical model parameters. 

However, on one hand, achieving the ``consensus'' incurs the difficulty in designing a method that enables efficient communication, high convergence rate, and consensus simultaneously. To address these issues, model compression can be combined with decentralized learning algorithms to reduce communication costs  ~\citep{CommCompforDecent,DecentStocOptimAndGossip,NIPS2018_7705,tang2020communication}. On the other hand, the Gossip architecture is limited to using symmetric communication, which inherently requires deadlock-avoidance and more synchronization, resulting in slower and more sensitivity to stragglers. Assran et al.~\citep{StocGradPush} proposed a solution that combines Stochastic Gradient Push (SGP)~\citep{SGPDirected} with PUSHSUM~\citep{8340193,gossip1}. The PUSHSUM provides an approximation of the distributed averaging, while the SGP makes each worker only send the gradient to its out-neighbors without waiting for the response from these neighbors. Thus, the overall system throughput is improved.


\subsection{Experimental Comparison}
We conduct experiments to compare BSP-SGD\footnote{From the algorithmic view, the BSP-SGD algorithm under the All-Reduce architecture is identical to that under the PS architecture. Thus, we choose one to compare the convergence performance.} and synchronous data-parallel SGD (DP-SGD) with the Gossip architecture. Table~\ref{tab:exp-cent} indicates that BSP-SGD and DP-SGD (Gossip) can achieve similar performance. DP-SGD (Gossip) exhibits a decrease in performance as the number of workers increases, consistent with the results presented in Table~\ref{tab:exp-number-worker}.

\begin{table}[!h]
\centering
\fontsize{8}{8}\selectfont
\caption{Test accuracy [\%] comparison of different communication architectures.}
\label{tab:exp-cent}
\begin{tabular}{cc|cccc|cccc}
\toprule
\multirow{2}{*}{Algorithm}           & \multirow{2}{*}{\# of Worker} & \multicolumn{4}{c|}{Resnet20} & 
\multicolumn{4}{c}{RNN} \\ \cline{3-10}
                                &                             &  \makecell[c]{$\gamma=$  \\  0.001}  & \makecell[c]{$\gamma=$  \\ 0.01}  & \makecell[c]{$\gamma=$  \\  0.1}  & \makecell[c]{$\gamma=$  \\ 0.3}  & \makecell[c]{$\gamma=$  \\ 0.03}  & \makecell[c]{$\gamma=$  \\ 0.1} & \makecell[c]{$\gamma=$  \\ 0.3} & \makecell[c]{$\gamma=$  \\  1.0} \\
\toprule
\multirow{2}{*}{BSP-SGD (PS)}    & 4                           & 84.16 & 89.70 & \textbf{91.25} & 90.36 & 52.03 & 55.24 & \textbf{56.28} & 47.70 \\
                              & 32                          & 65.42 & 85.16 & 89.21 & \textbf{89.77} & 53.74 & 53.73 & 52.16 & \textbf{54.34} \\ \cline{2-10}
\multirow{2}{*}{DP-SGD (Gossip)} & 4                           & 84.60 & 89.38 & \textbf{91.08} & 91.01 & \textbf{55.41} & 55.19 & 52.92 & 48.13 \\
                              & 32                          & 64.20 & 85.12 & 88.99 & \textbf{89.71} & 53.55 & 51.71 & \textbf{53.77} & 46.50 \\
\toprule
\end{tabular}
\end{table}

\section{Quantization methods}\label{sec:quantization}


Compression methods aim to reduce communication data by compressing gradients or model parameters that need to be transmitted between workers or servers. However, most of these methods use lossy compression, which prevents the receiver from fully recovering the original gradients or model parameters. As a result, convergence may be impacted due to the reduced amount of information available. It has been an active research direction to reduce the communication traffic with little impact on the convergence. 
\begin{figure}[!th]
	\centering
	\includegraphics[width=0.99\linewidth]{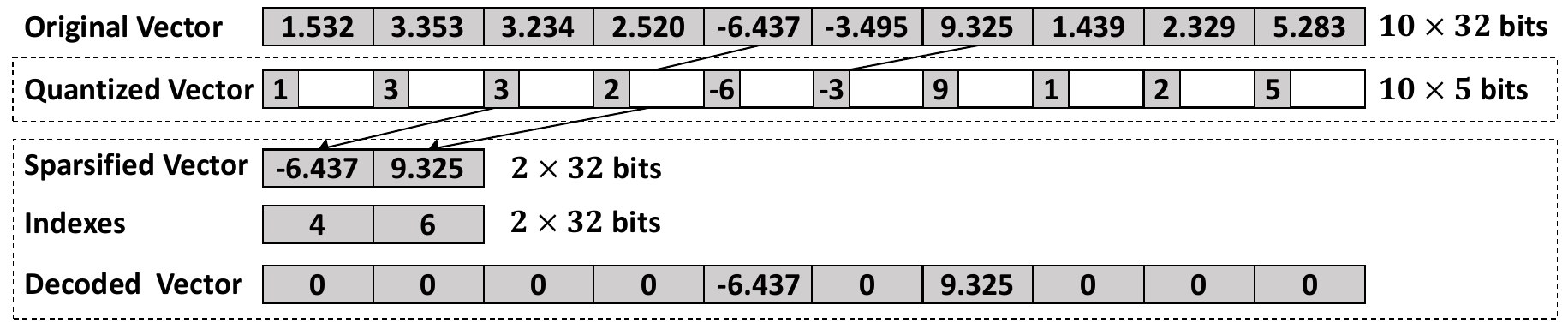}
	\caption{Comparison of Quantization and Sparsification.}
	\label{fig:compression}
\end{figure}

Quantization is a one of popular compression schemes that uses lower bits to represent data originally represented by 32 bits on each dimension of the transmitted gradient. As shown in Fig. \ref{fig:compression}, the element in the quantized gradient is coded by fewer bits, resulting in low-precision gradients after quantization. Low-precision gradients are beneficial to DL as higher-speed calculations and lower memory are needed for training DNNs on CPUs and GPUs. Many researchers have studied DL convergence under low-precision gradients with different quantization methods~\citep{ImproSpeedNN,DLwithlimited,ProbroundingNN,DorefaNet,Hubara2017}.
The process of quantized SGD can be formulated by as following equations:
\begin{equation}
    G_{i,t}^{quant}(\mathbf{x}_{t}) = Quant(G_{i,t}(\mathbf{x}_{t}) + \delta_{i,t}(\mathbf{x}_{t})) \label{1bitSGD1}
\end{equation} 
\begin{equation}
    \delta_{i,t}(\mathbf{x}_{t}) = G_{i,t}(\mathbf{x}_t) - {Quant}^{-1}(G_{i,t}^{quant}(\mathbf{x}_t)) \label{1bitSGD2}
\end{equation}
\begin{equation}
    \mathbf{x}_{t+1} = \mathbf{x}_{t} - {\gamma}\frac{1}{n} \sum_{i=1}^n G_{i,t}^{quant}(\mathbf{x}_{t}) \label{1bitSGD3}
\end{equation}
where $Quant(\cdot)$ denotes the quantization function 
that encodes gradients, and $Quant(\cdot)$ the unquantizer that decodes quantized gradients. This approach reduces communication costs as $G_{i,t}^{quant}(\mathbf{x_t})$, compared to the approach described in Eq. (\ref{eq:ssgd_gra}).

In ~\citep{DisMeanEst} and \citep{DisMeanEst}, the communication of gradients in BSP-SGD actually is  regarded as a distributed mean estimation problem~\citep{DisMeanEst,RandDisMean}. They ~\citep{DisMeanEst,RandDisMean} analyzed the communication-efficient algorithms for distributed mean estimation. They used the Mean Squared Error (MSE) to measure how accurate the quantization methods are. And then, they proposed coding strategies to achieve the best MSE for a given communication cost, considering that increasing the number of quantization levels increases the communication cost. To reduce the communication cost, Suresh et al.~\citep{DisMeanEst} proposed two methods, Stochastic Rotated Quantization (SRQ) and Variable Length Coding (VLC). In SRQ, all clients and the central server generate a global random rotation matrix and then try to find an orthogonal matrix $\mathbb{R}$ that achieves low MSE. The VLC uses arithmetic of Huffman Coding corresponding to the number of times each quantized value has appeared.

To achieve a higher compression ratio, Sei et al.~\citep{1bit} proposed 1-bit SGD to reduce the transmitted data volumes, and successfully trained the deep model on traditional speech applications with a 10$\times$ speedup. They reduced each gradient element to one bit and quantified a gradient $G_{i,t}(\mathbf{x}_t)$ while keeping the quantization error $\delta_{i,t}(\mathbf{x}_t)$ not large at the same time. The key idea of 1-bit quantization is also used in~\citep{scalableDisDNN}. Different from 1-bit SGD, Strom et al.~\citep{scalableDisDNN} chose a fixed threshold, and then encoded the gradient elements higher than $T$ with the value $1$, those less than $–T$ with value 0. The absolute value of a gradient element less than $T$ would not be sent, which is similar to the sparsification methods that will be discussed in \S\ref{sec:sparsification}.

Some researchers further proposed adaptive quantization methods ~\citep{7544448,7526802} that incorporate inexact proximal gradients. However, these methods lack empirical validation with respect to deep learning models. To achieve both communication efficiency and good optimization convergence, Alistarh et al.~\citep{QSGD} presented a family of algorithms using quantization rather than just one quantization method. This family of quantization algorithms, named Quantized SGD (QSGD), allows for the trade-off between the number of communicated bits and the variance of the compressed gradient. They evaluated the performance of QSGD by training DNNs on ImageNet~\citep{Imagenet} and CIFAR-10 datasets, achieving accuracy rates comparable to those of original SGD. Specifically, QSGD exploits a probabilistic approach to quantify a vector. Given any vector $\mathbf{v} \in \mathbb{R}^N$, every $j$-th element $Quant_s({v}_j)$ of quantization gradient $Quant_s(\mathbf{v})$ is quantized by $Quant_s(\cdot)$, corresponding to the element ${v}_j$ of the original gradient $\mathbf{v}$. The stochastic quantization function is defined as
\begin{equation}
    Quant_s(v_j) = \left \| \mathbf{v}	\right \|_2 \cdot sgn(v_j) \cdot \psi_j(\mathbf{v}, s),
\end{equation}
where $sgn(v_j) \in	\left \{ -1, +1 \right\}$ denotes the sign of $v_j$ and $sgn(0)=1$. $\psi_j(\mathbf{v}, s)$ is defined as
\begin{equation}\label{QSGDequation}  
    \psi_j(\mathbf{v}, s) = \left \{
    \begin{array}{ll} l/s     &\text{with probability}\ 1-p( \frac{\left| v_j \right|}{\left \| \mathbf{v} \right\|_2}  ,s ) , \\ 
    (l+1)/s  & \text{otherwise}
    \end{array}                
    \right .
\end{equation}
$l$ is an integer such that $\left| v_i\right|/ \left \| \mathbf{v} \right\|_2 \in \left[ l/s, (l+1)/s \right]$ s.t. $0<l<s$, and $p(a,s) = as-l$ for any $a \in \left [ 0, 1 \right ] $. This is called Standard (random) dithering in ~\citep{Horvath2019NaturalCF}, and is used in PCM coding ~\citep{6773262,1057702}. Note that if $\mathbf{v}=\mathbf{0}$, the algorithm sets $Quant_s(\mathbf{v},s)=\mathbf{0}$. The expectation $\psi_j(\mathbf{v},s)$ satisfies $\mathbb{E}\left[ \psi_j(\mathbf{v},s) \right]=\left| v_i  \right| / \left\| \mathbf{v}  \right\|_2$. Then it is obvious that QSGD can assure that the quantized gradient is the unbiased estimation of the original vector, i.e., $\mathbb{E} \left[ Quant_s(\mathbf{v}) \right]=\mathbf{v}$, and facilitates convergence of training.

Wen et al.~\citep{wen2017terngrad} proposed Terngrad, which differs from the Parameter Server architecture. Each worker stores a copy of parameters locally, and the communication of parameters in the floating-point form is changed to the transfer of quantized gradients. This results in smaller server-to-worker traffic since only the quantized gradients are pulled from servers. To minimize the number of levels when workers are transferring different scalars, Terngrad presented \textit{Scalar Sharing}, which selects the maximum scalar among all scalars and shares it across all workers. However, in a large deep model, a maximum element could be significantly larger than most gradients, which could weaken the approximation. To address this issue, Terngrad proposed layer-wise ternarizing and \textit{Gradient Clipping}. Layer-wise ternarizing utilizes the layer-wise scalar in each layer instead of a large global maximum scalar. And A series of works~\citep{faghri2020adaptive, jhunjhunwala2021adaptive,ElasticQuant} proposed to utilize adaptive quantization bits, to enhance the convergence with compressed gradients.

Sign-SGD is another kind of quantization method~\citep{signSGD}. In Sign-SGD, every worker quantifies the gradient to a binary value, which is the sign of each coordinate of the gradient vector. Bernstein et al. ~\citep{signSGDwithVote} provided an extensive theoretical analysis of Sign-SGD for non-convex optimization. They proved that when gradients are as dense or denser than stochasticity and curvature, Sign-SGD can converge with a theoretical rate. They also proposed a new algorithm named Sign-SGD with a majority vote. After workers exchange the sign of their gradient vector to a server, the overall update is decided by a majority vote. Mishchenko et al. ~\citep{DisLearningDIANA} introduced a novel method DIANA, which extends the methods of~\citep{QSGD,TernGrad} by splitting the whole gradient vector into some sub-vectors. And then, they quantified each sub-vector individually.

\section{Sparsification Methods}\label{sec:sparsification}
Quantization methods are limited in their compression rates, with a maximum compression rate of 32$\times$ achievable over commonly used SGD with single-precision floating-point arithmetic. However, reducing the number of transmitted elements can increase the compression rate even further. A set of algorithms that follow this approach are known as sparsification methods, where only a subset of elements are selected and transmitted~\citep{meProp,ZipML,EffiUseofLimitMemory,NIPS2019_9610,10.1145/3452296.3472904,li2022on,shi2021towards,li2022near,zhang2023evaluation}.


The key idea of sparsification is that only ``significant'' gradients are essential for the SGD update to ensure the convergence of the training process~\citep{DGC}. As illustrated in Fig. \ref{fig:compression}, a substantial proportion of the coordinates in the gradient vector can be zeroed-out such that the zero-valued elements are no need to transmit. Gradient sparsification is a more aggressive compression method than quantization, enabling much more significant reductions in communication traffic.

The antecedents of sparsification methods can be traced back to the Truncated Gradient method introduced by Langford et al.~\citep{SparOLviaTrunc} in the context of online learning algorithms. Truncated Gradient addresses memory space and computation constraints by inducing sparsity in the gradient. Rather than directly setting small coordinates to zero, Truncated Gradient keeps large coordinates with their original values and discards small coordinates that fall below a threshold. This method was the first sparsification technique developed for large-scale learning. After that, there are extensive studies to further improve sparsification in distributed DL.
The sparsification methods can be broadly classified into four main types: coordinate descent, random sparsification, deterministic sparsification, and proximal methods.

\subsection{Random Sparsification}
Random sparsification randomly selects a subset of entries from the gradient vector for communication and updating. This approach is also referred to as random-$k$, where $k$ denotes the number of selected elements.

In~\citep{FLStrategy}, Konecny et al. proposed Random Mask and Subsampling. In Random Mask, a pre-defined random sparsity pattern, to convert the update parameters $\mathbf{H}$ into a sparse matrix $\hat{\mathbf{H}}$. This random pattern is regenerated in each training iteration. In addition, it can be initialized independently by each client or created by the server and then distributed to the workers. In the former case, each worker must transmit both the indices and values of the non-zero entries of $\mathbf{H}$. In the latter case, workers need only to send the values of the non-zero entries of $\mathbf{H}$, because all indices of non-zero entries in every worker are the same as others. In Subsampling, unlike Random Mask, the sparse communication matrix $\hat{\mathbf{H}}$ is scaled to ensure that $\mathbf{E}(\hat{\mathbf{H}}) = \mathbf{H}$, making it an unbiased estimator of the true average.

In ~\citep{GradSparforDisOptim}, Wangni et al. proposed to randomly drop out coordinates with a probability vector $\mathbf{p}$ and amplifying the non-zero coordinates from $g_j$ to $g_j/p_j$. Formally, if one wants to compress original vector $\mathbf{g} =\left[ g_1, g_2, \cdots, g_N \right]$, given a probability vector $\mathbf{p} = \left[ p_1, p_2, \cdots, p_N\right]$, and the final sparsified vector is  $\mathbf{Q}_{spar}(\mathbf{g}) = \left[ Z_1\frac{g_1}{p_1}, Z_2\frac{g_2}{p_2}, \cdots, Z_N\frac{g_N}{p_N}\right]$, in which $Z_i$ represents selector, i.e. 0 or 1. Each item $\mathbf{Q}_{spar}(\mathbf{g})_i$ has the unbiased expectation: $\mathbb{E}\left[ \mathbf{Q}_{spar}(\mathbf{g})_i\right] = p_i \times \frac{g_i}{p_i} + (1 - p_i) \times 0 = g_i$, similar to~\citep{QSGD, FLStrategy}.


\subsection{Deterministic Sparsification}
Different from Random Sparsification, the sparse properties are guaranteed by Deterministic Sparsification~\citep{TonotopicANN,SparConnec}, in which most weights of DNNs can be close to zero. Due to the sparse weights, most of the gradients in each iteration are also around zero so that the zeroed gradients are no need for communication to reduce the communication traffic. There are mainly two ways to sparsify the gradients: Fixed Threshold and Adaptive Threshold. 

\subsubsection{Fixed Threshold}
Strom et al.~\citep{scalableDisDNN} introduced a new method to solve the communication bottleneck problem. In this algorithm, those gradient elements with absolute values less than a pre-defined threshold will be discarded. Because not all gradient elements are transmitted, the server must know which gradients to be transmitted and the indices of them. Strom et al. constructed key-value maps where keys are the indices and values are the corresponding gradient elements. The main drawback of the fixed threshold method is that it is non-trivial to choose a suitable threshold for different deep models or different iterations. Even worse, when Error Feedback (\S\ref{sec:auxiliary}) techniques are used, the fixed threshold method may result in the transmission of a large number of gradients.

\subsubsection{Adaptive Threshold}
To address the problem of Fixed Threshold, Top-$k$ sparsification algorithms~\citep{SparSGDwithMemory,ConvSparGrad,DGC,shi2020layer,shi2021towards} select the top-$k$ gradients (in terms of absolute values) at each iteration. 

Dryden et al.~\citep{CommQuantforDataPara} proposed an adaptive threshold, which uses a fixed proportion $\pi$ to indicate the proportion of gradient elements to transmit. In every iteration, the algorithm determines a positive threshold and a negative one to satisfy the desired proportion. This method is able to make sure that the compressing ratio will not change during the training process. Despite that this technique requires the sorting of the whole gradient vector elements and costs extra computing time, it still reduces the wall-lock time a lot. Aji \& Heafield~\citep{SparCommforDisGD} proposed another adaptive threshold method that uses a single threshold to drop all gradients whose absolute value is smaller than the threshold. However, parameters and their corresponding gradients may have varying scales across different parts of the model. As comparing all gradients with one global threshold may result in the loss of some small-scale gradients, Aji \& Heafield exploited layer normalization ~\citep{LayNorm} to make a global threshold work. 

To account for local gradient activity, Chen et al.~\citep{Adacomp} proposed a novel gradient compression scheme, AdaComp, which can self-adapt the compression rate. They showed that most of the gradient compression techniques do not work well for convolutional layers, as different types of neural layers, mini-batch sizes, and other factors may affect the compression rate. AdaComp automatically determines the sparsification level, adapting to all variations.

Scattler et al. ~\citep{sparsebinarycompression} combined sparsification and quantization methods to propose Sparse Binary Compression (SBC), a new compression algorithm. SBC discards elements of low absolute value, averages positive and negative values to obtain positive and negative means $\mu^{+}$ and $\mu^{-}$ respectively, and discards negative elements if $\mu^{+}$ is greater than $\mu^{-}$, and sets all positive elements to $\mu^{+}$ and vice versa. SBC then quantizes the non-zero elements, further reducing the communication cost by a factor of $\times3$. Following SBC, Sattler et al. ~\citep{SparseTernaryCompressionSTC} further exploited the combination of top-$k$ sparsification and Ternary quantization to develop a new method named Sparse Ternary Compression (STC). Unlike SBC, STC is particularly suitable for the Federated Learning~\citep{kairouz2019advances}. Experimental results in their paper demonstrate that sparsification methods achieve better convergence than quantization methods.

In many distributed training algorithms, workers pull the latest update from the PS before training to make their models not deviate too much from the global model. However, in most top-$k$ sparsification methods, the size of the latest update depends on the number of workers, and many indices of chosen elements of a gradient vector differ among workers. As a result, when all gradients are collected together, the elements of the global gradient vector increase almost linearly with the number of workers, leading to ineffective sparsity of master-to-workers communication. To address this problem, Shi et al.~\citep{GTopk} proposed a novel sparsification method gTop-$k$. After aggregating all gradients, gTop-$k$ sparsifies the global gradient vector once again, reducing the master-to-workers communication load and attaining convergence simultaneously~\citep{ijcai2019473}. Furthermore, by exploiting the properties of gradients that are empirically proved to follow bell-shaped distributions~\citep{shi2019understanding,shi2021towards}, the computing-efficient approximation of Top-$k$ can be further exploited. Adaptive selection of the number of gradients or model parameters for communication can also help reduce overall training time~\citep{han2020adaptive,9834260,pmlr-v151-wang22e,9721697,shi2020layerwise}.

The compression operation itself may need the computation costs. This is often overlooked by much of previous works. Some statistical approaches~\citep{m2021efficient,shi2019understanding,shi2021towards}  are also typically required to estimate an accurate threshold for compressing gradients with linear computation complexity in the size of model parameters.

\subsection{Coordinate Descent}
Coordinate descent is a kind of optimization method ~\citep{Passcode,AsynCoordD,blockCoordD,ConverBlockCoordD,DisBlockCoordD,ProxBlockCoordD,Qu2015QRD2969239} that splits all variables into multiple blocks, and then updates one block while fixing the remaining blocks. In an iteration, all blocks will be updated one by one ~\citep{blockCoordD}. Despite the success of gradient-based methods, they may still suffer from the vanishing gradient problem for training deep neural networks ~\citep{DL}. Gradient-free methods have been recently proposed to address the vanishing gradient problem, including BCD methods~\citep{ProxBlockCoordD,ConverBlockCoordD}. In distributed DL scenarios, BCD can be easily implemented in a distributed and parallel manner~\citep{DisBlockCoordD}. BCD has a property that only a subset of variables would be updated in each round, similar to the sparsification of distributed SGD. 

Lau et al.~\citep{ProxBlockCoordD} proposed an efficient BCD algorithm for DL and provided its convergence analysis. They highlighted  three major advantages of BCD: (1) higher efficiency per epoch compared to SGD at early stages; (2) good scalability; and (3) gradient-free optimization. In~\citep{GlobalconverBlockCoordD}, Zeng et al. presented a general methodology to establish provable convergence guarantees when using BCD in DL. 

Mishchenko et al. ~\citep{Mishchenko201999OP} developed a new algorithm named Independent Block Coordinate Descent (IBCD), which allows each worker to sample an independent subset of blocks.  They proved that the optimal number of blocks to be updated by each of $n$ units in every iteration is equal to $m$, where $m$ is the total number of blocks. Specifically, this means that when $n = 100$ parallel units are used, 99\% of work is a waste of time.

\subsection{Proximal Methods}
Proximal methods involve  two kinds of sparsity-inducing regularization in learning models and solving the optimization problem with proximal-based algorithms. These methods are utilized for sparsity learning to reduce the number of parameters in deep learning models.  Additionally, the communication of distributed deep learning can benefit from the sparsity. $L0$-and $L1$-norms of parameters are commonly  used for these methods ~\citep{CommEffiDisSpar,AsynCoordD}. 

Grishchenko et al.~\citep{AsynDisMLspar} firstly combined the Proximal method, coordinate descent, and random sparsification together. The workers compute a local update using a randomly selected subset of coordinates, while the master aggregates and averages all updates from workers and computes the proximity operator of the regularizer at the averaged updates.

To reduce communication overhead, Tsuzuku et al. ~\citep{VarianceGradCompression} developed a novel method that sends only gradient elements with small enough variance during training. They observed that some gradient elements are ambiguous, with low-amplitude but high-variance values, which may lead to futile updates to the global model. By controlling hyperparameters, their algorithm achieves high-rate sparsity.


\subsection{Experimental Comparison}

We conduct experiments of different compression schemes in the context of BSP, Local SGD, FedAVG and DPSGD algorithms. The compression schemes include QSGD quantization, Topk and error-feedback (EF) TopK sparsification. The overall experiment configuration is described in detail in Section~\ref{sec:overall-exp-conf}. The results are presented in Table~\ref{tab:exp-compression}.

As the compression ratio increased, almost algorithms suffered from varying degrees of performance drop. In the case of TopK compression, the Local SGD suffers less performance drop than BSP. Intuitively, we suppose the reason is that the Local SGD transmits the model weights rather than the gradients, which makes workers still can update all parameters. Interestingly, results show that the EF-TopK can successfully converge well even at extremely high compression ratio, i.e. 1000. This is because EF-TopK communicates parameters that are left by compression in the future iterations. Thus, all parameters of the model can still be updated finally.

\begin{table}[]
\centering
\fontsize{8}{8}\selectfont
\caption{Experiments of comparing test accuracy of different communication compression algorithms.}
\vspace{-10pt}
\label{tab:exp-compression}
\resizebox{\linewidth}{!}{
\begin{tabular}{ccc|cccc|cccc}
\toprule
\multirow{2}{*}{Algo}                & \multirow{2}{*}{Compress Ratio} & \multirow{2}{*}{Client Num} & \multicolumn{4}{c|}{Resnet20}  & \multicolumn{4}{c}{RNN}       \\ \cline{4-11}
                                & &                             &  \makecell[c]{$\gamma=$  \\  0.001}  & \makecell[c]{$\gamma=$  \\ 0.01}  & \makecell[c]{$\gamma=$  \\  0.1}  & \makecell[c]{$\gamma=$  \\ 0.3}  & \makecell[c]{$\gamma=$  \\ 0.03}  & \makecell[c]{$\gamma=$  \\ 0.1} & \makecell[c]{$\gamma=$  \\ 0.3} & \makecell[c]{$\gamma=$  \\  1.0} \\
\toprule
\multirow{2}{*}{BSP}                & \multirow{2}{*}{None}           & 4                           & 84.16 & 89.70 & \textbf{91.25} & 90.36 & 52.03 & 55.24 & \textbf{56.28} & 47.70 \\
                                     &                                 & 32                          & 65.42 & 85.16 & 89.21 & \textbf{89.77} & 53.74 & 53.73 & 52.16 & \textbf{54.34} \\ \cline{2-11}
\multirow{4}{*}{BSP quant}  & 
\multirow{2}{*}{2}             & 4                           & 84.28 & 89.33 & \textbf{90.52} & 0.31  & \textbf{55.19} & 55.06 & 52.87 & 48.13 \\
                                     &                                 & 32                          & 64.65 & 85.12 & 89.34 & \textbf{89.95} & 53.75 & 51.73 & 52.26 & \textbf{54.67} \\
                                     & \multirow{2}{*}{16}              & 4                           & 80.31 & 83.45 & \textbf{86.44} & 0.15  & \textbf{48.03} & 42.39 & 12.80 & 13.51 \\
                                     &                                 & 32                          & 64.34 & 83.56 & \textbf{85.37} & 84.41 & 51.84 & \textbf{52.62} & 0.62  & 8.68  \\
                                     \cline{2-11}
\multirow{6}{*}{BSP Topk}           & \multirow{2}{*}{10}             & 4                           & 76.68 & 86.58 & \textbf{88.30} & 88.29 & \textbf{53.53} & 52.72 & 50.73 & 46.02 \\
                                     &                                 & 32                          & 55.61 & 78.88 & \textbf{86.75} & 85.89 & 51.18 & 48.68 & 54.28 & \textbf{54.62} \\
                                     & \multirow{2}{*}{100}            & 4                           & 59.47 & \textbf{77.20} & 76.94 & 76.16 & 34.00 & \textbf{48.00} & 46.08 & 40.83 \\
                                     &                                 & 32                          & 39.59 & 62.80 & \textbf{77.66} & 75.88 & 41.74 & 44.33 & \textbf{48.93} & 33.41 \\
                                     & \multirow{2}{*}{1000}           & 4                           & 43.11 & 59.52 & \textbf{62.66} & 58.48 & 39.17 & \textbf{39.89} & 38.17 & 34.19 \\
                                     &                                 & 32                          & 29.43 & 45.04 & \textbf{61.98} & 56.99 & 30.70 & 36.74 & 39.40 & \textbf{39.81} \\  \cline{2-11}
\multirow{6}{*}{BSP eftopk}         & \multirow{2}{*}{10}             & 4                           & 83.74 & 89.16 & 90.47 & \textbf{90.48} & \textbf{55.25} & 55.05 & 53.07 & 47.94 \\
                                     &                                 & 32                          & 64.74 & 84.80 & 88.65 & \textbf{88.94} & \textbf{54.00} & 51.58 & 52.18 & 46.60 \\
                                     & \multirow{2}{*}{100}            & 4                           & 83.78 & 88.91 & \textbf{90.90} & 89.28 & 51.73 & 55.36 & \textbf{55.74} & 47.66 \\
                                     &                                 & 32                          & 64.52 & 84.33 & 88.08 & \textbf{88.69} & \textbf{53.98} & 51.62 & 52.14 & 46.88 \\
                                     & \multirow{2}{*}{1000}           & 4                           & 84.02 & 88.94 & \textbf{90.88} & 89.89 & \textbf{55.29} & 55.20 & 53.36 & 53.55 \\
                                     &                                 & 32                          & 62.82 & 83.56 & \textbf{87.76} & 86.95 & 52.35 & \textbf{54.68} & 52.87 & 46.27 \\ \cline{2-11}
\multirow{2}{*}{Local SGD $\tau=2$}       & \multirow{2}{*}{None}           & 4                           & 84.04 & 88.97 & \textbf{92.01} & 88.86 & \textbf{55.03} & 54.94 & 52.87 & 46.72 \\
                                     &                                 & 32                          & 65.41 & 84.96 & 89.35 & \textbf{90.11} & \textbf{54.11} & 51.72 & 52.30 & 46.17 \\  \cline{2-11}
\multirow{2}{*}{Local SGD $\tau=2$ topk}  & \multirow{2}{*}{100}            & 4                           & 61.04 & 81.45 & 87.26 & \textbf{87.40} & 49.23 & 50.84 & \textbf{50.89} & 49.70 \\
                                     &                                 & 32                          & 39.67 & 63.12 & 81.22 & \textbf{82.09} & 40.88 & 44.39 & \textbf{50.79} & 46.54 \\  \cline{2-11}
\multirow{2}{*}{Local SGD $\tau=4$}       & \multirow{2}{*}{None}           & 4                           & 83.85 & 89.61 & \textbf{90.90} & 90.00 & 51.92 & 55.05 & \textbf{56.14} & 47.36 \\
                                     &                                 & 32                          & 65.55 & 85.09 & 89.44 & \textbf{89.85} & \textbf{54.36} & 51.72 & 53.40 & 47.15 \\ \cline{2-11}
\multirow{2}{*}{Local SGD $\tau=4$ topk}  & \multirow{2}{*}{100}            & 4                           & 60.73 & 81.18 & \textbf{85.20} & 83.83 & 50.54 & 50.78 & \textbf{52.64} & 50.25 \\
                                     &                                 & 32                          & 39.75 & 62.86 & 79.11 & \textbf{81.55} & 40.75 & 44.62 & \textbf{50.40} & 46.64 \\ \cline{2-11}
\multirow{2}{*}{Local SGD  $\tau=8$}       & \multirow{2}{*}{None}           & 4                           & 84.58 & 89.32 & \textbf{91.20} & 90.86 & \textbf{55.25} & 55.19 & 53.35 & 48.55 \\
                                     &                                 & 32                          & 64.48 & 84.82 & 89.41 & \textbf{90.09} & 54.07 & 51.72 & \textbf{54.91} & 47.30 \\ \cline{2-11}
\multirow{2}{*}{Local SGD  $\tau=8$ topk}  & \multirow{2}{*}{100}            & 4                           & 61.82 & 82.19 & \textbf{85.46} & 84.77 & 49.02 & 50.53 & \textbf{51.25} & 50.32 \\
                                     &                                 & 32                          & 39.71 & 61.93 & 77.55 & \textbf{78.73} & 39.43 & 44.81 & \textbf{49.83} & 46.10 \\ \cline{2-11}
\multirow{2}{*}{Local SGD  $\tau=16$}      & \multirow{2}{*}{None}           & 4                           & 84.02 & 89.25 & \textbf{90.99} & 90.56 & 55.32 & \textbf{55.41} & 53.43 & 48.84 \\
                                     &                                 & 32                          & 64.74 & 84.69 & 89.57 & \textbf{90.15} & 53.85 & 51.72 & \textbf{56.69} & 46.93 \\ \cline{2-11}
\multirow{2}{*}{Local SGD  $\tau=16$ topk} & \multirow{2}{*}{100}            & 4                           & 61.13 & 81.60 & 84.68 & \textbf{83.27} & 49.27 & 50.30 & \textbf{50.60} & 49.36 \\
                                     &                                 & 32                          & 39.31 & 61.62 & 71.72 & \textbf{76.07} & 40.06 & 44.44 & \textbf{49.06} & 46.08 \\ \cline{2-11}
\multirow{2}{*}{FedAvg}              & \multirow{2}{*}{None}           & 4                           & 62.41 & 84.37 & 89.81 & \textbf{90.10} & 52.20 & \textbf{55.23} & 54.91 & 54.92 \\
                                     &                                 & 32                          & 40.23 & 64.42 & 83.45 & \textbf{86.65} & 28.77 & 43.70 & 51.91 & \textbf{52.02} \\ \cline{2-11}
\multirow{2}{*}{FedAvg topk}         & \multirow{2}{*}{4}              & 4                           & 59.26 & 82.23 & 89.00 & \textbf{89.84} & 51.08 & 54.88 & \textbf{55.06} & 54.85 \\
                                     &                                 & 32                          & 38.36 & 60.63 & 82.15 & \textbf{85.14} & 27.90 & 42.26 & 50.29 & \textbf{51.32} \\ \cline{2-11}
\multirow{2}{*}{DPSGD}               & \multirow{2}{*}{None}           & 4                           & 84.60 & 89.38 & \textbf{91.08} & 91.01 & \textbf{55.41} & 55.19 & 52.92 & 48.13 \\
                                     &                                 & 32                          & 64.20 & 85.12 & 88.99 & \textbf{89.71} & 53.55 & 51.71 & \textbf{53.77} & 46.50 \\ \cline{2-11}
\multirow{2}{*}{DCD-PSGD}           & \multirow{2}{*}{4}              & 4                           & 68.39 & 86.32 & 89.79 & \textbf{90.33} & 54.70 & 55.13 & \textbf{55.45} & 54.07 \\
                                     &                                 & 32                          & 44.66 & 70.57 & 85.78 & \textbf{87.17} & 37.73 & 47.26 & \textbf{54.38} & 51.83 \\ \cline{2-11}
\multirow{2}{*}{CHOCO-SGD}          & \multirow{2}{*}{100}            & 4                           & 84.25 & 88.82 & \textbf{91.27} & 88.77 & \textbf{55.21} & 55.19 & 52.73 & 48.10 \\
                                     &                                 & 32                          & 64.52 & 84.71 & \textbf{89.00} & \textbf{89.00} & 53.16 & 51.75 & \textbf{54.49} & 47.15 \\
\toprule
\end{tabular}
}
\vspace{-20pt}
\end{table}

\section{Scheduling of Communication and Computing}\label{sec:scheduling}


DL models have a layer-wise structure that allows communication and computation tasks to be carried out in parallel during training~\cite{zhang2017poseidon}. The parallelism of communication and computing can effectively hide the communication time and reduce the overall training time. The communication and computation tasks can be formulated into a general directed acyclic graph (DAG)~\citep{shi2018adag}, pipelining or scheduling can be achieved to minimize iteration time.

In 2017, wait-free backward propagation (WFBP)~\citep{zhang2017poseidon,awan2017s} was proposed to pipeline the gradient communication of layer $l$ and the gradient computation of layer $l-1$, as they are independent. WFBP is naively supported in current DL frameworks (e.g., TensorFlow, PyTorch, Horovod, etc.). However, for small tensors of gradients, the latency term (startup time) can dominate the communication cost particularly on extremely large-scale clusters. To address the problem, merged-gradient (or tensor fusion) techniques (e.g., MG-WFBP~\citep{Shi2018MGWFBPED,shi2021mgj}) are proposed to alleviate the negative impact of the startup time. For layer-wise gradient sparsification~\cite{shi2020layer,shi2020communication}, three types of tasks (gradient computation, gradient sparsification, and gradient communication) can be pipelined. However, for the large tensors, the long communication time can cause their previous small tensors to wait. To minimize the waiting time of different tasks, communication and computation tasks can be scheduled by changing their execution order. Some studies~\citep{jayarajan2018priority,hashemi2018tictac,harlap2018pipedream,peng2019generic,zhang2023accelerating,shi2023pipemoe} have proposed scheduling communication tasks and computation tasks by changing the order of execution. Peng et al.~\citep{peng2019generic} proposed tensor partitioning to communication scheduling (even feed-forward computations can be paralleled with communications) to further reduce the communication cost. To prevent multiple communication tasks from affecting training performance, Wang et al. \citep{wang2020contention} proposed communication contention aware scheduling of multiple deep learning training jobs on GPU clusters. The All-Reduce operation can be decoupled to two continuous operations such that they are possible to be overlapped with feed-forward and backward computations~\cite{zhang2023accelerating}.

DL frameworks commonly employ DAGs to schedule compute operations. However, this approach presents challenges in managing gradient communication between workers~\citep{276984}. Specifically, if each worker simply uses blocking All-Reduce to communicate gradients based on their production order, the resulting mismatch in the orders of produced gradients between workers can lead to issues such as deadlock, data corruption, or communication inefficiency~\citep{276984}. To mitigate such issues, it is necessary to schedule a global order for collective communication of tensors~\citep{276984}. Notably, a recent benchmark study~\citep{MLSYS2022_cedebb6e} compares various compression methods with and without overlapping, and demonstrates that the latter approach can significantly reduce training time.

\section{Convergence Analysis}\label{sec:convergence}
There are some commonly used assumptions for the convergence analysis:
\begin{enumerate}
\item \textbf{Lipschitzian continuous gradient:} All function $f_i(\cdot)$ have $L$-Lipschitzian gradients: $||\nabla f_i(\mathbf{x}) - \nabla f_i(\mathbf{y})|| \leq L  \left\|\mathbf{x} - \mathbf{y}\right\|, \ \forall \mathbf{x}, \mathbf{y} \in \mathbb{R}^n.$


\item \textbf{Unbiased stochastic gradient:} The gradient calculated at  every iteration provides an unbiased estimation of the gradient of $f_i(\mathbf{x})$: $G_i(\mathbf{x}):= \mathbb{E}_{\xi \sim \mathcal{D}_i}  \nabla F_i(\mathbf{x};\xi) = \nabla f_i(\mathbf{x}), \  \forall \mathbf{x}\in \mathbb{R}^n,$ in which the $\mathcal{D}_i$ is the data distribution on $i$-th worker.


\item \textbf{Bounded variance:} The variance of the stochastic gradient is bounded: $\mathbb{E}_{\xi \sim \mathcal{D}_i} \| \nabla F_i(\mathbf{x};\xi)-\nabla f_i(\mathbf{x}) \|^2 \leq \sigma^2 , \forall i, \forall \mathbf{x} \in \mathbb{R}^n.$


\item \textbf{Bounded stochastic gradient:} The second moment of the stochastic gradients is bounded: $    \mathbb{E}_{\xi \sim \mathcal{D}_i} \| \nabla F_i(\mathbf{x};\xi) \|^2 \leq M^2 , \forall i, \forall \mathbf{x} \in \mathbb{R}^n.$

\item  \textbf{$\mu$-Strongly convex function:} $f(\mathbf{x}) \ge f(\mathbf{y}) + 	\left \langle \nabla f(\mathbf{y}), \mathbf{x} - \mathbf{y} \right \rangle + \frac{\mu}{2}\left\| \mathbf{x} -\mathbf{y} \right\|^2.$

\end{enumerate}

For the gossip (peer-to-peer) algorithms, there are some extra assumptions ~\citep{CommCompforDecent}:
\begin{enumerate}
\item \textbf{Symmetric doubly stochastic matrix:} The communication topology among workers is represented by a weighted matrix $W$ that is a real symmetric doubly stochastic matrix satisfying $W = W^T$ and $W\mathbf{1}=W$.

\item \textbf{Spectral gap:} Given the symmetric doubly stochastic matrix $W$, the spectral gap is defined as $\rho:= \text{max}\left\{\| \lambda_2(W)\|, \|\lambda_n(W) \| \right\}$ where $\lambda_2(W)$ represent the second  largest eigenvalues of $W$. The condition $\rho < 1$ must hold.
\end{enumerate}

For the compression methods, there are also some assumptions:
\begin{enumerate}
\item \textbf{$k$-contraction ~\citep{SparSGDwithMemory}:} For a parameter $0<d<n$, a $k$-contraction operator is a (possibly randomized) operator $C(\cdot)$: $\mathbb{R}^n \rightarrow \mathbb{R}^n$ that satisfies the contraction property: $\mathbb{E}\left\| \mathbf{x} - C(\mathbf{x})\right\|^2 \leq \left( 1 - \frac{d}{n}\right)\left\|\mathbf{x} \right\|^2,  \forall \mathbf{x} \in \mathbb{R}^n.$

\item \textbf{Unbiased compression ~\citep{CommCompforDecent}:} The stochastic compression operator $C(\cdot)$ is unbiased for any $\mathbf{x}$: $\mathbb{E}[C(\mathbf{x})] = \mathbf{x},$ and the compression operators are independent on different workers or at different iterations.


\end{enumerate}

\subsection{Centralized Architecture (PS or All-Reduce)}

\subsubsection{BSP-SGD}
PS and All-Reduce architectures have the same iteration equations because the All-Reduce algorithm only changes the way of implementing global synchronization. Therefore, the convergence analysis of BSP-SGD with the PS architecture can be applied to the All-Reduce architecture.

For quantization methods, Christopher et al.~\citep{Tamingwild} provided a convergence analysis with the martingale-based approach under both convex and non-convex objectives. In the case of QSGD, Alistarh et al. ~\citep{QSGD} not only proposed a family of quantization algorithms, but also provided a convergence analysis. They proved that QSGD can achieve convergence both for convex and non-convex objectives. They also proved that QSGD satisfies $\frac{1}{L} \mathbb{E}\left[ \left\| \nabla f(\mathbf{x}) \right\|^2_2  \right] \leq O\left(  \frac{\sqrt{L(f(\mathbf{x}_1) - f^*)}}{T} + \frac{min(n/s,\sqrt{n}/s)B}{L} \right) $, where $L$ represents Lipschitzian constant, $s$ and $B$ are hyper-parameters.

When implementing error accumulation in quantization methods, the variance bound of quantized gradients exceed that of QSGD~\citep{ECQSGD}. Wu et al.~\citep{ECQSGD} provided a convergence analysis in this scenario, but their analysis is limited by the requirement of unbiased gradient compression.

In the case of sparsification methods, Alistarh et al. ~\citep{ConvSparGrad} theoretically proved that the Top-$k$ algorithm can achieve good convergence even with biased estimation and non-convex objectives. This analysis is subject to deceleration proportional to $k$. Extending the convergence analysis of Top-$k$ methods to a more general range of sparsification methods, such as random-$k$ or $k$-sparsification methods, Stich et al.~\citep{SparSGDwithMemory} proved that this scheme preserves the same order of convergence rate as vanilla SGD, i.e. $O\left( \frac{G^2}{\mu T}\right)$. Shi et al.~\citep{ijcai2019473} analyzed the gTop-k sparsification method~\citep{GTopk} theoretically. They proved that gTop-k S-SGD provides convergence
guarantees for non-convex problems and has the same theoretical convergence rate as the mini-batch SGD.

\subsubsection{SSP/Asynchronous}
Ho et al.~\citep{MoreEffDMLviaStaleSync} have established $O(1/\sqrt{T})$ time for SGD under SSP. Lian et al.~\citep{10.5555/2969442.2969545} proposed an ergodic convergence rate $O(1/\sqrt{T})$ and proved that the linear speedup can be achieved if the number of workers is bounded by $\sqrt{T}$. Alistarh et al. ~\citep{Alistarh2018TheCO} provided the convergence bounds for lock-free SGD. Moreover, they exhibit a fundamental trade-off between the delay of the system and the convergence rate. Zhang et al.~\citep{zhang2018taming} also provided a convergence rate of asynchronous SGD under a non-convex object function and established a unifying condition for asynchronous SGD.

\subsubsection{Local SGD}
Stich~\citep{Sebastian2019} and Yu et al. ~\citep{Yu2018ParallelRS} have provided a concise convergence analysis for Local-SGD, demonstrating that this method has convergence guarantees with reducing communication costs. In~\citep{NIPS2019_9288}, Haddadpour et al. strengthened the convergence analysis for Local-SGD and showed that it can be far less expensive and more generally applicable than the current theory suggests. They proved that for loss functions that satisfy the Polyak-Łojasiewicz condition, $O((pT )1/3)$ rounds of communication suffice to achieve linear speedup with an error of $O(1/pT)$.

Patel and Dieuleveut~\citep{NIPS2019_9512} proposed a non-asymptotic error analysis that enables the comparison to one-shot averaging and mini-batch averaging, providing adaptive lower bounds on the communication frequency. They showed that Local-SGD can reduce communication by a factor of $O(\frac{\sqrt{T}}{N^{3/2}})$.
Artin et. al. ~\citep{spiridonoff2021communicationefficient} provided an interesting convergence analysis of Local-SGD under a strong convexity assumption. They proved that Local-SGD only needs $O(N)$ communication rounds.


\subsection{Decentralized Architecture (Gossip Communication)}
In contrast to the centralized architecture, the gossip architecture involves each worker possessing an individual model. Consequently, the convergence analysis of the gossip architecture differs from that of the centralized architecture. In order to ensure consensus in both the convergence analysis and algorithm design of the gossip architecture, certain considerations must be made.

\subsubsection{BSP}
Lian et al.~\citep{CanDecent} were the first to provide a theoretical analysis demonstrating that decentralized SGD algorithms for both convex and non-convex objectives can be faster than centralized SGD with less communication on the busiest node. They proved the convergence rate of decentralized SGD is $O\left( \frac{1}{T} + \frac{1}{\sqrt{nT}}\right)$, where $T$ and $n$ represents the number of iterations and workers respectively. When $T$ is sufficiently large, the $\frac{1}{\sqrt{nK}}$ term becomes dominant. In this scenario, the convergence rate can be approximated as $\frac{1}{\sqrt{nK}}$, indicating linear speedup achieved with the number of workers. 

Considering communication compression, Tang et al. ~\citep{CommCompforDecent} proposed two algorithms, ECD-PSGD and DCD-PSGD, with detailed convergence analysis on both convex and non-convex problems. The convergence rate of DCD-PSGD is $O\left( \frac{\sigma}{\sqrt{nT}} + \frac{\zeta^{\frac{2}{3}}}{T^{\frac{2}{3}}} + \frac{1}{T} \right)$, in which $\sigma$ is the variance of stochastic gradient estimator, $\zeta$ is the variance of gradient divergence between single worker and all workers. For ECD-PSGD, they proved its convergence rate is $O\left( \frac{\sigma \left(1+ \frac{{\tilde{\sigma}}^2 logT}{n} \right)}{\sqrt{nT}} + \frac{\zeta^{\frac{2}{3}} \left(1+ \frac{{\tilde{\sigma}}^2 logT}{n}\right)}{T^{\frac{2}{3}}} + \frac{1}{T} + \frac{{\tilde{\sigma}}^2 logT}{T} \right)$, which is slightly worse than DCD-PSGD, due to the extra terms $O\left( \frac{\sigma {\tilde{\sigma}}^2 logT}{n\sqrt{nT}} + \frac{\zeta^{\frac{2}{3}} {\tilde{\sigma}}^2 logT}{nT^{\frac{2}{3}}} + \frac{{\tilde{\sigma}}^2 logT}{T} \right)$. 

It should be noted that the convergence analysis of ECD-PSGD and DCD-PSGD is restricted to unbiased compression operators. Koloskova et al. ~\citep{DecentStocOptimAndGossip} proposed CHOCO-SGD, the first method that is capable of handling biased compression operators. For convex objectives, their algorithm was found to converge with a rate of $O\left( 1/(nT) + 1/(T\delta^2\omega) \right)$, where $\delta$ denotes the eigenvalue gap of the gossip matrix and $\omega \leq 1$ represents the compression ratio. Furthermore, Koloskova et al.~\citep{Koloskova2019DecentralizedDL} demonstrated that CHOCO-SGD~\citep{DecentStocOptimAndGossip} can achieve convergence at a rate of $O(1/\sqrt{nT}+n/\left(\rho^4\sigma^2T \right) )$ for non-convex smooth functions, where $n$ denotes the number of nodes, $T$ the number of iterations, $\rho$ the spectral gap of the mixing matrix and $\sigma$ the compression ratio. Moreover, they proved that CHOCO-SGD can converge with arbitrarily high compression operators and achieve linear speedup.


\subsubsection{Asynchronous}
Jeff et al.~\citep{DBLP:journals/corr/abs-1803-05880} provided a sketch of proof of convergence without the convergence rate. Lian et al. ~\citep{pmlr-v80-lian18a} proposed a theoretical analysis of asynchronous gossip SGD with the non-convex objective function, which converges with the $O(1/\sqrt{T})$ rate and has linear speedup with respect to the number of workers. Assran et al.~\citep{StocGradPush} provided theoretical guarantees for another asynchronous gossip SGD algorithm that achieves similar convergence of~\citep{pmlr-v80-lian18a}. 

Table \ref{tab:summary_of_methods} provides a comparison of different algorithms. Most algorithms show $O(\frac{1}{T})$ convergence rate for convex object functions and $O(\frac{1}{T})$ for non-convex object functions. However, the communication costs vary depending on the architecture and algorithm used.

\begin{table*}[htbp]
\centering
\fontsize{6}{6}\selectfont
\caption{Summary of distributed learning algorithms}
\label{tab:summary_of_methods}
\addtolength{\tabcolsep}{-5pt}
\resizebox{\linewidth}{!}{
\begin{tabular}{cccccccc}
\hline
\multirow{2}{*}{Arch.} & \multirow{2}{*}{Comm.} & \multirow{2}{*}{Compression} & \multicolumn{2}{c}{Communication Cost in Big $O$} & \multicolumn{2}{c}{Convergence in Big $O$} & \multirow{2}{*}{References of Method}  \\
 & & & Server & Workers & convex & non-convex &  \\ \hline\hline
\multirow{12}{*}{PS} & \multirow{3}{*}{BSP} & Null &$O(32nNT)$  & $O(32NT)$ & $O(\frac{1}{T})$~\citep{Bottou2016OptimizationMF} & $O(\frac{1}{\sqrt{T}})$~\citep{Bottou2016OptimizationMF} &~\citep{Krizhevsky2014OneWT,Bradley2011,552669,375451} \\
 &  & Quant. & $O(32nNT)$ & $O(bNT)$ &$O(\frac{1}{T})$~\citep{DisLearningDIANA} ~\citep{AlinearSpeedupAnalysis} & \shortstack[c]{$O(\frac{1}{\sqrt{T}})$  ~\citep{AlinearSpeedupAnalysis,QSGD}} & \shortstack[c]{~\citep{3LC} ~\citep{FLStrategy,DisLearningDIANA,signSGDwithVote,TernGrad,Horvath2019NaturalCF} ~\citep{signSGD}\\ ~\citep{ATOMO,NIPS2019_8598,Mishchenko201999OP,QSGD,scalableDisDNN,CommQuantforDataPara,EFsignSGD}}  \\ 
 &  & Spars.  &$O(32nNT)$  & $O(k(\log N+32)T)$ &$O(\frac{1}{T})$~\citep{NIPS2019_9473,SparSGDwithMemory} &  $O(\frac{1}{\sqrt{T}})$ ~\citep{ConvSparGrad,ijcai2019473} &\shortstack[c]{~\citep{3LC,SparCommforDisGD} ~\citep{FLStrategy,NIPS2019_9473,Horvath2019NaturalCF,ATOMO}\\ ~\citep{GradSparforDisOptim,DGC,CommQuantforDataPara,Zhao2019GlobalMC,GTopk,VarianceGradCompression}} \\ \cline{2-8} 
 & \multirow{3}{*}{SSP}  & Null & $O(32N\sum_i^n T_i)$ & $O(32NT_i)$ & - & $O(\frac{1}{\sqrt{T}})$ ~\citep{MoreEffDMLviaStaleSync} & ~\citep{RevistSynSGD,MoreEffDMLviaStaleSync}\\
 &  & Quant. &  - & - & - & - & -\\
 &  & Spars.  & - & - & - & - & -\\ \cline{2-8} 
 & \multirow{3}{*}{ASP}  & Null & $O(32N\sum_i^n T_i)$  &  $O(32NT_i)$&$O(\frac{1}{T})$~\citep{Sebastian2019,10.5555/2969442.2969545} &$O(\frac{1}{\sqrt{T}})$ ~\citep{zhang2018taming}  &~\citep{Sebastian2019,dean2012large,Meng2016AAS3060832} \\
 &  & Quant. & $O(32N\sum_i^n T_i)$ &  $O(bNT_i)$ &$O(\frac{1}{T})$~\citep{NIPS2019_8694} & $O(\frac{1}{\sqrt{T}})$~\citep{QSGD,1bit} & ~\citep{NIPS2019_8694,QSGD,1bit,NIPS2019_9610}\\
 &  & Spars.  & $O(32N\sum_i^n T_i)$ &  $O(k(\log N+32)T_i)$ &$O(\frac{1}{T})$~\citep{GradSparforDisOptim} & - & ~\citep{GradSparforDisOptim,AsynDisMLspar,Meng2016AAS3060832,NIPS2019_9610}\\ \cline{2-8} 
 & \multirow{3}{*}{L-SGD}  & Null  & $O(32N\frac{T}{\tau})$ & $O(32N\frac{T}{\tau})$ &$O(\frac{1}{T})$~\citep{Sebastian2019} &$O(\frac{1}{\sqrt{T}})$ ~\citep{NIPS2019_9288,pmlrv97yu19d,pmlrv97yu19c}  & ~\citep{PSGD,Sebastian2019} ~\citep{NIPS2019_9288,Lin2018DontUL,pmlrv97yu19d,pmlrv97yu19c}\\
 &  & Quant. & $O(32N\frac{T}{\tau})$ & $O(bN\frac{T}{\tau})$ & - &  $O(\frac{1}{\sqrt{T}})$~\citep{AlinearSpeedupAnalysis}& ~\citep{SparseTernaryCompressionSTC,AlinearSpeedupAnalysis,NIPS2019_9610}\\
 &  & Spars.  & $O(32N\frac{T}{\tau})$ & $O(k(\log N+32)\frac{T}{\tau})$ & - & - & ~\citep{SparseTernaryCompressionSTC,NIPS2019_9610}\\ \cline{1-8} 
 
\multirow{6}{*}{A.R.} & \multirow{3}{*}{BSP} & Null & -  & $O(32NT)$ &$O(\frac{1}{T})$~\citep{Bottou2016OptimizationMF} & $O(\frac{1}{\sqrt{T}})$~\citep{Bottou2016OptimizationMF} &~\citep{Krizhevsky2014OneWT,Bradley2011,552669,375451} \\
 &  & Quant. & -  & $O(bNT)$ &$O(\frac{1}{T})$~\citep{DisLearningDIANA} ~\citep{AlinearSpeedupAnalysis} & \shortstack[c]{$O(\frac{1}{\sqrt{T}})$  ~\citep{AlinearSpeedupAnalysis,QSGD}} & \shortstack[c]{~\citep{3LC} ~\citep{FLStrategy,DisLearningDIANA,signSGDwithVote,TernGrad,Horvath2019NaturalCF} ~\citep{signSGD}\\ ~\citep{ATOMO,NIPS2019_8598,Mishchenko201999OP,QSGD,scalableDisDNN,CommQuantforDataPara,EFsignSGD}}  \\ 
 &  & Spars.  &  - & $O(kn(\log N+32)T)$ &$O(\frac{1}{T})$~\citep{NIPS2019_9473,SparSGDwithMemory} &  $O(\frac{1}{\sqrt{T}})$ ~\citep{ConvSparGrad,ijcai2019473} &\shortstack[c]{~\citep{3LC,SparCommforDisGD} ~\citep{FLStrategy,NIPS2019_9473,Horvath2019NaturalCF,ATOMO}\\ ~\citep{GradSparforDisOptim,DGC,CommQuantforDataPara,Zhao2019GlobalMC,GTopk,VarianceGradCompression}} \\ \cline{2-8} 
 & \multirow{3}{*}{L-SGD}  & Null  & -  & $O(32N\frac{T}{\tau})$ &$O(\frac{1}{T})$~\citep{Sebastian2019} &$O(\frac{1}{\sqrt{T}})$ ~\citep{NIPS2019_9288,pmlrv97yu19d,pmlrv97yu19c}  & ~\citep{PSGD,Sebastian2019} ~\citep{NIPS2019_9288,pmlrv97yu19d,Lin2018DontUL,pmlrv97yu19d,pmlrv97yu19c}\\
 &  & Quant. &  - & $O(bN\frac{T}{\tau})$ & - &  $O(\frac{1}{\sqrt{T}})$~\citep{AlinearSpeedupAnalysis}& ~\citep{SparseTernaryCompressionSTC,AlinearSpeedupAnalysis,NIPS2019_9610}\\
 &  & Spars.  & -  & $O(kn(\log N+32)\frac{T}{\tau})$& - & - & ~\citep{SparseTernaryCompressionSTC,NIPS2019_9610}\\ \cline{1-8} 
 
\multirow{9}{*}{Gossip} & \multirow{3}{*}{BSP} & Null & - & $O(32Nn_{peers}T)$ & &$O(\frac{1}{\sqrt{T}})$~\citep{CanDecent}  &~\citep{CanDecent,StocGradPush} \\ 
 &  & Quant. & - & $O(nbN_{peers}T)$ & - & $O(\frac{1}{\sqrt{T}})$~\citep{ECQSGD} &~\citep{CommCompforDecent,DecentStocOptimAndGossip,7544448,NIPS2018_7705,ECQSGD,NIPS2019_9047} \\
 &  & Spars.  & - & $O(k(\log N+32)n_{peers}T)$ & - & - & ~\citep{DecentStocOptimAndGossip,tang2020communication}\\ \cline{2-8} 
 & \multirow{3}{*}{ASP}  & Null & - & $O(32Nn_{peers}T_i)$  & - & $O(\frac{1}{\sqrt{T}})$ ~\citep{pmlr-v80-lian18a} &~\citep{pmlr-v80-lian18a} \\ 
 &  & Quant. & - & - & - & - & - \\
 &  & Spars.  & - & - & - & - & - \\ \cline{2-8} 
 & \multirow{3}{*}{L-SGD}  & Null  &  - &$O(32Nn_{peers}\frac{T}{\tau})$  &$O(\frac{1}{T})$ ~\citep{Jianyu180807576} & $O(\frac{1}{\sqrt{T}})$ ~\citep{Jianyu180807576} & ~\citep{Jianyu180807576}\\ 
 &  & Quant. & - & - & - & - & - \\
 &  & Spars.  & - & - & - & - & - \\ \cline{1-8} 
\multicolumn{3}{c}{Pipelining} & \multicolumn{5}{c}{~\citep{zhang2017poseidon,awan2017s,Shi2018MGWFBPED,shi2020layer,shi2020communication,aritra2020discrepancy}} \\
\multicolumn{3}{c}{Scheduling} & \multicolumn{5}{c}{ ~\citep{Shi2018MGWFBPED,hashemi2018tictac,harlap2018pipedream,peng2019generic,jayarajan2018priority,DBLP:journals/corr/abs-1810-08313,8884800}} \\ \cline{1-8}
\end{tabular}
}
\begin{tablenotes}  
    \footnotesize 
    \item Notes: 1) The ``Arch.'' indicates the architecture supported in the original paper, ``A.R.'' represents All-Reduce, ``Quant.'' quantization and ``Spars.'' sparsification. The ``Comm.'' column indicates the communication scheme which includes ``BSP'' (Bulk Synchronous Parallel), ``SSP'' (Stale Synchronous Parallel), ``ASP'' (ASynchronous Parallel), and ``L-SGD''(Local SGD), and $\tau$ in ``L-SGD'' means the local steps.
    2) Some methods use both compression techniques and both Asyc and Local SGD. 3) Some methods also optimize download communication, we have listed them together with upload communication. 4) The communication complexity and convergence rate of some paper maybe different, we just list out the common ones. 5) Pipelining and scheduling can be used in many methods, so we only list methods that uses pipeline without classifying it into any category.
\end{tablenotes}          
\end{table*}

\section{Auxiliary Technologies}\label{sec:auxiliary}
Communication compression methods can achieve convergence and reduce communication loads with the aid of auxiliary technologies. 

\subsection{Error Accumulation}
In 1-bit SGD ~\citep{1bit}, the current quantization error is added to the next mini-batch gradient before quantization in the next iteration through error accumulation. This error-feedback mechanism ensures that all gradients are eventually added up into the model with full accuracy, although with some delays. This process is described by Eq. (\ref{1bitSGD1}), (\ref{1bitSGD2}) and (\ref{1bitSGD3}). Karimireddy et al. ~\citep{EFsignSGD} proposed EF-SIGNSGD (SIGNSGD with Error-Feedback), which also uses error accumulation by storing the error locally and adding it to the next step.

Numerous formulations of error accumulation have been proposed in the literature ~\citep{3LC,SparCommforDisGD,sparsebinarycompression,Adacomp,CommQuantforDataPara,DisLearningDIANA,ECQSGD,EFsignSGD,SparseTernaryCompressionSTC,scalableDisDNN,SparCommforDisGD,VarianceGradCompression,DEF,9442310,pmlr-v139-tang21a,EF21}. In summary, error accumulation involves incorporating the quantization error into the subsequent gradient computations, which improves the accuracy of the final model. This technique can be formulated as following steps: (1) gradient compression: $C_{i,t} = Sparse(v_{i,t-1} + \nabla_{i,t})$; (2) error accumulation $v_{i,t} = \nabla_{i,t} - C_{i,t}$; (3) update weight: $x_{t+1} = x_{t} - \gamma\frac{1}{n} \sum_{i=1}^n C_{i,t}$, where $C_{i,t}$ represents the updates which are compressed by any compression method $Sparse(\cdot)$, $\nabla$ the gradient, at $t$-th iteration and in worker $i$.


Wu et al.~\citep{ECQSGD} proposed ECQ-SGD (Error Compensated Quantized SGD). This method differs from other compression techniques as it considers not only the compression error in the current iteration but also all previous errors by accumulating them. Tang et al. proposed 1-bit Adam~\citep{pmlr-v139-tang21a}, which incorporates error compensation into distributed Adam with 1-bit compression.

\subsection{Momentum Correction}
Lin et al.~\citep{DGC} proposed the Momentum Correction technique to assist DGC in utilizing Momentum SGD. This approach involves the application of vanilla momentum SGD~\citep{vanillaMomentum} and error accumulation to the sparse gradient. The update process is as follows:
(1) momentum accumulation: $u_{i,t} = mu_{i,t-1} + \nabla_{i,t}$; (2) error accumulation $v_{i,t} = v_{i,t-1} + u_{i,t}$; (3) update weight: $x_{t+1} = x_{t} - \gamma\sum_{i=1}^n sparse(v_{i,t})$, where $m$ denotes the coefficient of momentum and $u_{i,t}$ denotes the momentum at $t$-th iteration on worker $i$. Momentum Correction has also been exploited in  ~\citep{sparsebinarycompression,DisLearningDIANA,SparseTernaryCompressionSTC}.


Zhao et al.~\citep{Zhao2019GlobalMC} proposed Global Momentum Compression, of which the update process becomes:
(1) momentum accumulation: $u_{i,t} = \nabla_{i,t} - m(x_t - x_{t-1})$;
(2) error accumulation: $v_{i,t} = v_{i,t-1} + u_{i,t} - sparse(v_{i,t-1} + u_{i,t})$;
(3) update weight: $x_{t+1} = x_{t} - \gamma\sum_{i=1}^n sparse(v_{i,t-1} + u_{i,t})$. Thus, the gradient and them momentum are both compressed.


\subsection{Low-rank Decomposition}
The low-rand decomposition of matrix means to decompose a large matrix into the multiplication of small matrices. Thus, senders can reduce communication costs by sending smaller matrices, and recovering original matrix on receivers.

Wang et al.~\citep{ATOMO} developed a new method named Atomic Sparsification (ATOMO). They demonstrated that gradient sparsification and quantization are parts of a general approach that sparsifies the gradients in some atomic decomposition, such as entry-wise methods like QSGD, singular value decomposition (SVD), Fourier decomposition, etc. ATOMO aims to minimize the variance of the sparsified gradient that is sparse on the atomic basis and maintain it as an unbiased estimator of the original gradient. They illustrate that the 1-bit QSGD and TernGrad are special cases of ATOMO. Furthermore, they improved ATOMO with SVD, named as Spectral-ATOMO. In their experiments, Spectral-ATOMO reduces the training time by a factor of $2\times$ and $3\times$, compared to QSGD and TernGrad, respectively.

Ivkin et al. ~\citep{NIPS2019_9473} exploited a technique that is widely adopted in distributed systems, Count Sketch ~\citep{Charikar2002FFI}. It compresses a gradient vector $G$ into a sketch $S(G)$ of size $O(1/\epsilon \log n)$., which can approximate every coordinate of $G$ and the $l_2$ norm of the entire gradient. Every worker sends this sketched gradient to the server, and the server recovers the $d$ largest coordinates of the sum of gradients and then performs the update.

\subsection{Local Gradient Clipping} 
The Gradient Clipping ~\citep{GradientClipping} is a widely used method in vanilla SGD to avoid the exploding gradient problem. This technique involves clipping all gradients that have values exceeding a user-defined threshold. For the BSP-SGD with gradient sparsification, Lin et al.~\citep{DGC} modified it as Local Gradient Clipping, which is performed before adding the error accumulation term and the gradient in current iteration. The $k$-th worker has a threshold $thr\left(G_k\right)$ for its local gradient $ G_k $, and the aggregation of gradients has a threshold $thr\left(G\right)$ for the global gradient $G := \sum_{k=1}^N G_k$.

If we assume that all $N$ workers have an independent and identically distributed (IID) gradient distribution with variance $\sigma^2$, then the aggregation of all gradients have the variance $N\sigma^2$. So there are
\begin{equation}
    E\left[\left\| G_k \right\|_2\right] \approx \sigma, \ E\left[\left\| G \right\|_2\right] \approx N^{1/2}\sigma.
\end{equation}
Local Gradient Clipping can restore the original variance of the aggregation models by adjusting the clipping threshold with respect to the number of workers.

\subsection{Warm-up Training}
Lin et al.\citep{DGC} utilized Warm-up Training~\citep{imagenet1hour} in DGC to overcome the problem of rapidly varying neural network behavior in the first few epochs of training when the gradient values are excessively large. This technique involves dividing the training process into two periods: the warm-up period and the normal training period. During the warm-up period, the algorithm trains the model using a less aggressive learning rate and less aggressive gradient sparsity. This reduces the number of extreme gradients being delayed. In addition, the gradient sparsity increases exponentially from a small value to the final value. Subsequently, the algorithm trains the model using high sparsity and a decreasing learning rate, similar to vanilla SGD.


\section{Conclusion and Future Directions}\label{sec:conclusion}
In this survey, we provided a comprehensive introduction to the communication-efficient distributed DL algorithms. We summarized the taxonomy of distributed DL and classified the communication-efficient distributed training algorithms into four main dimensions: 1) synchronous schemes, 2) system architectures, 3) compression techniques, and 4) parallelism of communication and computing tasks. For each dimension, related techniques that address communication problems were introduced comprehensively. Furthermore, we provided a review of convergence bounds of different algorithms and some auxiliary techniques that help accelerate the training speed.

Below, we summarize some challenges and future directions:
\begin{enumerate}
\item \textbf{Will current communication-efficient methods work in training foundation models?} With current rapid developments of large language models, efficient training methods are crucially important for developing new techniques, democratizing them and energy saving. Thus, verifying current and designing new communication-efficient distributed training methods will be valuable.

\item \textbf{Higher compression level.} Is it possible to implement an extremely high compression level without sacrificing training performance? While the current quantization method can reduce data size by a factor of 32 and sparsification by a factor of 100-1000, achieving a higher compression ratio while maintaining model accuracy remains a challenging question. 

\item \textbf{Adaptive Compression.} Gradient/model compression can reduce communication size and time. However, achieving a very high compression ratio typically requires a larger number of iterations to reach the target optimization error, making it challenging to balance compression ratio and convergence speed. Can different compression ratios be set for different layers/tensors in a deep model or for peers with varying network bandwidth to achieve optimal system throughput?

\item \textbf{Fault-tolerant algorithms.} While the algorithm runs smoothly in a stable computing cluster, uncertainty arises when a large number of heterogeneous devices are used to train deep models. Factors such as severe stragglers, network congestion, and worker failure may cause interference. Developing more fault-tolerant algorithms is an important direction for increasing the reliability of training.

\end{enumerate}




\bibliographystyle{abbrv}

\bibliography{cites}

\begin{thebibliography}{100}

\bibitem{AsystemLarScalSupML}
M.~Abadi, P.~Barham, J.~Chen, Z.~Chen, A.~Davis, J.~Dean, M.~Devin,
  S.~Ghemawat, G.~Irving, M.~Isard, M.~Kudlur, et~al.
\newblock Tensorflow: A system for large-scale machine learning.
\newblock In {\em OSDI}, 2016.

\bibitem{MLSYS2022_cedebb6e}
S.~Agarwal, H.~Wang, S.~Venkataraman, and D.~Papailiopoulos.
\newblock On the utility of gradient compression in distributed training
  systems.
\newblock In {\em MLSys}, 2022.

\bibitem{Ahmed2012}
A.~Ahmed, M.~Aly, J.~Gonzalez, S.~Narayanamurthy, and A.~J. Smola.
\newblock Scalable inference in latent variable models.
\newblock In {\em WSDM}, 2012.

\bibitem{pmlrv32ahn14}
S.~Ahn, B.~Shahbaba, and M.~Welling.
\newblock Distributed stochastic gradient mcmc.
\newblock In {\em ICML}, 2014.

\bibitem{SparCommforDisGD}
A.~F. Aji and K.~Heafield.
\newblock Sparse communication for distributed gradient descent.
\newblock In {\em EMNLP}, 2017.

\bibitem{ConvSparGrad}
D.~Alistarh, T.~Hoefler, M.~Johansson, S.~Khirirat, N.~Konstantinov, and
  C.~Renggli.
\newblock The convergence of sparsified gradient methods.
\newblock In {\em NeurIPS}, 2018.

\bibitem{QSGD}
D.~Alistarh, J.~Li, R.~Tomioka, and M.~Vojnovic.
\newblock Qsgd: Randomized quantization for communication-optimal stochastic
  gradient descent.
\newblock {\em ArXiv}, abs/1610.02132, 2016.

\bibitem{Alistarh2018TheCO}
D.~Alistarh, C.~D. Sa, and N.~Konstantinov.
\newblock The convergence of stochastic gradient descent in asynchronous shared
  memory.
\newblock In {\em PODC '18}, 2018.

\bibitem{deepspeech}
D.~Amodei, S.~Ananthanarayanan, R.~Anubhai, et~al.
\newblock Deep speech 2: End-to-end speech recognition in english and mandarin.
\newblock In {\em ICML}, 2016.

\bibitem{StocGradPush}
M.~Assran, N.~Loizou, N.~Ballas, and M.~Rabbat.
\newblock Stochastic gradient push for distributed deep learning.
\newblock In {\em ICML}, 2019.

\bibitem{awan2017s}
A.~A. Awan, K.~Hamidouche, J.~M. Hashmi, and D.~K. Panda.
\newblock S-caffe: Co-designing mpi runtimes and caffe for scalable deep
  learning on modern {GPU} clusters.
\newblock In {\em Acm Sigplan Notices}, volume~52, pages 193--205. ACM, 2017.

\bibitem{7942055DistributedLinearized}
N.~S. {Aybat}, Z.~{Wang}, T.~{Lin}, and S.~{Ma}.
\newblock Distributed linearized alternating direction method of multipliers
  for composite convex consensus optimization.
\newblock {\em IEEE Transactions on Automatic Control}, 63(1):5--20, Jan 2018.

\bibitem{LayNorm}
J.~Ba, J.~R. Kiros, and G.~E. Hinton.
\newblock Layer normalization.
\newblock {\em ArXiv}, abs/1607.06450, 2016.

\bibitem{NIPS2019_9610}
D.~Basu, D.~Data, C.~Karakus, and S.~Diggavi.
\newblock {Qsparse-local-SGD}: Distributed {SGD} with quantization,
  sparsification and local computations.
\newblock In {\em NeurIPS}. 2019.

\bibitem{10.1145/3320060}
T.~Ben-Nun and T.~Hoefler.
\newblock Demystifying parallel and distributed deep learning: An in-depth
  concurrency analysis.
\newblock {\em ACM Comput. Surv.}, 52(4), Aug. 2019.

\bibitem{GradientClipping}
Y.~{Bengio}, P.~{Simard}, and P.~{Frasconi}.
\newblock Learning long-term dependencies with gradient descent is difficult.
\newblock {\em IEEE Transactions on Neural Networks}, 5(2):157--166, March
  1994.

\bibitem{signSGD}
J.~Bernstein, Y.~Wang, K.~Azizzadenesheli, and A.~Anandkumar.
\newblock {SIGNSGD:} compressed optimisation for non-convex problems.
\newblock In {\em ICML}, 2018.

\bibitem{signSGDwithVote}
J.~Bernstein, J.~Zhao, K.~Azizzadenesheli, and A.~Anandkumar.
\newblock signsgd with majority vote is communication efficient and byzantine
  fault tolerant.
\newblock {\em ArXiv}, abs/1810.05291, 2018.

\bibitem{bijral2016data}
A.~S. Bijral, A.~D. Sarwate, and N.~Srebro.
\newblock On data dependence in distributed stochastic optimization.
\newblock {\em arXiv: Optimization and Control}, 2016.

\bibitem{Bottou2016OptimizationMF}
L.~Bottou, F.~E. Curtis, and J.~Nocedal.
\newblock Optimization methods for large-scale machine learning.
\newblock {\em SIAM Review}, 60:223--311, 2016.

\bibitem{1498447Gossipalgorithms}
S.~{Boyd}, A.~{Ghosh}, B.~{Prabhakar}, and D.~{Shah}.
\newblock Gossip algorithms: design, analysis and applications.
\newblock In {\em Proceedings IEEE 24th Annual Joint Conference of the IEEE
  Computer and Communications Societies.}, 2005.

\bibitem{gossip3}
S.~Boyd, A.~Ghosh, B.~Prabhakar, and D.~Shah.
\newblock Randomized gossip algorithms.
\newblock {\em IEEE/ACM Trans. Netw.}, 14(SI):2508--2530, June 2006.

\bibitem{Bradley2011}
J.~K. Bradley, A.~Kyrola, D.~Bickson, and C.~Guestrin.
\newblock Parallel coordinate descent for l1-regularized loss minimization.
\newblock In {\em ICML}, 2011.

\bibitem{gpt3}
T.~B. Brown, B.~Mann, N.~Ryder, M.~Subbiah, J.~Kaplan, P.~Dhariwal,
  A.~Neelakantan, P.~Shyam, G.~Sastry, A.~Askell, et~al.
\newblock Language models are few-shot learners.
\newblock {\em arXiv preprint arXiv:2005.14165}, 2020.

\bibitem{642949}
J.~{Bruck}, {Ching-Tien Ho}, S.~{Kipnis}, E.~{Upfal}, and D.~{Weathersby}.
\newblock Efficient algorithms for all-to-all communications in multiport
  message-passing systems.
\newblock {\em IEEE TPDS}, 8(11):1143--1156, Nov 1997.

\bibitem{Carli2010}
R.~Carli, F.~Fagnani, P.~Frasca, and S.~Zampieri.
\newblock Gossip consensus algorithms via quantized communication.
\newblock {\em Automatica}, 46(1):70--80, Jan. 2010.

\bibitem{Charikar2002FFI}
M.~Charikar, K.~Chen, and M.~Farach-Colton.
\newblock Finding frequent items in data streams.
\newblock In {\em Proceedings of the 29th International Colloquium on Automata,
  Languages and Programming}, ICALP '02, 2002.

\bibitem{375451}
T.~{Cheatham}, A.~{Fahmy}, D.~C. {Stefanescu}, and L.~G. {Valiant}.
\newblock Bulk synchronous parallel computing-a paradigm for transportable
  software.
\newblock In {\em Proceedings of the Twenty-Eighth Annual Hawaii International
  Conference on System Sciences}, volume~2, pages 268--275 vol.2, Jan 1995.

\bibitem{Adacomp}
C.~Chen, J.~Choi, D.~Brand, A.~Agrawal, W.~Zhang, and K.~Gopalakrishnan.
\newblock Adacomp : Adaptive residual gradient compression for data-parallel
  distributed training.
\newblock In {\em AAAI}, 2018.

\bibitem{8737587}
C.~{Chen}, W.~{Wang}, and B.~{Li}.
\newblock Round-robin synchronization: Mitigating communication bottlenecks in
  parameter servers.
\newblock In {\em IEEE INFOCOM}, 2019.

\bibitem{RevistSynSGD}
J.~Chen, R.~Monga, S.~Bengio, and R.~Jozefowicz.
\newblock Revisiting distributed synchronous sgd.
\newblock In {\em ICLR Workshop Track}, 2016.

\bibitem{chen2016scalable}
K.~Chen and Q.~Huo.
\newblock Scalable training of deep learning machines by incremental block
  training with intra-block parallel optimization and blockwise model-update
  filtering.
\newblock In {\em ICASSP-2016}, March 2016.

\bibitem{LAG}
T.~Chen, G.~Giannakis, T.~Sun, and W.~Yin.
\newblock Lag: Lazily aggregated gradient for communication-efficient
  distributed learning.
\newblock In {\em NeurIPS}, 2018.

\bibitem{chetlur2014cudnn}
S.~Chetlur, C.~Woolley, P.~Vandermersch, J.~Cohen, J.~Tran, B.~Catanzaro, and
  E.~Shelhamer.
\newblock {cuDNN}: Efficient primitives for deep learning.
\newblock {\em arXiv preprint arXiv:1410.0759}, 2014.

\bibitem{padam}
T.~Chilimbi, Y.~Suzue, J.~Apacible, and K.~Kalyanaraman.
\newblock Project adam: Building an efficient and scalable deep learning
  training system.
\newblock In {\em OSDI}, 2014.

\bibitem{chu2017efficient}
C.-H. Chu, X.~Lu, A.~A. Awan, H.~Subramoni, J.~Hashmi, B.~Elton, and D.~K.
  Panda.
\newblock Efficient and scalable multi-source streaming broadcast on {GPU}
  clusters for deep learning.
\newblock In {\em ICPP}, 2017.

\bibitem{gossip4}
I.~Colin, A.~Bellet, J.~Salmon, and S.~Cl{\'e}men\c{c}on.
\newblock Gossip dual averaging for decentralized optimization of pairwise
  functions.
\newblock In {\em ICML}, 2016.

\bibitem{Cui2014EBS}
H.~Cui, J.~Cipar, Q.~Ho, J.~K. Kim, S.~Lee, A.~Kumar, J.~Wei, W.~Dai, G.~R.
  Ganger, P.~B. Gibbons, G.~A. Gibson, and E.~P. Xing.
\newblock Exploiting bounded staleness to speed up big data analytics.
\newblock In {\em Proceedings of the 2014 USENIX Conference on USENIX Annual
  Technical Conference}, USENIX ATC'14, pages 37--48, Berkeley, CA, USA, 2014.
  USENIX Association.

\bibitem{Cui2014IP67}
H.~Cui, A.~Tumanov, J.~Wei, L.~Xu, W.~Dai, J.~Haber-Kucharsky, Q.~Ho, G.~R.
  Ganger, P.~B. Gibbons, G.~A. Gibson, and E.~P. Xing.
\newblock Exploiting iterative-ness for parallel ml computations.
\newblock In {\em ACM SOCC}, 2014.

\bibitem{Cui2016}
H.~Cui, H.~Zhang, G.~R. Ganger, P.~B. Gibbons, and E.~P. Xing.
\newblock Geeps: Scalable deep learning on distributed gpus with a
  gpu-specialized parameter server.
\newblock In {\em EuroSys}, 2016.

\bibitem{DBLP:journals/corr/abs-1803-05880}
J.~Daily, A.~Vishnu, C.~Siegel, T.~Warfel, and V.~Amatya.
\newblock Gossipgrad: Scalable deep learning using gossip communication based
  asynchronous gradient descent.
\newblock {\em CoRR}, abs/1803.05880, 2018.

\bibitem{mpi4py}
L.~Dalcín, R.~Paz, M.~Storti, and J.~D’Elía.
\newblock Mpi for python: Performance improvements and mpi-2 extensions.
\newblock {\em Journal of Parallel and Distributed Computing}, 2008.

\bibitem{dean2012large}
J.~Dean, G.~Corrado, R.~Monga, K.~Chen, M.~Devin, M.~Mao, M.~Ranzato,
  A.~Senior, P.~Tucker, K.~Yang, et~al.
\newblock Large scale distributed deep networks.
\newblock In {\em NeurIPS}, 2012.

\bibitem{NIPS2019_8452}
A.~Defazio and L.~Bottou.
\newblock On the ineffectiveness of variance reduced optimization for deep
  learning.
\newblock In {\em NeurIPS}. 2019.

\bibitem{Imagenet}
J.~{Deng}, W.~{Dong}, R.~{Socher}, L.~{Li}, {Kai Li}, and {Li Fei-Fei}.
\newblock Imagenet: A large-scale hierarchical image database.
\newblock In {\em CVPR}, 2009.

\bibitem{bert}
J.~Devlin, M.-W. Chang, K.~Lee, and K.~Toutanova.
\newblock {BERT}: Pre-training of deep bidirectional transformers for language
  understanding.
\newblock In {\em NAACL}, 2019.

\bibitem{NIPS2019_9512}
A.~Dieuleveut and K.~K. Patel.
\newblock Communication trade-offs for local-sgd with large step size.
\newblock In {\em NeurIPS}. 2019.

\bibitem{CommQuantforDataPara}
N.~Dryden, S.~A. Jacobs, T.~Moon, and B.~Van~Essen.
\newblock Communication quantization for data-parallel training of deep neural
  networks.
\newblock In {\em Proceedings of the Workshop on MLHPC}, 2016.

\bibitem{EffiUseofLimitMemory}
C.~D{\"{u}}nner, T.~P. Parnell, and M.~Jaggi.
\newblock Efficient use of limited-memory accelerators for linear learning on
  heterogeneous systems.
\newblock In {\em NeurIPS}, 2017.

\bibitem{aritra2020discrepancy}
A.~Dutta, E.~H. Bergou, A.~M. Abdelmoniem, C.-Y. Ho, A.~N. Sahu, M.~Canini, and
  P.~Kalnis.
\newblock On the discrepancy between the theoretical analysis and practical
  implementations of compressed communication for distributed deep learning.
\newblock In {\em AAAI}, 2020.

\bibitem{faghri2020adaptive}
F.~Faghri, I.~Tabrizian, I.~Markov, D.~Alistarh, D.~M. Roy, and
  A.~Ramezani-Kebrya.
\newblock Adaptive gradient quantization for data-parallel sgd.
\newblock 2020.

\bibitem{4497789Randomizedconsensus}
F.~{Fagnani} and S.~{Zampieri}.
\newblock Randomized consensus algorithms over large scale networks.
\newblock {\em JSAC}, 2008.

\bibitem{10.1145/3452296.3472904}
J.~Fei, C.-Y. Ho, A.~N. Sahu, M.~Canini, and A.~Sapio.
\newblock Efficient sparse collective communication and its application to
  accelerate distributed deep learning.
\newblock In {\em ACM SIGCOMM}, 2021.

\bibitem{6773262}
W.~M. {Goodall}.
\newblock Television by pulse code modulation.
\newblock {\em The Bell System Technical Journal}, 30(1):33--49, Jan 1951.

\bibitem{DL}
I.~Goodfellow, Y.~Bengio, and A.~Courville.
\newblock {\em Deep Learning}.
\newblock The MIT Press, 2016.

\bibitem{goyal2017accurate}
P.~Goyal, P.~Doll{\'a}r, R.~Girshick, P.~Noordhuis, L.~Wesolowski, A.~Kyrola,
  A.~Tulloch, Y.~Jia, and K.~He.
\newblock Accurate, large minibatch sgd: Training imagenet in 1 hour.
\newblock {\em arXiv preprint arXiv:1706.02677}, 2017.

\bibitem{imagenet1hour}
P.~Goyal, P.~Doll{\'a}r, R.~B. Girshick, P.~Noordhuis, L.~Wesolowski,
  A.~Kyrola, A.~Tulloch, Y.~Jia, and K.~He.
\newblock Accurate, large minibatch sgd: Training imagenet in 1 hour.
\newblock {\em ArXiv}, abs/1706.02677, 2017.

\bibitem{AsynDisMLspar}
D.~Grishchenko, F.~Iutzeler, J.~Malick, and M.-R. Amini.
\newblock Asynchronous distributed learning with sparse communications and
  identification.
\newblock {\em ArXiv}, abs/1812.03871, 2018.

\bibitem{guo2009bcube}
C.~Guo, G.~Lu, D.~Li, H.~Wu, X.~Zhang, Y.~Shi, C.~Tian, Y.~Zhang, and S.~Lu.
\newblock {BCube}: a high performance, server-centric network architecture for
  modular data centers.
\newblock In {\em ACM SIGCOMM}, 2009.

\bibitem{DBLP:abs-1808-04752}
Y.~Guo.
\newblock A survey on methods and theories of quantized neural networks.
\newblock {\em CoRR}, abs/1808.04752, 2018.

\bibitem{DLwithlimited}
S.~Gupta, A.~Agrawal, K.~Gopalakrishnan, and P.~Narayanan.
\newblock Deep learning with limited numerical precision.
\newblock In {\em ICML}, 2015.

\bibitem{NIPS2019_9288}
F.~Haddadpour, M.~M. Kamani, M.~Mahdavi, and V.~Cadambe.
\newblock Local sgd with periodic averaging: Tighter analysis and adaptive
  synchronization.
\newblock In {\em NeurIPS}, pages 11080--11092. 2019.

\bibitem{han2020adaptive}
P.~Han, S.~Wang, and K.~K. Leung.
\newblock Adaptive gradient sparsification for efficient federated learning: An
  online learning approach.
\newblock {\em arXiv preprint arXiv:2001.04756}, 2020.

\bibitem{harlap2018pipedream}
A.~Harlap, D.~Narayanan, A.~Phanishayee, V.~Seshadri, N.~Devanur, G.~Ganger,
  and P.~Gibbons.
\newblock Pipedream: Fast and efficient pipeline parallel {DNN} training.
\newblock {\em arXiv preprint arXiv:1806.03377}, 2018.

\bibitem{hashemi2018tictac}
S.~H. Hashemi, S.~A. Jyothi, and R.~H. Campbell.
\newblock Tictac: Accelerating distributed deep learning with communication
  scheduling.
\newblock In {\em In Proceedings of Systems and Machine Learning (SysML)},
  2018.

\bibitem{chaoyanghe2020fedml}
C.~He, S.~Li, J.~So, M.~Zhang, H.~Wang, X.~Wang, P.~Vepakomma, A.~Singh,
  H.~Qiu, L.~Shen, P.~Zhao, Y.~Kang, Y.~Liu, R.~Raskar, Q.~Yang, M.~Annavaram,
  and S.~Avestimehr.
\newblock Fedml: A research library and benchmark for federated machine
  learning.
\newblock {\em arXiv preprint arXiv:2007.13518}, 2020.

\bibitem{resnet}
K.~{He}, X.~{Zhang}, S.~{Ren}, and J.~{Sun}.
\newblock Deep residual learning for image recognition.
\newblock In {\em CVPR}, 2016.

\bibitem{NIPS2018_7705}
L.~He, A.~Bian, and M.~Jaggi.
\newblock Cola: Decentralized linear learning.
\newblock In S.~Bengio, H.~Wallach, H.~Larochelle, K.~Grauman, N.~Cesa-Bianchi,
  and R.~Garnett, editors, {\em NeurIPS}, pages 4536--4546. 2018.

\bibitem{MoreEffDMLviaStaleSync}
Q.~Ho, J.~Cipar, H.~Cui, J.~K. Kim, S.~Lee, P.~B. Gibbons, G.~A. Gibson, G.~R.
  Ganger, and E.~P. Xing.
\newblock More effective distributed ml via a stale synchronous parallel
  parameter server.
\newblock In {\em NeurIPS}, 2013.

\bibitem{Hoefler:2010}
T.~Hoefler, W.~Gropp, R.~Thakur, and J.~L. Tr\"{a}ff.
\newblock Toward performance models of mpi implementations for understanding
  application scaling issues.
\newblock In {\em EuroMPI'10}, 2010.

\bibitem{Hoffer2017TLG32947713294936}
E.~Hoffer, I.~Hubara, and D.~Soudry.
\newblock Train longer, generalize better: Closing the generalization gap in
  large batch training of neural networks.
\newblock In {\em NeurIPS}, 2017.

\bibitem{ProbroundingNN}
M.~H{\"o}hfeld and S.~E. Fahlman.
\newblock Probabilistic rounding in neural network learning with limited
  precision.
\newblock {\em Neurocomputing}, 4:291--299, 1992.

\bibitem{Horvath2019NaturalCF}
S.~Horvath, C.-Y. Ho, L.~Horvath, A.~N. Sahu, M.~Canini, and P.~Richt{\'a}rik.
\newblock Natural compression for distributed deep learning.
\newblock {\em ArXiv}, abs/1905.10988, 2019.

\bibitem{Passcode}
C.-J. Hsieh, H.-F. Yu, and I.~S. Dhillon.
\newblock Passcode: Parallel asynchronous stochastic dual co-ordinate descent.
\newblock In {\em ICML}, 2015.

\bibitem{huang2019gpipe}
Y.~Huang, Y.~Cheng, A.~Bapna, O.~Firat, D.~Chen, M.~Chen, H.~Lee, J.~Ngiam,
  Q.~V. Le, Y.~Wu, et~al.
\newblock Gpipe: Efficient training of giant neural networks using pipeline
  parallelism.
\newblock {\em NeurIPS}, 2019.

\bibitem{Hubara2017}
I.~Hubara, M.~Courbariaux, D.~Soudry, R.~El-Yaniv, and Y.~Bengio.
\newblock Quantized neural networks: Training neural networks with low
  precision weights and activations.
\newblock {\em J. Mach. Learn. Res.}, 18(1):6869--6898, Jan. 2017.

\bibitem{AsynDisOptimRandADMM}
F.~{Iutzeler}, P.~{Bianchi}, P.~{Ciblat}, and W.~{Hachem}.
\newblock Asynchronous distributed optimization using a randomized alternating
  direction method of multipliers.
\newblock In {\em 52nd IEEE Conference on Decision and Control}, 2013.

\bibitem{NIPS2019_9473}
N.~Ivkin, D.~Rothchild, E.~Ullah, V.~braverman, I.~Stoica, and R.~Arora.
\newblock Communication-efficient distributed sgd with sketching.
\newblock In {\em NeurIPS}. 2019.

\bibitem{jayarajan2018priority}
A.~Jayarajan, J.~Wei, G.~Gibson, A.~Fedorova, and G.~Pekhimenko.
\newblock Priority-based parameter propagation for distributed dnn training.
\newblock In {\em In Proceedings of Systems and Machine Learning (SysML)},
  2018.

\bibitem{jhunjhunwala2021adaptive}
D.~Jhunjhunwala, A.~Gadhikar, G.~Joshi, and Y.~C. Eldar.
\newblock Adaptive quantization of model updates for communication-efficient
  federated learning.
\newblock In {\em ICASSP}. IEEE, 2021.

\bibitem{jia2018highly}
X.~Jia, S.~Song, S.~Shi, W.~He, Y.~Wang, H.~Rong, F.~Zhou, L.~Xie, Z.~Guo,
  Y.~Yang, L.~Yu, T.~Chen, G.~Hu, and X.~Chu.
\newblock Highly scalable deep learning training system with mixed-precision:
  Training {ImageNet} in four minutes.
\newblock In {\em Proc. of Workshop on Systems for ML and Open Source Software,
  collocated with NeurIPS 2018}, 2018.

\bibitem{AlinearSpeedupAnalysis}
P.~Jiang and G.~Agrawal.
\newblock A linear speedup analysis of distributed deep learning with sparse
  and quantized communication.
\newblock In {\em NeurIPS}, 2018.

\bibitem{jiang2022pisces}
Z.~Jiang, W.~Wang, B.~Li, and B.~Li.
\newblock Pisces: efficient federated learning via guided asynchronous
  training.
\newblock In {\em SoCC}, 2022.

\bibitem{jouppi2017datacenter}
N.~P. Jouppi, C.~Young, N.~Patil, D.~Patterson, G.~Agrawal, R.~Bajwa, S.~Bates,
  S.~Bhatia, N.~Boden, A.~Borchers, et~al.
\newblock In-datacenter performance analysis of a tensor processing unit.
\newblock In {\em ISCA}, 2017.

\bibitem{kairouz2019advances}
P.~Kairouz, H.~B. McMahan, B.~Avent, A.~Bellet, M.~Bennis, A.~N. Bhagoji,
  K.~Bonawitz, Z.~Charles, G.~Cormode, R.~Cummings, et~al.
\newblock Advances and open problems in federated learning.
\newblock {\em arXiv preprint arXiv:1912.04977}, 2019.

\bibitem{EFsignSGD}
S.~P. Karimireddy, Q.~Rebjock, S.~Stich, and M.~Jaggi.
\newblock Error feedback fixes {S}ign{SGD} and other gradient compression
  schemes.
\newblock In {\em ICML}, 2019.

\bibitem{gossip1}
D.~{Kempe}, A.~{Dobra}, and J.~{Gehrke}.
\newblock Gossip-based computation of aggregate information.
\newblock In {\em 44th Annual IEEE Symposium on Foundations of Computer
  Science, 2003. Proceedings.}, pages 482--491, Oct 2003.

\bibitem{DBLPconficlrKeskarMNST17}
N.~S. Keskar, D.~Mudigere, J.~Nocedal, M.~Smelyanskiy, and P.~T.~P. Tang.
\newblock On large-batch training for deep learning: Generalization gap and
  sharp minima.
\newblock In {\em ICLR}, 2017.

\bibitem{Koloskova2019DecentralizedDL}
A.~Koloskova, T.~Lin, S.~U. Stich, and M.~Jaggi.
\newblock Decentralized deep learning with arbitrary communication compression.
\newblock {\em ArXiv}, abs/1907.09356, 2019.

\bibitem{DecentStocOptimAndGossip}
A.~Koloskova, S.~Stich, and M.~Jaggi.
\newblock Decentralized stochastic optimization and gossip algorithms with
  compressed communication.
\newblock In {\em ICML}, 2019.

\bibitem{FLStrategy}
J.~Konečný, H.~B. McMahan, F.~X. Yu, P.~Richtarik, A.~T. Suresh, and
  D.~Bacon.
\newblock Federated learning: Strategies for improving communication
  efficiency.
\newblock In {\em NeurIPS Workshop on Private Multi-Party Machine Learning},
  2016.

\bibitem{RandDisMean}
J.~Konečný and P.~Richtárik.
\newblock Randomized distributed mean estimation: Accuracy vs. communication.
\newblock {\em Frontiers in Applied Mathematics and Statistics}, 4:62, 2018.

\bibitem{552669}
R.~{Krizanc} and A.~{Saarimaki}.
\newblock Bulk synchronous parallel: practical experience with a model for
  parallel computing.
\newblock In {\em Proceedings of the 1996 Conference on Parallel Architectures
  and Compilation Technique}, pages 208--217, Oct 1996.

\bibitem{Krizhevsky2014OneWT}
A.~Krizhevsky.
\newblock One weird trick for parallelizing convolutional neural networks.
\newblock {\em ArXiv}, abs/1404.5997, 2014.

\bibitem{krizhevsky2010cifar}
A.~Krizhevsky, V.~Nair, and G.~Hinton.
\newblock Cifar-10 (canadian institute for advanced research).
\newblock {\em URL http://www.cs.toronto.edu/kriz/cifar.html}, 2010.

\bibitem{CommEffDenctStoc}
G.~Lan, S.~Lee, and Y.~Zhou.
\newblock Communication-efficient algorithms for decentralized and stochastic
  optimization.
\newblock {\em CoRR}, abs/1701.03961, 2017.

\bibitem{SparOLviaTrunc}
J.~Langford, L.~Li, and T.~Zhang.
\newblock Sparse online learning via truncated gradient.
\newblock {\em J. Mach. Learn. Res.}, 2009.

\bibitem{ProxBlockCoordD}
T.~T. Lau, J.~Zeng, B.~Wu, and Y.~Yao.
\newblock A proximal block coordinate descent algorithm for deep neural network
  training.
\newblock In {\em ICLR}, 2018.

\bibitem{8884800}
W.~{Lee}, Y.~{Lee}, J.~S. {Jeong}, G.~{Yu}, J.~Y. {Kim}, H.~J. {Park},
  B.~{Jeon}, W.~{Song}, G.~{Kim}, M.~{Weimer}, B.~{Cho}, and B.~{Chun}.
\newblock Automating system configuration of distributed machine learning.
\newblock In {\em ICDCS}, 2019.

\bibitem{lepikhin2021gshard}
D.~Lepikhin, H.~Lee, Y.~Xu, D.~Chen, O.~Firat, Y.~Huang, M.~Krikun, N.~Shazeer,
  and Z.~Chen.
\newblock {\{}GS{\}}hard: Scaling giant models with conditional computation and
  automatic sharding.
\newblock In {\em ICLR}, 2021.

\bibitem{ScalDMLwithPS}
M.~Li, D.~G. Andersen, J.~W. Park, A.~J. Smola, A.~Ahmed, V.~Josifovski,
  J.~Long, E.~J. Shekita, and B.-Y. Su.
\newblock Scaling distributed machine learning with the parameter server.
\newblock In {\em OSDI}, 2014.

\bibitem{communicationDisMLwithPS}
M.~Li, D.~G. Andersen, A.~Smola, and K.~Yu.
\newblock Communication efficient distributed machine learning with the
  parameter server.
\newblock In {\em NeurIPS}, 2014.

\bibitem{PSforDisML}
M.~Li, L.~Zhou, Z.~Yang, A.~Q. Li, F.~Xia, D.~G. Andersen, and A.~J. Smola.
\newblock Parameter server for distributed machine learning.
\newblock In {\em In Big Learning NeurIPS Workshop}, 2013.

\bibitem{li2021chimera}
S.~Li and T.~Hoefler.
\newblock Chimera: efficiently training large-scale neural networks with
  bidirectional pipelines.
\newblock In {\em SC}, 2021.

\bibitem{li2022near}
S.~Li and T.~Hoefler.
\newblock Near-optimal sparse allreduce for distributed deep learning.
\newblock In {\em Proceedings of the 27th ACM SIGPLAN Symposium on Principles
  and Practice of Parallel Programming}, pages 135--149, 2022.

\bibitem{li2022on}
X.~Li, B.~Karimi, and P.~Li.
\newblock On distributed adaptive optimization with gradient compression.
\newblock In {\em ICLR}, 2022.

\bibitem{10.5555/2969442.2969545}
X.~Lian, Y.~Huang, Y.~Li, and J.~Liu.
\newblock Asynchronous parallel stochastic gradient for nonconvex optimization.
\newblock In {\em NeurIPS}, 2015.

\bibitem{CanDecent}
X.~Lian, C.~Zhang, H.~Zhang, C.-J. Hsieh, W.~Zhang, and J.~Liu.
\newblock Can decentralized algorithms outperform centralized algorithms? a
  case study for decentralized parallel stochastic gradient descent.
\newblock In {\em NeurIPS}, 2017.

\bibitem{pmlr-v80-lian18a}
X.~Lian, W.~Zhang, C.~Zhang, and J.~Liu.
\newblock Asynchronous decentralized parallel stochastic gradient descent.
\newblock In {\em ICML}, 2018.

\bibitem{3LC}
H.~Lim, D.~G. Andersen, and M.~Kaminsky.
\newblock 3lc: Lightweight and effective traffic compression for distributed
  machine learning.
\newblock {\em ArXiv}, abs/1802.07389, 2019.

\bibitem{Lin2018DontUL}
T.~Lin, S.~U. Stich, and M.~Jaggi.
\newblock Don't use large mini-batches, use local sgd.
\newblock {\em ArXiv}, abs/1808.07217, 2018.

\bibitem{Lin2020Dont}
T.~Lin, S.~U. Stich, K.~K. Patel, and M.~Jaggi.
\newblock Don't use large mini-batches, use local sgd.
\newblock In {\em ICLR}, 2020.

\bibitem{DGC}
Y.~Lin, S.~Han, H.~Mao, Y.~Wang, and B.~Dally.
\newblock Deep gradient compression: Reducing the communication bandwidth for
  distributed training.
\newblock In {\em ICLR}, 2018.

\bibitem{gossip2}
{Lin Xiao} and S.~{Boyd}.
\newblock Fast linear iterations for distributed averaging.
\newblock In {\em 42nd IEEE International Conference on Decision and Control},
  2003.

\bibitem{DisGraphlab}
Y.~Low, D.~Bickson, J.~Gonzalez, C.~Guestrin, A.~Kyrola, and J.~M. Hellerstein.
\newblock Distributed graphlab: A framework for machine learning and data
  mining in the cloud.
\newblock {\em Proc. VLDB Endow.}, 5(8):716--727, Apr. 2012.

\bibitem{luo2020plink}
L.~Luo, P.~West, J.~Nelson, A.~Krishnamurthy, and L.~Ceze.
\newblock {PLink}: Efficient cloud-based training with topology-aware dynamic
  hierarchical aggregation.
\newblock In {\em MLSys}, 2020.

\bibitem{m2021efficient}
A.~M~Abdelmoniem, A.~Elzanaty, M.-S. Alouini, and M.~Canini.
\newblock An efficient statistical-based gradient compression technique for
  distributed training systems.
\newblock {\em MLSys}, 2021.

\bibitem{muli2013}
D.~G.~A. M.~Li and A.~Smola.
\newblock Distributed delayed proximal gradient methods.
\newblock In {\em In NeurIPS Workshop on Optimization for Machine Learning},
  Lake Tahoe, CA, 2013.

\bibitem{DisBlockCoordD}
D.~Mahajan, S.~S. Keerthi, and S.~Sundararajan.
\newblock A distributed block coordinate descent method for training
  l1regularized linear classifiers.
\newblock {\em J. Mach. Learn. Res.}, 18(1):3167--3201, Jan. 2017.

\bibitem{martens2015optimizing}
J.~Martens and R.~Grosse.
\newblock Optimizing neural networks with kronecker-factored approximate
  curvature.
\newblock In {\em International conference on machine learning}, pages
  2408--2417. PMLR, 2015.

\bibitem{3154842}
R.~Mayer, C.~Mayer, and L.~Laich.
\newblock The tensorflow partitioning and scheduling problem: It’s the
  critical path!
\newblock In {\em Proceedings of the 1st Workshop on Distributed
  Infrastructures for Deep Learning}, DIDL ’17, New York, NY, USA, 2017.

\bibitem{NIPS2009_3881}
R.~Mcdonald, M.~Mohri, N.~Silberman, D.~Walker, and G.~S. Mann.
\newblock Efficient large-scale distributed training of conditional maximum
  entropy models.
\newblock In {\em NeurIPS}. 2009.

\bibitem{McDonald2010DistributedTS}
R.~T. McDonald, K.~B. Hall, and G.~Mann.
\newblock Distributed training strategies for the structured perceptron.
\newblock In {\em HLT-NAACL}, 2010.

\bibitem{7526802}
N.~{McGlohon} and S.~{Patterson}.
\newblock Distributed semi-stochastic optimization with quantization
  refinement.
\newblock In {\em 2016 American Control Conference (ACC)}, pages 7159--7164,
  July 2016.

\bibitem{FederatedLearning}
B.~McMahan, E.~Moore, D.~Ramage, S.~Hampson, and B.~A. y~Arcas.
\newblock Communication-efficient learning of deep networks from decentralized
  data.
\newblock In {\em AISTATS}, pages 1273--1282, 2017.

\bibitem{Meng2016AAS3060832}
Q.~Meng, W.~Chen, J.~Yu, T.~Wang, Z.-M. Ma, and T.-Y. Liu.
\newblock Asynchronous accelerated stochastic gradient descent.
\newblock In {\em IJCAI}, 2016.

\bibitem{mikami2018massively}
H.~Mikami, H.~Suganuma, Y.~Tanaka, Y.~Kageyama, et~al.
\newblock Massively distributed {SGD}: {ImageNet/ResNet-50} training in a
  flash.
\newblock {\em arXiv preprint arXiv:1811.05233}, 2018.

\bibitem{Mirhoseini2017}
A.~Mirhoseini, H.~Pham, Q.~V. Le, B.~Steiner, R.~Larsen, Y.~Zhou, N.~Kumar,
  M.~Norouzi, S.~Bengio, and J.~Dean.
\newblock Device placement optimization with reinforcement learning.
\newblock In {\em ICML}, 2017.

\bibitem{NEURIPS2022_029df12a}
K.~Mishchenko, F.~Bach, M.~Even, and B.~E. Woodworth.
\newblock Asynchronous sgd beats minibatch sgd under arbitrary delays.
\newblock In {\em NeurIPS}, 2022.

\bibitem{DisLearningDIANA}
K.~Mishchenko, E.~A. Gorbunov, M.~Tak{\'a}c, and P.~Richt{\'a}rik.
\newblock Distributed learning with compressed gradient differences.
\newblock {\em ArXiv}, abs/1901.09269, 2019.

\bibitem{Mishchenko201999OP}
K.~Mishchenko, F.~Hanzely, and P.~Richt{\'a}rik.
\newblock 99\% of parallel optimization is inevitably a waste of time.
\newblock {\em ArXiv}, abs/1901.09437, 2019.

\bibitem{sparknet}
P.~Moritz, R.~Nishihara, I.~Stoica, and M.~I. Jordan.
\newblock Sparknet: Training deep networks in spark.
\newblock In {\em ICLR}, 2016.

\bibitem{Dadmm}
J.~F.~C. {Mota}, J.~M.~F. {Xavier}, P.~M.~Q. {Aguiar}, and M.~{Püschel}.
\newblock D-admm: A communication-efficient distributed algorithm for separable
  optimization.
\newblock {\em IEEE Transactions on Signal Processing}, 61(10):2718--2723, May
  2013.

\bibitem{DisSubGradMultiOptim}
A.~{Nedic} and A.~{Ozdaglar}.
\newblock Distributed subgradient methods for multi-agent optimization.
\newblock {\em IEEE Transactions on Automatic Control}, 54(1):48--61, Jan 2009.

\bibitem{SGPDirected}
A.~{Nedić} and A.~{Olshevsky}.
\newblock Stochastic gradient-push for strongly convex functions on
  time-varying directed graphs.
\newblock {\em IEEE Transactions on Automatic Control}, 61(12):3936--3947, Dec
  2016.

\bibitem{8340193}
A.~{Nedić}, A.~{Olshevsky}, and M.~G. {Rabbat}.
\newblock Network topology and communication-computation tradeoffs in
  decentralized optimization.
\newblock {\em Proceedings of the IEEE}, 106(5):953--976, May 2018.

\bibitem{hogwild}
F.~Niu, B.~Recht, C.~Re, and S.~J. Wright.
\newblock Hogwild!: A lock-free approach to parallelizing stochastic gradient
  descent.
\newblock In {\em NeurIPS}, 2011.

\bibitem{4118472ConsensusandCooperation}
R.~{Olfati-Saber}, J.~A. {Fax}, and R.~M. {Murray}.
\newblock Consensus and cooperation in networked multi-agent systems.
\newblock {\em Proceedings of the IEEE}, 95(1):215--233, Jan 2007.

\bibitem{Ooi2015SD}
B.~C. Ooi, K.-L. Tan, S.~Wang, W.~Wang, Q.~Cai, G.~Chen, J.~Gao, Z.~Luo, A.~K.
  Tung, Y.~Wang, Z.~Xie, M.~Zhang, and K.~Zheng.
\newblock Singa: A distributed deep learning platform.
\newblock In {\em ACM International Conference on Multimedia}, 2015.

\bibitem{HowToScaleDDL}
F.~N.~I. P.~H.~Jin, Q.~Yuan and K.~Keutzer.
\newblock How to scale distributed deep learning?
\newblock In {\em ML Systems Workshop at NeurIPS}, 2016.

\bibitem{peng2019generic}
Y.~Peng, Y.~Zhu, Y.~Chen, Y.~Bao, B.~Yi, C.~Lan, C.~Wu, and C.~Guo.
\newblock A generic communication scheduler for distributed {DNN} training
  acceleration.
\newblock In {\em Proceedings of the 27th ACM Symposium on Operating Systems
  Principles}, 2019.

\bibitem{Peteiro2013}
D.~Peteiro-Barral and B.~Guijarro-Berdi{\~{n}}as.
\newblock A survey of methods for distributed machine learning.
\newblock {\em Progress in Artificial Intelligence}, 2:1--11, 2013.

\bibitem{Power2010PBF}
R.~Power and J.~Li.
\newblock Piccolo: Building fast, distributed programs with partitioned tables.
\newblock In {\em OSDI}, 2010.

\bibitem{7544448}
Y.~{Pu}, M.~N. {Zeilinger}, and C.~N. {Jones}.
\newblock Quantization design for distributed optimization.
\newblock {\em IEEE Transactions on Automatic Control}, 62(5):2107--2120, May
  2017.

\bibitem{vanillaMomentum}
N.~Qian.
\newblock On the momentum term in gradient descent learning algorithms.
\newblock {\em Neural Netw.}, 12(1):145--151, Jan. 1999.

\bibitem{Qu2015QRD2969239}
Z.~Qu, P.~Richt\'{a}rik, and T.~Zhang.
\newblock Quartz: Randomized dual coordinate ascent with arbitrary sampling.
\newblock In {\em NeurIPS}, 2015.

\bibitem{MultiagentMirrorDescentDenct}
M.~{Rabbat}.
\newblock Multi-agent mirror descent for decentralized stochastic optimization.
\newblock In {\em 2015 IEEE 6th International Workshop on CAMSAP}, 2015.

\bibitem{OptimizationofCollective}
R.~Rabenseifner.
\newblock Optimization of collective reduction operations.
\newblock In M.~Bubak, G.~D. van Albada, P.~M.~A. Sloot, and J.~Dongarra,
  editors, {\em Computational Science - ICCS 2004}, 2004.

\bibitem{Ram2008DistributedSS}
S.~S. Ram, A.~Nedic, and V.~V. Veeravalli.
\newblock Distributed stochastic subgradient projection algorithms for convex
  optimization.
\newblock {\em Journal of Optimization Theory and Applications}, 147:516--545,
  2008.

\bibitem{ml1991}
T.~Reinartz.
\newblock {\em Focusing Solutions for Data Mining: Analytical Studies and
  Experimental Results in Real-world Domains}.
\newblock Springer-Verlag, Berlin, Heidelberg, 1999.

\bibitem{NIPS2019_9047}
A.~Reisizadeh, H.~Taheri, A.~Mokhtari, H.~Hassani, and R.~Pedarsani.
\newblock Robust and communication-efficient collaborative learning.
\newblock In {\em NeurIPS}. 2019.

\bibitem{CommEffiDisSpar}
J.~{Ren}, X.~{Li}, and J.~{Haupt}.
\newblock Communication-efficient distributed optimization for sparse learning
  via two-way truncation.
\newblock In {\em 2017 IEEE 7th International Workshop on CAMSAP}, 2017.

\bibitem{EF21}
P.~Richt{\'a}rik, I.~Sokolov, and I.~Fatkhullin.
\newblock Ef21: A new, simpler, theoretically better, and practically faster
  error feedback.
\newblock {\em NeurIPS}, 2021.

\bibitem{3PC}
P.~Richtarik, I.~Sokolov, E.~Gasanov, I.~Fatkhullin, Z.~Li, and E.~Gorbunov.
\newblock 3{PC}: Three point compressors for communication-efficient
  distributed training and a better theory for lazy aggregation.
\newblock In {\em ICML}, 2022.

\bibitem{1057702}
L.~{Roberts}.
\newblock Picture coding using pseudo-random noise.
\newblock {\em IRE Transactions on Information Theory}, 8(2):145--154, February
  1962.

\bibitem{276984}
J.~Romero, J.~Yin, N.~Laanait, B.~Xie, M.~T. Young, S.~Treichler,
  V.~Starchenko, A.~Borisevich, A.~Sergeev, and M.~Matheson.
\newblock Accelerating collective communication in data parallel training
  across deep learning frameworks.
\newblock In {\em NSDI}, 2022.

\bibitem{Russakovsky2015ILS28465472846559}
O.~Russakovsky, J.~Deng, H.~Su, J.~Krause, S.~Satheesh, S.~Ma, Z.~Huang,
  A.~Karpathy, A.~Khosla, M.~Bernstein, A.~C. Berg, and L.~Fei-Fei.
\newblock Imagenet large scale visual recognition challenge.
\newblock {\em Int. J. Comput. Vision}, 2015.

\bibitem{Tamingwild}
C.~D. Sa, C.~Zhang, K.~Olukotun, and C.~R{\'e}.
\newblock Taming the wild: A unified analysis of hogwild-style algorithms.
\newblock {\em NeurIPS}, 2015.

\bibitem{sanders2009two}
P.~Sanders, J.~Speck, and J.~L. Tr{\"a}ff.
\newblock Two-tree algorithms for full bandwidth broadcast, reduction and scan.
\newblock {\em Parallel Computing}, 35(12):581--594, 2009.

\bibitem{In-network}
A.~Sapio, M.~Canini, C.-Y. Ho, J.~Nelson, P.~Kalnis, C.~Kim, A.~Krishnamurthy,
  M.~Moshref, D.~Ports, and P.~Richtarik.
\newblock Scaling distributed machine learning with {In-Network} aggregation.
\newblock In {\em NSDI 21}, pages 785--808, 2021.

\bibitem{sarvotham2001connection}
S.~Sarvotham, R.~Riedi, and R.~Baraniuk.
\newblock Connection-level analysis and modeling of network traffic.
\newblock In {\em Proceedings of ACM SIGCOMM Workshop on IMW}. ACM, 2001.

\bibitem{SparseTernaryCompressionSTC}
F.~Sattler, S.~Wiedemann, K.~M{\"{u}}ller, and W.~Samek.
\newblock Robust and communication-efficient federated learning from non-iid
  data.
\newblock {\em CoRR}, abs/1903.02891, 2019.

\bibitem{sparsebinarycompression}
F.~Sattler, S.~Wiedemann, K.-R. M{\"u}ller, and W.~Samek.
\newblock Sparse binary compression: Towards distributed deep learning with
  minimal communication.
\newblock {\em IJCN}, 2018.

\bibitem{OptimNonsmoothDisOptim}
K.~Scaman, F.~Bach, S.~Bubeck, Y.~T. Lee, and L.~Massouli{\'e}.
\newblock Optimal algorithms for non-smooth distributed optimization in
  networks.
\newblock In {\em NeurIPS}, 2018.

\bibitem{4434671Adistributedconsensus}
L.~{Schenato} and G.~{Gamba}.
\newblock A distributed consensus protocol for clock synchronization in
  wireless sensor network.
\newblock In {\em 2007 46th IEEE Conference on Decision and Control}, pages
  2289--2294, Dec 2007.

\bibitem{OptimDisOptim}
K.~Seaman, F.~Bach, S.~Bubeck, Y.~T. Lee, and L.~Massouli{\'e}.
\newblock Optimal algorithms for smooth and strongly convex distributed
  optimization in networks.
\newblock In {\em ICML}, 2017.

\bibitem{1bit}
F.~Seide, H.~Fu, J.~Droppo, G.~Li, and D.~Yu.
\newblock 1-bit stochastic gradient descent and its application to
  data-parallel distributed training of speech {DNNs}.
\newblock In {\em INTERSPEECH}, 2014.

\bibitem{DBLPjournalscorrabs181103600}
C.~J. Shallue, J.~Lee, J.~M. Antognini, J.~Sohl{-}Dickstein, R.~Frostig, and
  G.~E. Dahl.
\newblock Measuring the effects of data parallelism on neural network training.
\newblock {\em CoRR}, abs/1811.03600, 2018.

\bibitem{shi2019understanding}
S.~Shi, X.~Chu, K.~C. Cheung, and S.~See.
\newblock Understanding top-k sparsification in distributed deep learning.
\newblock {\em arXiv preprint arXiv:1911.08772}, 2019.

\bibitem{Shi2018MGWFBPED}
S.~Shi, X.~Chu, and B.~Li.
\newblock {MG-WFBP}: Efficient data communication for distributed synchronous
  {SGD} algorithms.
\newblock In {\em IEEE INFOCOM}, pages 172--180. IEEE, 2019.

\bibitem{shi2021mgj}
S.~Shi, X.~Chu, and B.~Li.
\newblock {MG-WFBP}: Merging gradients wisely for efficient communication in
  distributed deep learning.
\newblock {\em TPDS}, 2021.

\bibitem{shi2023pipemoe}
S.~Shi, X.~Pan, X.~Chu, and B.~Li.
\newblock {PipeMoE}: Accelerating mixture-of-experts through adaptive
  pipelining.
\newblock In {\em IEEE INFOCOM}, 2023.

\bibitem{shi2020quantitative}
S.~Shi, Z.~Tang, X.~Chu, C.~Liu, W.~Wang, and B.~Li.
\newblock A quantitative survey of communication optimizations in distributed
  deep learning.
\newblock {\em IEEE Network}, 35(3):230--237, 2020.

\bibitem{shi2020layer}
S.~Shi, Z.~Tang, Q.~Wang, K.~Zhao, and X.~Chu.
\newblock Layer-wise adaptive gradient sparsification for distributed deep
  learning with convergence guarantees.
\newblock In {\em ECAI}, 2020.

\bibitem{shi2018performance}
S.~Shi, Q.~Wang, and X.~Chu.
\newblock Performance modeling and evaluation of distributed deep learning
  frameworks on gpus.
\newblock In {\em 2018 IEEE 4th Intl Conf on Big Data Intelligence and
  Computing}, pages 949--957. IEEE, 2018.

\bibitem{shi2018adag}
S.~Shi, Q.~Wang, X.~Chu, and B.~Li.
\newblock A {DAG} model of synchronous stochastic gradient descent in
  distributed deep learning.
\newblock In {\em ICPADS}, 2018.

\bibitem{shi2020communication}
S.~Shi, Q.~Wang, X.~Chu, B.~Li, Y.~Qin, R.~Liu, and X.~Zhao.
\newblock Communication-efficient distributed deep learning with merged
  gradient sparsification on gpus.
\newblock In {\em IEEE INFOCOM}, 2020.

\bibitem{shi2016benchmarking}
S.~Shi, Q.~Wang, P.~Xu, and X.~Chu.
\newblock Benchmarking state-of-the-art deep learning software tools.
\newblock In {\em 2016 7th International Conference on Cloud Computing and Big
  Data (CCBD)}, pages 99--104. IEEE, 2016.

\bibitem{GTopk}
S.~Shi, Q.~Wang, K.~Zhao, Z.~Tang, Y.~Wang, X.~Huang, and X.~Chu.
\newblock A distributed synchronous {SGD} algorithm with global top-k
  sparsification for low bandwidth networks.
\newblock In {\em ICDCS}, pages 2238--2247, 2019.

\bibitem{shi2021accelerating}
S.~Shi, L.~Zhang, and B.~Li.
\newblock Accelerating distributed k-fac with smart parallelism of computing
  and communication tasks.
\newblock In {\em ICDCS}, 2021.

\bibitem{ijcai2019473}
S.~Shi, K.~Zhao, Q.~Wang, Z.~Tang, and X.~Chu.
\newblock A convergence analysis of distributed {SGD} with
  communication-efficient gradient sparsification.
\newblock In {\em IJCAI}, 2019.

\bibitem{shi2021towards}
S.~Shi, X.~Zhou, S.~Song, X.~Wang, Z.~Zhu, X.~Huang, X.~Jiang, F.~Zhou, Z.~Guo,
  L.~Xie, et~al.
\newblock Towards scalable distributed training of deep learning on public
  cloud clusters.
\newblock {\em MLSys}, 3:401--412, 2021.

\bibitem{shoeybi2019megatron}
M.~Shoeybi, M.~Patwary, R.~Puri, P.~LeGresley, J.~Casper, and B.~Catanzaro.
\newblock Megatron-lm: Training multi-billion parameter language models using
  model parallelism.
\newblock {\em arXiv preprint arXiv:1909.08053}, 2019.

\bibitem{LENA}
H.~Shokri~Ghadikolaei, S.~Stich, and M.~Jaggi.
\newblock Lena: Communication-efficient distributed learning with
  self-triggered gradient uploads.
\newblock In {\em AISTATS}, 2021.

\bibitem{AnArchforParallel}
A.~Smola and S.~Narayanamurthy.
\newblock An architecture for parallel topic models.
\newblock In {\em Proc. VLDB Endow.}, 2010.

\bibitem{spiridonoff2021communicationefficient}
A.~Spiridonoff, A.~Olshevsky, and I.~Paschalidis.
\newblock Communication-efficient {SGD}: From local {SGD} to one-shot
  averaging.
\newblock In A.~Beygelzimer, Y.~Dauphin, P.~Liang, and J.~W. Vaughan, editors,
  {\em NeurIPS}, 2021.

\bibitem{Sebastian2019}
S.~U. Stich.
\newblock Local {SGD} converges fast and communicates little.
\newblock In {\em ICLR}, 2019.

\bibitem{SparSGDwithMemory}
S.~U. Stich, J.-B. Cordonnier, and M.~Jaggi.
\newblock Sparsified {SGD} with memory.
\newblock In {\em NeurIPS}, 2018.

\bibitem{TonotopicANN}
N.~{Strom}.
\newblock A tonotopic artificial neural network architecture for phoneme
  probability estimation.
\newblock In {\em 1997 IEEE Workshop on Automatic Speech Recognition and
  Understanding Proceedings}, pages 156--163, Dec 1997.

\bibitem{scalableDisDNN}
N.~Strom.
\newblock Scalable distributed {DNN} training using commodity {GPU} cloud
  computing.
\newblock In {\em INTERSPEECH}, 2015.

\bibitem{SparConnec}
N.~Ström.
\newblock Sparse connection and pruning in large dynamic artificial neural
  networks.
\newblock In {\em EUROSPEECH}, 1997.

\bibitem{ElasticQuant}
B.~Sudharsan, D.~Sheth, S.~Arya, F.~Rollo, P.~Yadav, P.~Patel, J.~G. Breslin,
  and M.~I. Ali.
\newblock Elasticl: Elastic quantization for communication efficient
  collaborative learning in iot.
\newblock In {\em SenSys}, 2021.

\bibitem{7816979}
N.~{Sukhija}, M.~{Tatineni}, N.~{Brown}, M.~V. {Moer}, P.~{Rodriguez}, and
  S.~{Callicott}.
\newblock Topic modeling and visualization for big data in social sciences.
\newblock In {\em 2016 Intl IEEE Conferences on Ubiquitous Intelligence
  Computing, Advanced and Trusted Computing, Scalable Computing and
  Communications, Cloud and Big Data Computing, Internet of People, and Smart
  World Congress (UIC/ATC/ScalCom/CBDCom/IoP/SmartWorld)}, 2016.

\bibitem{8237359}
C.~{Sun}, A.~{Shrivastava}, S.~{Singh}, and A.~{Gupta}.
\newblock Revisiting unreasonable effectiveness of data in deep learning era.
\newblock In {\em ICCV}, 2017.

\bibitem{NIPS2019_8598}
J.~Sun, T.~Chen, G.~Giannakis, and Z.~Yang.
\newblock Communication-efficient distributed learning via lazily aggregated
  quantized gradients.
\newblock In {\em NeurIPS}. 2019.

\bibitem{AsynCoordD}
T.~Sun, R.~Hannah, and W.~Yin.
\newblock Asynchronous coordinate descent under more realistic assumption.
\newblock In {\em NeurIPS}, 2017.

\bibitem{meProp}
X.~Sun, X.~Ren, S.~Ma, and H.~Wang.
\newblock me{P}rop: Sparsified back propagation for accelerated deep learning
  with reduced overfitting.
\newblock In {\em ICML}, 2017.

\bibitem{DisMeanEst}
A.~T. Suresh, F.~X. Yu, S.~Kumar, and H.~B. McMahan.
\newblock Distributed mean estimation with limited communication.
\newblock In {\em ICML}, 2017.

\bibitem{sze2017efficient}
V.~Sze, Y.-H. Chen, T.-J. Yang, and J.~S. Emer.
\newblock Efficient processing of deep neural networks: A tutorial and survey.
\newblock {\em Proceedings of the IEEE}, 105(12):2295--2329, 2017.

\bibitem{pmlr-v139-tang21a}
H.~Tang, S.~Gan, A.~A. Awan, S.~Rajbhandari, C.~Li, X.~Lian, J.~Liu, C.~Zhang,
  and Y.~He.
\newblock 1-bit adam: Communication efficient large-scale training with
  adam’s convergence speed.
\newblock In {\em ICML}, 2021.

\bibitem{CommCompforDecent}
H.~Tang, S.~Gan, C.~Zhang, T.~Zhang, and J.~Liu.
\newblock Communication compression for decentralized training.
\newblock In {\em NeurIPS}, 2018.

\bibitem{DecentTrainingoverDecentData}
H.~Tang, X.~Lian, M.~Yan, C.~Zhang, and J.~Liu.
\newblock Decentralized training over decentralized data.
\newblock In {\em ICML}, 2018.

\bibitem{tang2023fedml}
Z.~Tang, X.~Chu, R.~Y. Ran, S.~Lee, S.~Shi, Y.~Zhang, Y.~Wang, A.~Q. Liang,
  S.~Avestimehr, and C.~He.
\newblock Fedml parrot: A scalable federated learning system via
  heterogeneity-aware scheduling on sequential and hierarchical training.
\newblock {\em arXiv preprint arXiv:2303.01778}, 2023.

\bibitem{tang2020communication}
Z.~Tang, S.~Shi, and X.~Chu.
\newblock Communication-efficient decentralized learning with sparsification
  and adaptive peer selection.
\newblock {\em arXiv preprint arXiv:2002.09692}, 2020.

\bibitem{tang2022gossipfl}
Z.~Tang, S.~Shi, B.~Li, and X.~Chu.
\newblock Gossipfl: A decentralized federated learning framework with
  sparsified and adaptive communication.
\newblock {\em IEEE Transactions on Parallel and Distributed Systems},
  34(3):909--922, 2022.

\bibitem{tang2022virtual}
Z.~Tang, Y.~Zhang, S.~Shi, X.~He, B.~Han, and X.~Chu.
\newblock Virtual homogeneity learning: Defending against data heterogeneity in
  federated learning.
\newblock In {\em International Conference on Machine Learning}, pages
  21111--21132. PMLR, 2022.

\bibitem{Thakur2005}
R.~Thakur, R.~Rabenseifner, and W.~Gropp.
\newblock Optimization of collective communication operations in mpich.
\newblock {\em Int. J. High Perform. Comput. Appl.}, 19(1):49--66, Feb. 2005.

\bibitem{VarianceGradCompression}
Y.~Tsuzuku, H.~Imachi, and T.~Akiba.
\newblock Variance-based gradient compression for efficient distributed deep
  learning.
\newblock In {\em ICLR}, 2018.

\bibitem{ueno2019exhaustive}
Y.~Ueno and R.~Yokota.
\newblock Exhaustive study of hierarchical allreduce patterns for large
  messages between gpus.
\newblock In {\em Proceedings of the 18th IEEE/ACM International Symposium on
  Cluster, Cloud and Grid Computing}, 2019.

\bibitem{DualApprochforOptim}
C.~A. Uribe, S.~Lee, A.~Gasnikov, and A.~Nedic.
\newblock A dual approach for optimal algorithms in distributed optimization
  over networks.
\newblock {\em CoRR}, abs/1809.00710, 2018.

\bibitem{BSP7917379181}
L.~G. Valiant.
\newblock A bridging model for parallel computation.
\newblock {\em Commun. ACM}, 33(8):103–111, Aug. 1990.

\bibitem{ImproSpeedNN}
V.~Vanhoucke, A.~Senior, and M.~Z. Mao.
\newblock Improving the speed of neural networks on cpus.
\newblock In {\em Deep Learning and Unsupervised Feature Learning Workshop,
  NeurIPS 2011}, 2011.

\bibitem{attention}
A.~Vaswani, N.~Shazeer, N.~Parmar, J.~Uszkoreit, L.~Jones, A.~N. Gomez,
  L.~Kaiser, and I.~Polosukhin.
\newblock Attention is all you need.
\newblock In {\em NeurIPS}, 2017.

\bibitem{wang2020blink}
G.~Wang, S.~Venkataraman, A.~Phanishayee, J.~Thelin, N.~Devanur, and I.~Stoica.
\newblock Blink: Fast and generic collectives for distributed {ML}.
\newblock In {\em MLSys}, 2020.

\bibitem{9442310}
H.~Wang, S.~Guo, Z.~Qu, R.~Li, and Z.~Liu.
\newblock Error-compensated sparsification for communication-efficient
  decentralized training in edge environment.
\newblock {\em IEEE TPDS}, 2022.

\bibitem{ATOMO}
H.~Wang, S.~Sievert, Z.~Charles, S.~Liu, S.~Wright, and D.~Papailiopoulos.
\newblock Atomo: Communication-efficient learning via atomic sparsification.
\newblock In {\em NeurIPS}, 2018.

\bibitem{DBLP:journals/corr/abs-1810-08313}
J.~Wang and G.~Joshi.
\newblock Adaptive communication strategies to achieve the best error-runtime
  trade-off in local-update {SGD}.
\newblock {\em Proc. of Workshop on Systems for ML and Open Source Software,
  collocated with NeurIPS 2018}, 2018.

\bibitem{Jianyu180807576}
J.~Wang and G.~Joshi.
\newblock Cooperative {SGD:} {A} unified framework for the design and analysis
  of communication-efficient {SGD} algorithms.
\newblock {\em CoRR}, abs/1808.07576, 2018.

\bibitem{icdcsWangWL19}
L.~Wang, W.~Wang, and B.~Li.
\newblock {CMFL:} mitigating communication overhead for federated learning.
\newblock In {\em ICDCS}, 2019.

\bibitem{wang2020contention}
Q.~Wang, S.~Shi, C.~Wang, and X.~Chu.
\newblock Communication contention aware scheduling of multiple deep learning
  training jobs.
\newblock {\em arXiv preprint arXiv:2002.10105}, 2020.

\bibitem{wang2018bml}
S.~Wang, D.~Li, Y.~Cheng, J.~Geng, Y.~Wang, S.~Wang, S.-T. Xia, and J.~Wu.
\newblock {BML}: A high-performance, low-cost gradient synchronization
  algorithm for {DML} training.
\newblock In {\em NeurIPS}, 2018.

\bibitem{wang2019impact}
S.~Wang, D.~Li, J.~Geng, Y.~Gu, and Y.~Cheng.
\newblock Impact of network topology on the performance of dml: Theoretical
  analysis and practical factors.
\newblock In {\em IEEE INFOCOM}, 2019.

\bibitem{pmlr-v151-wang22e}
Y.~Wang, L.~Lin, and J.~Chen.
\newblock Communication-compressed adaptive gradient method for distributed
  nonconvex optimization.
\newblock In {\em AISTATS}, 2022.

\bibitem{wang2019performance}
Y.~Wang, Q.~Wang, S.~Shi, X.~He, Z.~Tang, K.~Zhao, and X.~Chu.
\newblock Benchmarking the performance and power of {AI} accelerators for {AI}
  training.
\newblock {\em arXiv preprint arXiv:1909.06842}, 2019.

\bibitem{GradSparforDisOptim}
J.~Wangni, J.~Wang, J.~Liu, and T.~Zhang.
\newblock Gradient sparsification for communication-efficient distributed
  optimization.
\newblock In {\em NeurIPS}, 2018.

\bibitem{convergeceAsynDisADMM}
E.~Wei and A.~E. Ozdaglar.
\newblock On the o(1=k) convergence of asynchronous distributed alternating
  direction method of multipliers.
\newblock In {\em 2013 IEEE Global Conference on Signal and Information
  Processing}, pages 551--554, Dec 2013.

\bibitem{TernGrad}
W.~Wen, C.~Xu, F.~Yan, C.~Wu, Y.~Wang, Y.~Chen, and H.~Li.
\newblock Terngrad: Ternary gradients to reduce communication in distributed
  deep learning.
\newblock In {\em NeurIPS}, 2017.

\bibitem{wen2017terngrad}
W.~Wen, C.~Xu, F.~Yan, C.~Wu, Y.~Wang, Y.~Chen, and H.~Li.
\newblock Terngrad: Ternary gradients to reduce communication in distributed
  deep learning.
\newblock In {\em NeurIPS}, 2017.

\bibitem{ECQSGD}
J.~Wu, W.~Huang, J.~Huang, and T.~Zhang.
\newblock Error compensated quantized {SGD} and its applications to large-scale
  distributed optimization.
\newblock In {\em ICML}, 2018.

\bibitem{disml}
E.~P. {Xing}, Q.~{Ho}, W.~{Dai}, J.~K. {Kim}, J.~{Wei}, S.~{Lee}, X.~{Zheng},
  P.~{Xie}, A.~{Kumar}, and Y.~{Yu}.
\newblock Petuum: A new platform for distributed machine learning on big data.
\newblock {\em IEEE Transactions on Big Data}, 1(2):49--67, June 2015.

\bibitem{XING2016179}
E.~P. Xing, Q.~Ho, P.~Xie, and D.~Wei.
\newblock Strategies and principles of distributed machine learning on big
  data.
\newblock {\em Engineering}, 2(2):179 -- 195, 2016.

\bibitem{DEF}
A.~Xu and H.~Huang.
\newblock Detached error feedback for distributed {SGD} with random
  sparsification.
\newblock In {\em ICML}, 2022.

\bibitem{xu2017performance}
P.~Xu, S.~Shi, and X.~Chu.
\newblock Performance evaluation of deep learning tools in docker containers.
\newblock In {\em 2017 3rd International Conference on Big Data Computing and
  Communications (BIGCOM)}, pages 395--403. IEEE, 2017.

\bibitem{blockCoordD}
Y.~Xu and W.~Yin.
\newblock A block coordinate descent method for regularized multiconvex
  optimization with applications to nonnegative tensor factorization and
  completion.
\newblock {\em SIAM J. Imaging Sciences}, 6:1758--1789, 2013.

\bibitem{yan2020optimizing}
D.~Yan, W.~Wang, and X.~Chu.
\newblock Optimizing batched winograd convolution on {GPUs}.
\newblock In {\em Proceedings of the 25th ACM SIGPLAN Symposium on Principles
  and Practice of Parallel Programming}, pages 32--44, 2020.

\bibitem{9834260}
G.~Yan, T.~Li, S.-L. Huang, T.~Lan, and L.~Song.
\newblock Ac-sgd: Adaptively compressed sgd for communication-efficient
  distributed learning.
\newblock {\em IEEE JSAC}, 2022.

\bibitem{yang2020survey}
S.~Yang, Y.~Wang, and X.~Chu.
\newblock A survey of deep learning techniques for neural machine translation.
\newblock {\em arXiv preprint arXiv:2002.07526}, 2020.

\bibitem{you2017scaling}
Y.~You, A.~Bulu{\c{c}}, and J.~Demmel.
\newblock Scaling deep learning on gpu and knights landing clusters.
\newblock In {\em Proceedings of the International Conference for High
  Performance Computing, Networking, Storage and Analysis}, pages 1--12, 2017.

\bibitem{you2019large}
Y.~You, J.~Li, S.~Reddi, J.~Hseu, S.~Kumar, S.~Bhojanapalli, X.~Song,
  J.~Demmel, K.~Keutzer, and C.-J. Hsieh.
\newblock Large batch optimization for deep learning: Training {BERT} in 76
  minutes.
\newblock In {\em ICLR}, 2020.

\bibitem{you2018imagenet}
Y.~You, Z.~Zhang, C.-J. Hsieh, J.~Demmel, and K.~Keutzer.
\newblock Imagenet training in minutes.
\newblock In {\em Proceedings of the 47th International Conference on Parallel
  Processing}, page~1. ACM, 2018.

\bibitem{pmlrv97yu19c}
H.~Yu and R.~Jin.
\newblock On the computation and communication complexity of parallel {SGD}
  with dynamic batch sizes for stochastic non-convex optimization.
\newblock In {\em ICML}, 2019.

\bibitem{pmlrv97yu19d}
H.~Yu, R.~Jin, and S.~Yang.
\newblock On the linear speedup analysis of communication efficient momentum
  {SGD} for distributed non-convex optimization.
\newblock In {\em ICML}, 2019.

\bibitem{Yu2018ParallelRS}
H.~Yu, S.~X. Yang, and S.~Zhu.
\newblock Parallel restarted sgd with faster convergence and less
  communication: Demystifying why model averaging works for deep learning.
\newblock In {\em AAAI}, 2018.

\bibitem{NIPS2019_8694}
Y.~Yu, J.~Wu, and L.~Huang.
\newblock Double quantization for communication-efficient distributed
  optimization.
\newblock In {\em NeurIPS}. 2019.

\bibitem{GlobalconverBlockCoordD}
J.~Zeng, T.~T.-K. Lau, S.~Lin, and Y.~Yao.
\newblock Global convergence of block coordinate descent in deep learning.
\newblock In {\em ICML}, 2019.

\bibitem{Zhang2014DSM}
C.~Zhang and C.~R{\'e}.
\newblock Dimmwitted: A study of main-memory statistical analytics.
\newblock {\em Proc. VLDB Endow.}, 7(12):1283--1294, Aug. 2014.

\bibitem{ZipML}
H.~Zhang, J.~Li, K.~Kara, D.~Alistarh, J.~Liu, and C.~Zhang.
\newblock {Z}ip{ML}: Training linear models with end-to-end low precision, and
  a little bit of deep learning.
\newblock In {\em ICML}, 2017.

\bibitem{zhang2017poseidon}
H.~Zhang, Z.~Zheng, S.~Xu, W.~Dai, Q.~Ho, X.~Liang, Z.~Hu, J.~Wei, P.~Xie, and
  E.~P. Xing.
\newblock Poseidon: An efficient communication architecture for distributed
  deep learning on {GPU} clusters.
\newblock In {\em USENIX ATC}, 2017.

\bibitem{Zhang2016ParallelSW}
J.~Zhang, C.~D. Sa, I.~Mitliagkas, and C.~R{\'e}.
\newblock Parallel sgd: When does averaging help?
\newblock {\em ArXiv}, abs/1606.07365, 2016.

\bibitem{zhang2023accelerating}
L.~Zhang, S.~Shi, X.~Chu, W.~Wang, B.~Li, and C.~Liu.
\newblock Accelerating distributed deep learning with fine-grained all-reduce
  pipelining.
\newblock In {\em IEEE ICDCS}, 2023.

\bibitem{zhang2023eva}
L.~Zhang, S.~Shi, and B.~Li.
\newblock {Eva}: Practical second-order optimization with kronecker-vectorized
  approximation.
\newblock In {\em The Eleventh International Conference on Learning
  Representations}, 2023.

\bibitem{zhang2023evaluation}
L.~Zhang, L.~Zhang, S.~Shi, X.~Chu, and B.~Li.
\newblock Evaluation and optimization of gradient compression for distributed
  deep learning.
\newblock In {\em IEEE ICDCS}, 2023.

\bibitem{AsynDisADMM}
R.~Zhang and J.~T. Kwok.
\newblock Asynchronous distributed admm for consensus optimization.
\newblock In {\em ICML}, 2014.

\bibitem{Zhang2015DLE296}
S.~Zhang, A.~Choromanska, and Y.~LeCun.
\newblock Deep learning with elastic averaging sgd.
\newblock In {\em NeurIPS}, 2015.

\bibitem{zhang2018taming}
X.~Zhang, J.~Liu, and Z.~Zhu.
\newblock Taming convergence for asynchronous stochastic gradient descent with
  unbounded delay in non-convex learning.
\newblock {\em arXiv preprint arXiv:1805.09470}, 2018.

\bibitem{6853589}
X.~{Zhang}, J.~{Trmal}, D.~{Povey}, and S.~{Khudanpur}.
\newblock Improving deep neural network acoustic models using generalized
  maxout networks.
\newblock In {\em ICASSP}, 2014.

\bibitem{ConverBlockCoordD}
Z.~Zhang and M.~Brand.
\newblock Convergent block coordinate descent for training tikhonov regularized
  deep neural networks.
\newblock In {\em NeurIPS}, 2017.

\bibitem{9721697}
Z.~Zhang and C.~Wang.
\newblock Mipd: An adaptive gradient sparsification framework for distributed
  dnns training.
\newblock {\em IEEE TPDS}, pages 3053--3066, 2022.

\bibitem{8644613}
Z.~{Zhang}, L.~{Yin}, Y.~{Peng}, and D.~{Li}.
\newblock A quick survey on large scale distributed deep learning systems.
\newblock In {\em ICPADS}, 2018.

\bibitem{Zhao2019GlobalMC}
S.-Y. Zhao, Y.~Xie, H.~Gao, and W.-J. Li.
\newblock Global momentum compression for sparse communication in distributed
  sgd.
\newblock {\em ArXiv}, abs/1905.12948, 2019.

\bibitem{zhao2020distributed}
W.~Zhao, D.~Xie, R.~Jia, Y.~Qian, R.~Ding, M.~Sun, and P.~Li.
\newblock Distributed hierarchical gpu parameter server for massive scale deep
  learning ads systems.
\newblock {\em MLSys}, 2020.

\bibitem{DorefaNet}
S.~Zhou, Y.~Wu, Z.~Ni, X.~Zhou, H.~Wen, and Y.~Zou.
\newblock Dorefa-net: Training low bitwidth convolutional neural networks with
  low bitwidth gradients.
\newblock {\em arXiv preprint arXiv:1606.06160}, 2016.

\bibitem{PSGD}
M.~A. Zinkevich, M.~Weimer, A.~Smola, and L.~Li.
\newblock Parallelized stochastic gradient descent.
\newblock In {\em NeurIPS}, 2010.

\bibitem{Zou2014MarianaTD}
Y.~Zou, X.~Jin, Y.~Li, Z.~Guo, E.~Wang, and B.~Xiao.
\newblock Mariana: Tencent deep learning platform and its applications.
\newblock {\em PVLDB}, 7:1772--1777, 2014.

\end{thebibliography}

\appendix

\end{document}